\documentstyle[aps,epsf,floats]{revtex}

\begin{document}
\draft

\title{Conversion of conventional gravitational-wave interferometers into QND
interferometers by modifying their input and/or output optics
}

\author{H.\ J.\ Kimble$^1$, Yuri Levin$^{2,5}$,
Andrey B.\ Matsko$^3$, Kip S.\ Thorne$^2$, and 
Sergey P.\ Vyatchanin$^4$} 

\address{$^1$Norman Bridge Laboratory of Physics 12-33, California Institute of
Technology, Pasadena, CA 91125}
\address{$^2$Theoretical Astrophysics, California Institute of
Technology, Pasadena, CA 91125}
\address{$^3$Department of Physics, Texas A\&M University, College Station, TX
77843-4242}
\address{$^4$Physics Faculty, Moscow State University, Moscow, Russia, 119899}
\address{$^5${\rm Current address:} Department of Astronomy, University of
California, Berkeley, CA 94720}

\date{Submitted to Physical Review D on 11 August 2000}

\maketitle

\begin{abstract}

The LIGO-II gravitational-wave interferometers 
($ca.\ 2006$--2008) are designed to have sensitivities near
the standard quantum limit (SQL) in the vicinity of 100 Hz.  This
paper describes and analyzes possible designs for subsequent, 
LIGO-III interferometers that can beat the SQL. 
These designs are identical to a 
conventional broad-band interferometer (without signal recycling),
except for 
new input and/or output optics.  Three designs are analyzed: (i) a
{\it squeezed-input interferometer} 
(conceived by Unruh
based on earlier work of Caves)
in which
squeezed vacuum with frequency-dependent (FD) squeeze angle is injected 
into the interferometer's dark port; (ii) a
{\it variational-output} interferometer (conceived in a
different form by Vyatchanin, Matsko and Zubova), in which
homodyne detection with FD homodyne phase is performed
on the output light; and (iii) a {\it
squeezed-variational interferometer} with squeezed input and
FD-homodyne output.  It is shown that the FD squeezed-input light can
be produced by sending ordinary squeezed light through two successive 
Fabry-Perot filter cavities before injection into the interferometer,
and FD-homodyne detection can be achieved by sending the output
light through two filter cavities before ordinary homodyne detection.  
With anticipated technology (power squeeze factor $e^{-2R}=0.1$ for input
squeezed vacuum and net fractional 
loss
of signal power in arm cavities
and output optical train $\epsilon_* = 0.01$) and using an input laser power 
$I_o$ in units of that required to reach the SQL (the planned LIGO-II power,
$I_{\rm SQL})$, the three types 
of interferometer could beat the amplitude SQL at 100 Hz by the following 
amounts 
$\mu \equiv \sqrt{S_h}/\sqrt{S_h^{\rm SQL}}$ and
with the following corresponding increase 
${\cal V} = 1/\mu^3$ in the volume of the
universe that can be searched for a given non-cosmological source:

\noindent
{\it Squeezed-Input} --- 
$\mu \simeq \sqrt{e^{-2R}} \simeq 0.3$ and ${\cal V} \simeq
1/0.3^3 \simeq 30$ using $I_o/I_{\rm SQL} = 1$.  

\noindent
{\it Variational-Output} --- 
$\mu \simeq \epsilon_*^{1/4} 
\simeq 0.3$ and ${\cal V} \simeq 30$ but only if the optics can handle a
ten times larger power: $I_o/I_{\rm SQL} \simeq 1/\sqrt{\epsilon_*} = 10$.

\noindent
{\it Squeezed-Varational} --- $\mu
= 1.3 (e^{-2R}\epsilon_*)^{1/4} \simeq 0.24$ and ${\cal V} \simeq 80$ using
$I_o/I_{\rm SQL} = 1$; and $\mu \simeq (e^{-2R}\epsilon_*)^{1/4}
\simeq 0.18$ and ${\cal V} \simeq 180$ using $I_o/I_{\rm SQL} =
\sqrt{e^{-2R}/\epsilon_*} \simeq 3.2$.   

\end{abstract}


\narrowtext
\twocolumn
\section{Introduction and summary}
\label{sec:Introduction}

In an interferometric gravitational-wave detector, laser light is used to
monitor the motions of mirror-endowed
test masses, which are driven by gravitational waves $h(t)$.
The light produces two types of noise: photon {\it shot noise}, which it
superposes on the interferometer's output signal, and fluctuating {\it
radiation-pressure noise}, by which it pushes the test masses in random
a manner that can mask their gravity-wave-induced motion.  
The shot-noise spectral density scales 
with the light power $I_o$ entering the interferometer
as $S_h^{\rm shot} \propto 1/I_o$; the
radiation-pressure noise scales as $S_h^{\rm rp} \propto I_o$.

In the first generation of kilometer-scale interferometers (e.g., LIGO-I,
2002--2003 \cite{Ligo}), the laser power
will be low enough that shot-noise dominates and radiation-pressure noise is
unimportant.  Tentative plans for the next generation interferometers
(LIGO-II, ca.\ 2006--2008) include increasing $I_o$ to the point that, 
$S_h^{\rm rp} = S_h^{\rm shot}$ at the interferometers' optimal 
gravitational-wave frequency, $\Omega /2\pi \sim 100$ Hz.  The resulting
net noise $S_h = S_h^{\rm rp} + S_h^{\rm shot} = 2 S_h^{\rm shot}$ is the
lowest that can be achieved with conventional interferometer designs.  Further
{\it increases} of light power will drive the radiation-pressure on upward,
increasing the net noise, while {\it reductions} of light power will drive 
the shot noise upward, also increasing the net noise.

This minimum achievable noise is called the ``Standard Quantum Limit'' (SQL)
\cite{SQL} and is denoted $S_h^{\rm SQL} \equiv h_{\rm SQL}^2$.  It can be 
regarded as arising 
from the effort of the
quantum properties of the light to enforce the Heisenberg uncertainty principle
on the interferometer test masses, in just the manner of the Heisenberg
microscope.  Indeed, a common derivation of the SQL is based on the uncertainty
principle for the test masses' position and momentum 
\cite{QuantumMeasurement}: 
The light makes a sequence of measurements of the difference $x$ of
test-mass positions. 
If a measurement is too accurate, then by state reduction
it will narrow the test-mass wave function so tightly ($\Delta x$ very small)
that the momentum becomes highly uncertain (large $\Delta p$), producing a
wave-function spreading 
that is so rapid as to create great
position uncertainty at the time of the next measurement. There is an
optimal accuracy for the first measurement---an accuracy that produces only a 
factor $\sqrt2$
spreading and results in optimal predictability for the next measurement.
This optimal accuracy corresponds to $h_{\rm SQL}$.

Despite this {\it apparent} intimate connection of the SQL to test-mass 
quantization, it turns out that the test-mass quantization has {\it no
influence whatsoever} on the output noise in gravitational-wave interferometers
\cite{TestMassQM}.  The 
sole forms of quantum noise in the output are photon
shot noise and photon radiation-pressure noise.\footnote{In brief, the
reasons for this are the following: 
The interferometer's measured output, in general,
is one quadrature of the electric field [the $b_\zeta$ of Eqs.\ 
(\ref{bzetaDef}) and (\ref{E1E2Def}) below], and this output observable
commutes with itself at different times by virtue of Eqs. (\ref{a12Commutator})
with $a \to b$.  This means that the digitized data points (collected at a rate
of 20 kHz) are mutually commuting Hermitian observables.  One consequence
of this is that reduction of the state of the interferometer
due to data collected at one moment of time will not influence the data
collected at any later moment of time. Another consequence is that,
when one Fourier analyzes the interferometer output, one puts all information
about the initial states of the test masses 
into data points near zero frequency,
and when one then filters the output to remove 
low-frequency noise (noise at $f = \Omega/2\pi
\alt 10$ Hz), one thereby removes from the data all information about the 
test-mass initial states; the only remaining test-mass information is that
associated with Heisenberg-picture
changes of the test-mass positions at $f\agt 10$ Hz, changes
induced by external forces: light pressure (which is quantized) and  
thermal- and seismic-noise forces (for which quantum effects are unimportant).
See Ref.\ \cite{TestMassQM} for further detail.}

Vladimir Braginsky (the person who first recognized the existence of the SQL
for gravitational-wave detectors and other high-precision measuring
devices \cite{SQLBraginsky}) realized, in the mid 1970s, that the SQL can be
overcome, but to do so would require significant modifications of the
experimental design.  Braginsky gave the name Quantum Nondemolition (QND)
to devices
that can beat the SQL; this name indicates the ability of QND devices to 
prevent their own 
quantum properties from demolishing the information one is 
trying to extract \cite{QNDBraginsky}. 

The LIGO-I interferometers are now being assembled at the LIGO sites, in
preparation for the first LIGO gravitational-wave searches.  In parallel,
the LIGO Scientific Community (LSC) is deeply immersed in R\&D for the
LIGO-II interferometers \cite{WhitePaper}, and a small portion of the 
LSC is attempting to invent practical designs for the third generation
of interferometers, LIGO-III.  
This paper is a contribution to the LIGO-III design effort.

In going from LIGO-II to LIGO-III, a large number of noise sources
must be reduced.  Perhaps the most serious are the photon shot noise and
radiation pressure noise (``optical noise''), and thermal noise
in the test masses and their suspensions \cite{WhitePaper,ThermalNoise}.  
In this paper
we shall deal solely with the shot noise and radiation pressure noise (and the
associated SQL); we shall tacitly assume that all other noise sources,
including thermal noise, can be reduced sufficiently to take full advantage
of the optical techniques that we propose and analyze.

Because LIGO-II 
is designed to
operate at the SQL, in moving to
LIGO-III there are just two ways to reduce the optical noise:
increase the masses $m$ of the mirrored test masses (it turns out that
$h_{\rm SQL}^2 \propto 1/m$), or redesign the interferometers so they can
perform QND.  The transition from LIGO-I to LIGO-II will already (probably)
entail a mass increase, from $m=11$ kg to $m=30$ kg, in large measure because
the SQL at 11 kg was unhappily constraining \cite{WhitePaper}.  Any 
large further mass increase would entail great danger of unacceptably
large noise due to energy coupling through the test-mass suspensions and into 
or from the overhead supports
(the seismic isolation system);
a larger mass would also entail practical 
problems
due to the increased test-mass dimensions.  Accordingly, there is strong
motivation for trying to pursue the QND route.

Our Caltech and Moscow-University research groups are jointly exploring three
approaches to QND interferometer design: 

\begin{itemize}
\item The conversion of conventional
interferometers into QND interferometers by modifying their input and/or output
optics [this paper].  
This approach
achieves QND by creating and manipulating correlations between
photon shot noise and radiation pressure noise; see below.  It is the
simplest of our three approaches,
but has one serious drawback: an
uncomfortably high light power, 
$W_{\rm circ} \agt 1$ MW,
that must
circulate inside the interferometers' arm cavities
\cite{KhaliliAmaldi}.  
It is not clear whether
the test-mass mirrors can be improved sufficiently to handle this high a
power
in a sufficiently noise-free way.

\item A modification of the interferometer design (including using two
optical cavities in each arm) so as to make its output signal be proportional
to the relative speeds of the test masses rather than their relative positions
\cite{SpeedMeterApp,Purdue}.  
Since the test-mass speed is proportional to momentum, and momentum
(unlike position) is very nearly conserved under free test-mass evolution
on gravity-wave timescales ($\sim 0.01$ sec), the relative speed is 
very nearly a ``QND observable'' \cite{QNDObservable}
and thus is beautifully suited to QND
measurements.  Unfortunately, the resulting {\it speed-meter interferometer},
like our input-output-modified interferometers, 
suffers from a high circulating light power 
\cite{KhaliliAmaldi}, 
$W_{\rm circ} \agt 1$ MW. 

\item Radical redesigns of the interferometer aimed at achieving QND
performance with $W_{\rm circ}$ well below 1 MW 
\cite{LowPowerQND}.  These, 
as currently conceived by Braginsky, Gorodetsky and Khalili, entail 
transfering the gravitational-wave signal to a
single, small test mass via light pressure, and using a local QND sensor to
read out the test mass's motions relative to a local inertial frame.  
\end{itemize}

In this paper we explore the first approach.  The foundation for this approach
is the realization that: (i) photon shot noise and radiation-pressure
noise together enforce the SQL {\it only if they are uncorrelated}; see,
e.g., Ref.\ \cite{TestMassQM}; (ii) whenever carrier light with side bands
reflects off a mirror
(in our case, the mirrors of an interferometer's arm cavities), the
reflection {\it ponderomotively squeezes} the light's side bands, thereby 
creating correlations between their radiation-pressure noise in one quadrature 
and shot noise in the other; (iii) these
correlations are not accessed by a conventional interferometer because
of the particular quadrature that its photodiode measures; 
(iv) however, these correlations
{\it can} be accessed by (conceptually) simple modifications of the
interferometer's input and/or output optics, and by doing so one can
beat the SQL.
These correlations were first noticed explicitly by Unruh \cite{Unruh},
but were present implicitly in Braginsky's earlier
identification of the phenomenon of ponderomotive squeezing
\cite{BraginskySqueeze,Rudenko}.  

\begin{figure}
\epsfxsize=3.2in\epsfbox{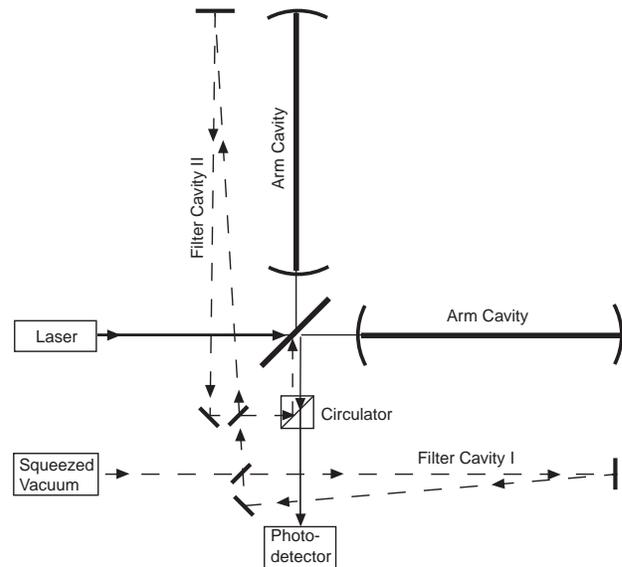}
\caption{
Schematic diagram of a squeezed-input interferometer.  
\label{fig:Fig1}
}
\end{figure}

In this paper we study three variants of QND interferometers that rely
on ponderomotive-squeeze correlations:

{\it (i) Squeezed-Input Interferometer:}  Unruh \cite{Unruh} 
(building on earlier work of Caves \cite{Caves}) 
invented
this design nearly 20 years ago, and since then it has been reanalyzed 
by several other researchers \cite{JaekelReynaud,PCW}.  In this design,
squeezed vacuum is sent into the dark port of the interferometer (``modified
input'') and the output light is monitored with a photodetector as in
conventional interferometers.  

For a broad-band squeezed-input  
interferometer, the squeeze angle must be a specified
function of frequency that changes significantly across the interferometer's
operating gravity-wave band.  
(This contrasts with past experiments 
employing squeezed light to enhance interferometry \cite{Xaio,Grangier},
where the squeeze angle was constant across the operating band.)
Previous papers on squeezed-input interferometers
have ignored the issue of how,
in practice, one might achieve the required frequency-dependent (FD) 
squeeze angle.  In Sec.\ \ref{sec:FDSqueeze}, we 
show that it can be produced via 
ordinary, frequency-independent
squeezing (e.g., by nonlinear optics \cite{JOSASqueeze}),
followed by filtration through
two Fabry-Perot cavities with suitably adjusted bandwidths and
resonant-frequency offsets from the light's carrier frequency.  
A schematic
diagram of the resulting squeezed-input interferometer is shown in Fig.\
\ref{fig:Fig1} and is discussed in detail below.  
Our predicted
performance for such an interferometer agrees with that of previous research.

\begin{figure}
\epsfxsize=3.2in\epsfbox{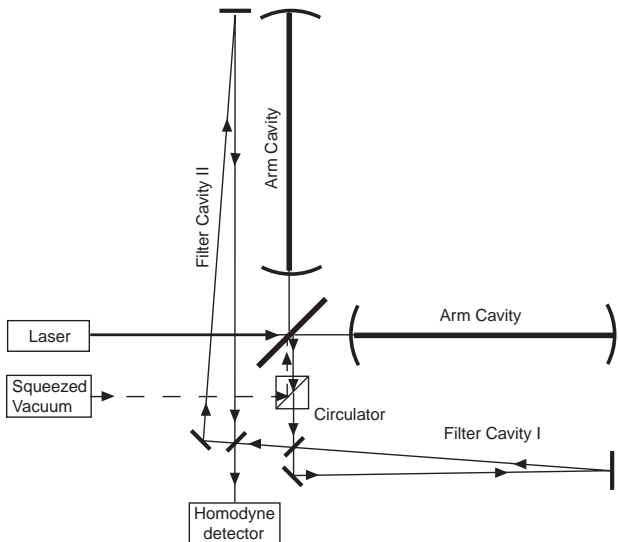}
\caption{
Schematic diagram of a squeezed-variational interferometer.
A variational-output interferometer differs from this solely by
replacing the input squeezed vacuum by ordinary vacuum.
\label{fig:Fig2}
}
\end{figure}

{\it (ii) Variational-Output Interferometer:}  Vyatchanin, Matsko and Zubova
invented this design conceptually in the early 1990's 
\cite{VMZ,VM1,VM2}.  
It entails a
conventional interferometer input (ordinary vacuum into the dark port), but a
modified output: instead of photodetection, one performs homodyne detection
with a homodyne phase that depends on frequency in essentially the same way as
the squeeze angle of a squeezed-input interferometer.  Vyatchanin, Matsko and
Zubova did not know how to achieve FD homodyne detection in practice, so
they proposed approximating it by homodyne detection with a time-dependent
(TD) homodyne phase.  Such TD homodyne detection can beat the SQL, 
but (by contrast with FD homodyne)
it is not well-suited to gravitational-wave searches, where little
is known in advance about the gravitational waveforms or their arrival 
times. 
In this paper (Sec.\ \ref{sec:FDHomodyne} and Appendix \ref{app:filters}),
we show that the desired FD homodyne detection can
be achieved by sending the interferometer's
output light through two successive Fabry-Perot cavities that are essentially
identical to those needed in our variant of a squeezed-input interferometer,
and by then performing conventional homodyne detection with fixed 
homodyne angle.  
A schematic diagram of the resulting variational-output
interferometer is shown in Fig.\ \ref{fig:Fig2}. 

{\it (iii) Squeezed-Variational Interferometer:}  This design (not considered
in the previous literature\footnote{A 
design similar to it has previously been proposed and analyzed \cite{VM1}
for a simple optical meter, in which the position of a movable mirror (test
mass) is monitored by measuring the phase or some other quadrature of a light
wave reflected from the mirror.  In this case it was shown that 
the SQL can be beat by a combination of phase-squeezed input light 
and TD homodyne detection.}) 
is the obvious
combination of the first two; one puts squeezed vacuum into the dark port and
performs FD homodyne detection on the output light. 
The optimal 
performance is achieved by squeezing the input at a fixed 
(frequency-independent) angle; filtration cavities are needed only at the
output (for the FD homodyne detection) and not at the input; 
cf.\ Fig.\ \ref{fig:Fig2}.  

In Sec.\ \ref{sec:LosslessPerformance} 
we compute the spectral density of the noise for all three designs, ignoring
the effects of optical 
losses.
We find (in agreement with previous
analyses \cite{JaekelReynaud,PCW}) that,
when the FD squeeze angle is optimized,
the squeezed-input
interferometer has its shot noise and radiation-pressure noise both reduced in
amplitude 
(at fixed light power)
by $e^{-R}$, where $R$ is the (frequency-independent)
squeeze factor; see Fig.\ 2 below.
This enables a lossless
squeezed-input interferometer to beat the SQL by a factor
$e^{-R}$ (when the power is optimized) but no more.
By contrast, the lossless, variational-output interferometer,
with optimized FD homodyne phase, can have its
radiation-pressure noise completely removed from the output signal, and its
shot noise will scale with light power as $1/\sqrt{I_o}$ as for a conventional
interferometer.  As a result, the lossless
variational-output interferometer can beat
the SQL in amplitude
by $\sqrt{I_{\rm SQL}/2I_o}$, where $I_{\rm SQL}$ is the light power
required by a conventional interferometer to reach the SQL. 
The optimized, lossless, squeezed-variational interferometer has its
radiation-pressure noise completely removed, and its shot noise reduced by 
$e^{-R}$, so it can beat the SQL in amplitude by 
$e^{-R} \sqrt{I_{\rm SQL}/2I_o}$.  

Imperfections in squeezing, in the filter cavities, and in the homodyne 
local-oscillator phase will produce errors 
$\Delta\lambda$ in the FD squeeze angle $\lambda(\Omega)$ of a squeezed-input
or squeezed-variational interferometer, and $\Delta\zeta$
in the FD homodyne phase $\zeta(\Omega)$ of a variational-output or
squeezed-variational interferometer.
At the end of Sec.\
\ref{sec:ComputationSpectra}, 
we shall show that, to keep these
errors from seriously compromising the 
most promising interferometer's performance,
$|\Delta\lambda|$ must be no larger than $\sim 0.05$ radian, and 
$|\Delta\zeta|$ must be no larger than $\sim 0.01$ radian.
This translates into constraints of order five percent on the accuracies
of the filter cavity
finesses and about 0.01 on their fractional frequency offsets
and  on the
homodyne detector's local-oscillator phase. 

The performance will be 
seriously constrained by unsqueezed vacuum that leaks into the 
interferometer's optical train at all locations where there 
are optical losses, whether those losses are fundamentally irreversible (e.g.\
absorption) or reversible (e.g.\ finite transmissivity of an arm cavity's end
mirror).  We explore the effects of such optical
losses 
in Sec.\ \ref{sec:LossyInterferometer}.
The dominant losses and
associated noise production occur in 
the interferometer's arm cavities and FD filter cavities. 
The filter cavities' net losses and noise will
dominate unless the number of bounces the light makes in them is minimized by
making them roughly as long as the arm cavities.  This suggests that they be
4km long and reside in the beam tubes alongside the interferometer's arm
cavities.  To separate the filters' inputs and outputs,
they might best be triangular cavities with two mirrors at the corner station
and one in the end station.  

Our loss calculations reveal the following:  

The {\it squeezed-input}
interferometer is little affected by losses in the interferometer's arm
cavities or in the output optical train. However, losses in the input optical 
train (most seriously the filter cavities and a circulator) 
influence the noise by constraining the net squeeze
factor $e^{-2R}$ of the light entering the arm cavities.  The resulting
noise, expressed in terms of $e^{-2R}$, is the same as in a lossless 
squeezed-input interferometer (discussed above):  With the light power
optimized so $I_o=I_{\rm SQL}$, the squeezed-input interferometer can beat the
amplitude SQL by a factor $\mu\equiv \sqrt{S_h}/\sqrt{S_h^{\rm SQL}}
\simeq \sqrt{e^{-2R}} \simeq 0.3$ (where $e^{-2R}\simeq 0.1$ is a likely
achievable value of the power squeeze factor).  

The {\it variational-output} and {\it squeezed-variational} interferometers are
strongly affected by losses in the interferometer's arm cavities and in the
output optical train (most seriously: a circulator, the two filter cavities,
the mixing with the homodyne detector's local-oscillator field, and the
photodiode inefficiency).  The net fractional loss $\epsilon_*$ of 
signal power 
and (for squeezed-variational) the squeeze factor $e^{-2R}$ for input power   
together determine the interferometer's optimized performance:   
The amplitude SQL can be beat by an amount
$\mu = (e^{-2R}\epsilon_*)^{1/4}$, and the input laser power required to
achieve this optimal performance is $I_o/I_{\rm SQL} \simeq
\sqrt{e^{-2R}/\epsilon_*}$.  
In particular, the variational-output interferometer
(no input squeezing; $e^{-2R}=1$), with the likely achievable loss level
$\epsilon_*=0.01$, can beat the SQL by the same amount as our estimate for the
squeezed-input interferometer, $\mu \simeq \epsilon_*^{1/4} \simeq
0.3$, but requires ten times higher
input optical power, $I_o/I_{\rm SQL} \simeq 1/\sqrt{\epsilon_*} 
\simeq 10$ --- which could be a very
serious problem.  By contrast, the squeezed-variational interferometer
with the above parameters has an optimized performance 
$\mu \simeq (0.1 \times 0.01)^{1/4} \simeq 0.18$ (substantially better than
squeezed-input or variational-output), and achieves this with an optimizing
input power $I_o/I_{\rm SQL} = \sqrt{0.1/0.01} \simeq 3.2$.  If the input power
is pulled down from this optimizing value to $I_o/I_{\rm SQL} = 1$ so it is the
same as for the squeezed-input interferometer, then the squeezed-variational
performance is debilitated by a factor 1.3, to $\mu \simeq 0.24$, which is
still somewhat better than for squeezed-input.

It will require considerable R\&D to actually achieve performances at the above
levels, and there could be a number of unknown pitfalls along the way.
For example, ponderomotive squeezing, which 
underlies all three of our QND configurations, has never yet been seen in the
laboratory and may entail unknown technical difficulties. 

Fortunately, the technology for producing squeezed vacuum via nonlinear optics
is rather well developed \cite{JOSASqueeze}
and has even been used to
enhance the performance of interferometers \cite{Xaio,Grangier}.  Moreover,
much effort is being invested in the development of low-loss test-mass
suspensions, and this gives the prospect for new (ponderomotive)
methods of generating
squeezed light that may perform better than traditional nonlinear optics.
These facts, plus the fact that, in a squeezed-input
configuration, the output signal is
only modestly squeezed and thus is not nearly so delicate as the
highly-squeezed output of an optimally performing 
squeezed-variational configuration,
make us feel more confident 
of success with squeezed-input interferometers than
with squeezed-variational ones.  

On the other hand,
the technology for a
squeezed-variational interferometer is not much different from that for 
a squeezed-input one: 
Both require input squeezing and both require filter cavities with roughly the 
same specifications; the only significant differences are the need for
conventional, frequency-independent homodyne detection in the
squeezed-variational interferometer, and its higher-degree of output
squeezing corresponding to higher sensitivity.
Therefore, the 
squeezed-variational
interferometer may turn out to be just as practical as the squeezed-input,
and may achieve significantly better overall performance at the same 
laser power.

This paper is organized as follows:  In Sec.\ \ref{InterferometerDescription}
we sketch our mathematical description of the interferometer, including our
use of the Caves-Schumaker \cite{CavesSchumaker,SchumakerCaves} formalism for
two-photon quantum optics, including light squeezing (cf.\ Appendix 
\ref{app:2photon}); and we write down 
the interferometer's input-output
relation in the absence of losses [Eq.\ (\ref{bjFromaj}); cf.\ Appendix
\ref{app:interferometer} for derivation].  In Sec.\ 
\ref{sec:ConventionalInterferometer}, relying on our general lossless 
input-output
relation (\ref{bjFromaj}), we derive the noise spectral density $S_h(f)$ for a
conventional interferometer and elucidate thereby the SQL.  In
Sec.\ \ref{sec:LosslessPerformance}, we describe mathematically our three
QND interferometer designs and, using
our lossless input-output relation (\ref{bjFromaj}), derive their lossless
noise
spectral densities.  In Sec.\ \ref{sec:FDHomodyne}, we show that FD homodyne
detection can be achieved by filtration followed by conventional homodyne
detection, and in Appendix \ref{app:filters} we show that the required
filtration can be achieved by sending the light through two successive
Fabry-Perot cavities with suitably chosen cavity parameters.  We list and
discuss the required cavity parameters in Sec.\ \ref{sec:FDHomodyne}.
In Sec.\ \ref{sec:LossyInterferometer}, we compute the effects of optical
losses on the interferometers' noise spectral density; 
our computation relies on an input-output
relation (\ref{bjFromajLossy}) and (\ref{DeltabjLossy})
derived in Appendix \ref{app:interferometer}.  In Sec.\ \ref{sec:Discussion}
we discuss and compare the noise performances of our three types of 
inteferometers.  
Finally, in Sec.\
\ref{sec:Conclusions} we briefly recapitulate and then list and briefly
discuss a number of issues that need study, as foundations for possibly
implementing these QND interferometers in LIGO-III.

This paper assumes that the reader is familiar with modern quantum optics 
and its theoretical tools as presented, for example, in Refs.\ \cite{QORefs}.

\section{Mathematical Description of the Interferometer}
\label{InterferometerDescription}

\subsection{Input and Output fields}

\begin{figure}
\epsfxsize=3.2in\epsfbox{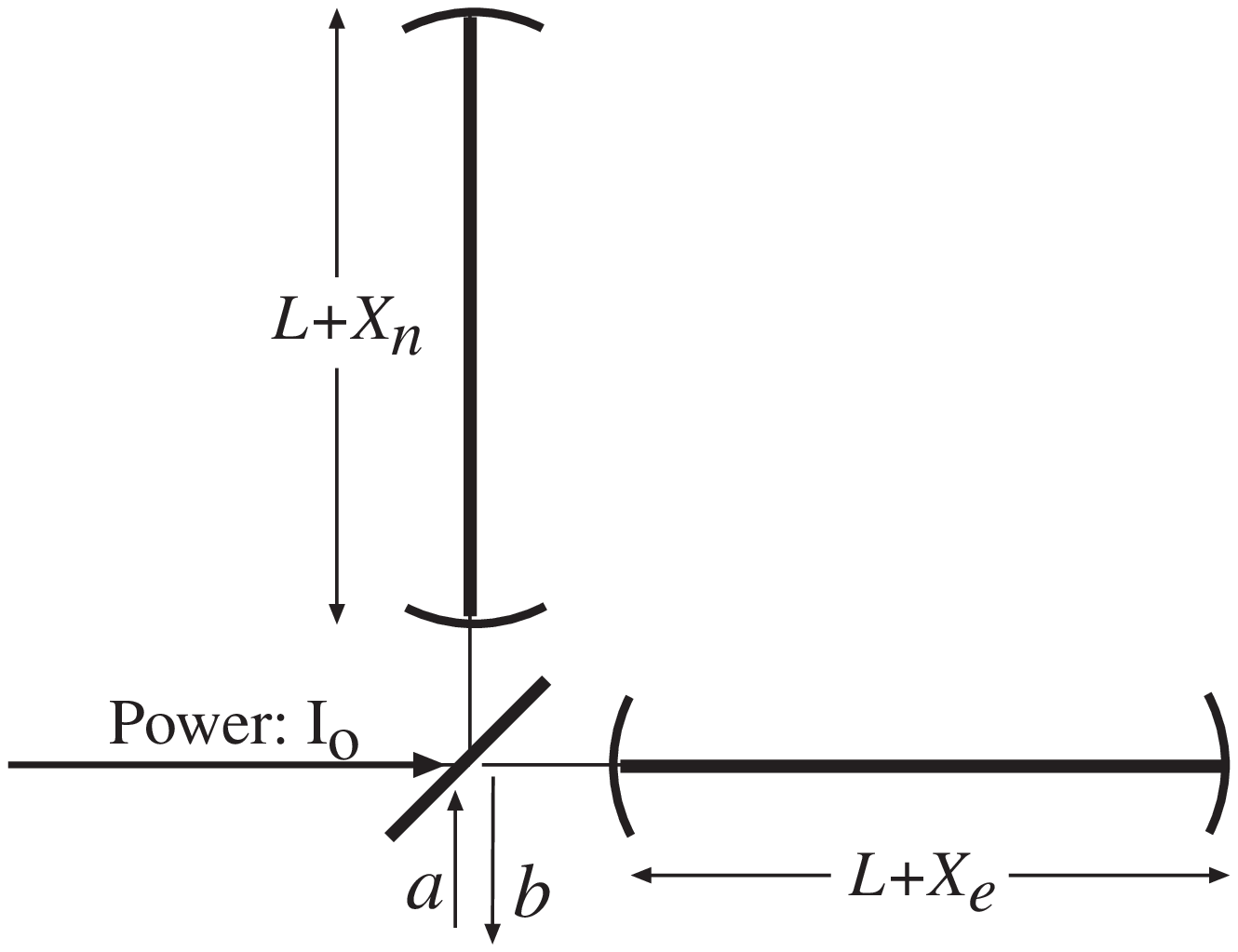}
\caption{Gravitational-wave interferometer with two inputs (the carrier which
has power $I_o$ entering the bright port, and quantum field $a$ entering the
dark port) and one relevant output 
(the quantum field $b$ leaving the dark
port). 
\label{fig:Fig3}
}
\end{figure}

Figure \ref{fig:Fig3} shows the standard configuration for a 
gravitational-wave interferometer.  In this subsection we focus on the 
beam splitter's input and output.  In our equations we idealize the beam
splitter as infinitesimally thin and write the input and output fields as
functions of time (not time and position) at the common centers of the beams as
they strike the splitter.  

At the beam splitter's bright port the input is a carrier field, presumed
to be in a perfectly coherent state with power
$I_o \sim 10$ kW (achieved via power recycling\cite{PowerRecycle}), 
angular frequency $\omega_o \simeq 1.78 \times 10^{15} {\rm sec}^{-1}$
(1.06 micron light), and excitation confined to the $\cos(\omega_o t)$
quadrature 
(i.e.,  the mean field arriving at the beam splitter
is proportional to $\cos(\omega_o t)$). 

At the dark port the input is a (quantized) 
electromagnetic field with the positive-frequency part of the electric field
given by the standard expression
\begin{equation}
E^{(+)}_{\rm in} 
= \int_0^\infty \sqrt{{2\pi\hbar\omega \over {\cal A}c}}\; a_\omega
e^{-i\omega t} {d\omega \over 2\pi}\;.
\label{EPlusIn}
\end{equation}
Here $\cal A$ is the effective cross sectional area of the beam and 
$a_\omega$ is the annihilation operator, whose commutation relations are
\begin{equation}
[a_\omega, a_{\omega'}] = 0\;, \quad [a_\omega, a_{\omega'}^{\dag} ] = 2\pi
\delta(\omega-\omega')\;.
\label{aomegaCommutator}
\end{equation}
Throughout this paper we use the Heisenberg Picture, so $E^{(+)}$ evolves with
time as indicated.  However, our creation and annihilation operators $a_\omega$
and $a_\omega^{\dag}$ are fixed in time, with their usual Heisenberg-Picture
time evolutions always factored out explicitly as in Eq.\ (\ref{EPlusIn}).  

We split the field (\ref{EPlusIn})
into side bands about the carrier frequency $\omega_o$, 
$\omega = \omega_o \pm \Omega$, with side-band frequencies $\Omega$ 
in the gravitational-wave range $\sim 60$ to $\sim 6000 {\rm sec}^{-1}$ (10 to
1000 Hz), and we define
\begin{equation}
a_+ \equiv a_{\omega_o + \Omega}\;, \quad a_- \equiv a_{\omega_o - \Omega}\;.
\label{apmDef}
\end{equation}  
As in Eq.\ (\ref{aomegaCommutator}), we continue to use a
prime on the subscript to denote frequency $\Omega'$: $a_{+'} 
\equiv a_{\omega_o
+ \Omega'}$.  Correspondingly, the commutation relations
(\ref{aomegaCommutator}) imply for the only nonzero commutators
\begin{equation}
[a_+, a_{+'}^{\dag}] = 2\pi \delta(\Omega-\Omega')\;, \;\;
[a_-, a_{-'}^{\dag}] = 2\pi \delta(\Omega-\Omega')\;;
\label{apmCommutator}
\end{equation}
and expression (\ref{EPlusIn}) for the dark-port input field becomes
\begin{equation}
E^{(+)}_{\rm in} = \sqrt{2\pi\hbar\omega_o\over{\cal A}c} e^{-i\omega_o t}
\int_0^\infty \left(a_+ e^{-i\Omega t} + a_- e^{+i\Omega t} \right)
{d\Omega\over2\pi}\;.
\label{EPlusIn1}
\end{equation}
Here (and throughout this paper) we approximate 
$\omega_0\pm\Omega \simeq \omega_o$ 
inside the square root, since 
$\Omega/\omega_o \sim 3\ 10^{-13}$  
is so small; and we formally extend the integrals over 
$\Omega$ to infinity, for ease of notation.

Because the radiation pressure in the optical cavities produces squeezing, and
because this ponderomotive squeezing is central to the operation of our 
interferometers, we shall find it convenient to think about the interferometer
not in terms of the single-photon modes, whose annihilation operators are $a_+$
and $a_-$, but rather in terms of the correlated
two-photon modes (Appendix A and 
Refs.\ \cite{CavesSchumaker,SchumakerCaves}) whose field amplitudes are
\begin{equation}
a_1 = {a_+ + a_-^{\dag} \over \sqrt2}\;, \quad 
a_2 = {a_+ - a_-^{\dag} \over \sqrt2 i}\;.
\label{a12Def}
\end{equation}
The commutation relations (\ref{apmCommutator}) imply the following values for
the commutators of these field amplitudes and their adjoints:
\begin{mathletters}
\begin{equation}
[a_1, a_{2'}^{\dag}] = - [a_2, a_{1'}^{\dag}]\; 
= i2\pi\delta(\Omega-\Omega') 
\end{equation}
and all others vanish (though some would be of order $(\Omega/\omega_o)$ if we
had not approximated $\omega_o\pm\Omega \simeq \omega_o$ inside the square root
in Eq.\ (\ref{EPlusIn1}); cf.\ \cite{CavesSchumaker,SchumakerCaves}):
\begin{equation}
\;[a_1,a_{1'}] = [a_1,a_{1'}^{\dag}] = [a_1^{\dag}, a_{1'}^{\dag}] 
= [a_1,a_{2'}] = [a_1^{\dag},a_{2'}^{\dag}] = 0\;, 
\end{equation}
\label{a12Commutator}
\end{mathletters}
and similarly with $1 \leftrightarrow 2$.  In terms of these two-photon 
amplitudes, 
Eq.\ (\ref{EPlusIn1}) and $E^{(-)} = E^{(+)\dag}$ imply that
the full electric field operator for the dark-port input is 
\begin{eqnarray}
E_{\rm in} &=& E^{(+)}_{\rm in} + E^{(-)}_{\rm in} \nonumber\\
&=& \sqrt{4\pi\hbar\omega_o\over{\cal A}c} \left[ 
\cos(\omega_o t)
\int_0^\infty \left(a_1 e^{-\i\Omega t} + a_1^{\dag} e^{+i\Omega t}
\right) {d\Omega\over2\pi}\right. \nonumber\\
&&\quad\quad + \left. \sin(\omega_o t) 
\int_0^\infty \left(a_2 e^{-\i\Omega t} + a_2^{\dag} e^{+i\Omega t}
\right) {d\Omega\over2\pi} \right]\;.
\label{EIn}
\end{eqnarray}
Thus, we see that $a_1$ is the field amplitude for photons in the
$\cos\omega_o t$ quadrature and $a_2$ is that for photons in the $\sin\omega_o
t$ quadrature \cite{CavesSchumaker,SchumakerCaves}.  These and other
quadratures will be central to our analysis.

The output field at the beam splitter's dark port is described by the same
equations as the input field, but with the annihilation operators $a$ replaced
by $b$; for example,
\begin{eqnarray}
E_{\rm out} 
&=& \sqrt{4\pi\hbar\omega_o\over{\cal A}c} \left[
\cos(\omega_o t)
\int_0^\infty \left(b_1 e^{-\i\Omega t} + b_1^{\dag} e^{+i\Omega t}
\right) {d\Omega\over2\pi}\right. \nonumber\\
&&\quad\quad + \left. \sin(\omega_o t)
\int_0^\infty \left(b_2 e^{-\i\Omega t} + b_2^{\dag} e^{+i\Omega t}
\right) {d\Omega\over2\pi} \right]\;.
\label{EOut}
\end{eqnarray}
We shall find it convenient to introduce explicitly the cosine and sine
quadratures of the output field, $E_1(t)$ and $E_2(t)$, defined by
\begin{eqnarray}
E_{\rm out} &=& E_1(t) \cos(\omega_o t) + E_2 (t) \sin(\omega_o t)\;; 
\nonumber\\
E_j(t) &=& \sqrt{4\pi\hbar\omega_o\over{\cal A}c} \int_0^\infty 
\left(b_j e^{-\i\Omega t} + b_j^{\dag} e^{+i\Omega t}
\right) {d\Omega\over2\pi}\;.
\label{E1E2Def}
\end{eqnarray}

\begin{table}
\caption{Interferometer parameters and their fiducial values. 
}
\vskip15pt
\begin{tabular}{lll}
Parameter&Symbol&Fiducial Value\\
\tableline
light frequency & $\omega_o$ & $1.8 \times 10^{15} {\rm s}^{-1}$ \\
arm cavity ${1\over2}$-bandwidth & $\gamma$ & $2\pi\times 100 {\rm s}^{-1}$ \\ 
grav'l wave frequency & $\Omega$ & --- \\
mirror mass & $m$ & 30 kg \\
arm length & $L$ & 4 km \\
light power to beam splitter & $I_o$ & --- \\
light power to reach SQL & $I_{\rm SQL}$ & $1.0 \times 10^4$ W \\
grav'l wave SQL & $h_{\rm SQL}$ & $2\times 10^{-24}(\gamma/\Omega) {\rm
Hz}^{-{1/2}}$ \\
opto-mech'l coupling const & ${\cal K}$ & 
${(I_o/I_{\rm SQL})2\gamma^4 \over \Omega^2(\gamma^2 + \Omega^2)}$ \\
fractional signal-power loss & $\epsilon_*$ &  0.01 \\ 
max power squeeze factor & $e^{-2R}$ & $0.1$ \\
\end{tabular}
\label{Parameters.tbl}
\end{table}

\subsection{Interferometer arms and gravitational waves}
\label{subsec:IFOarms}

LIGO's interferometers are generally optimized for 
the waves from inspiraling neutron-star and black-hole
binaries---sources that emit roughly equal power into all logarthmic frequency
intervals $\Delta \Omega/\Omega \sim 1$ in the LIGO band $\sim 10 \hbox{Hz}
\alt f \equiv \Omega/2\pi \alt 1000 \hbox{Hz}$.
Optimization turns out to entail making the lowest point in the 
interferometer's dimensionless
noise spectrum $f\times S_h(f)$ as low as possible.  Because of the relative 
contributions of
shot noise, radiation pressure noise, and thermal noise, this lowest
point turns out to be at $f \equiv \Omega/2\pi\simeq 100$ Hz.  To
minimize the noise at this frequency, one makes the end 
mirrors of the interferometer's
arm cavities (Fig.\ \ref{fig:Fig3}) 
as highly reflecting as possible (we shall idealize them as perfectly
reflecting until Sec.\ \ref{sec:LossyInterferometer}), and one gives
their corner mirrors transmisivities $T\simeq 0.033$, so the 
cavities' half bandwidths are
\begin{equation}
\gamma \equiv {T c \over 4L} \simeq 2\pi \times 100 {\rm Hz}\;.
\label{gammaDef}
\end{equation}
Here $L = 4$ km is the cavities' length (the interferometer ``arm length'').
We shall refer to $\gamma$ as the interferometer's {\it optimal frequency},
and when analyzing QND interferometers, we shall adjust their parameters so
as to beat the SQL by the maximum possible amount at $\Omega = \gamma$.  
In Table
\ref{Parameters.tbl} we list $\gamma$, $L$ and other parameters that
appear extensively in this paper, along with their fiducial numerical values.

In this and the next few sections we assume, for simplicity, that the mirrors
and beam splitter are lossless; we shall study the effects of losses in Sec.\
\ref{sec:LossyInterferometer} below.  
We assume that the carrier light (frequency $\omega_o$) exites
the arm cavities precisely on resonance.

We presume that all four mirrors (``test masses'') have masses $m \simeq 30$
kg, as is planned for LIGO-II.  

We label the two arms $n$ for north and $e$ for east, and denote by $X_n$ and
$X_e$ the changes in the lengths of the cavities induced by the 
test-mass motions.  We denote by
\begin{equation}
x \equiv X_n - X_e
\label{xDef}
\end{equation}
the changes in the arm-length difference, and we regard $x$ as a quantum
mechanical observable (though it could equally well be treated as classical 
\cite{TestMassQM}).  In the absence of external forces, we idealize $x$ as
behaving like a free mass (no pendular restoring forces). This 
idealization
could easily be relaxed, but because all signals below $\sim 10$ Hz are removed
in the data analysis, the pendular forces have no influence on the
interferometer's ultimate performance.  

The arm-length difference evolves in response to the gravitational wave and to
the back-action influence of the light's fluctuating radiation pressure.  
Accordingly, we can write it as
\begin{equation}
x(t) = x_o + {p_o\over m/4}t + \int_{-\infty}^{+\infty} (Lh + x_{\rm
BA})e^{-i\Omega t} {d\Omega\over 2\pi}\;.
\label{xhBA}
\end{equation}
Here $x_o$ is the initial value of $x$ when a particular segment of data begins
to be collected, $p_o$ is the corresponding initial generalized momentum, $m/4$
is the reduced mass\footnote{In each arm of the interferometer, the quantity
measured is the difference between the positions of the two mirrors' centers of
mass; this degree of freedom behaves like a free particle with reduced mass 
$m_r = m \times m/
(m+m) = m/2$.  The interferometer output is the difference, between the two
arms, of this free-particle degree of freedom; that difference behaves like a
free particle with reduced mass $m_r/2 = m/4$.}
associated with the test-mass degree of freedom $x$, 
$h$ is the Fourier transform of the gravitational-wave field
\begin{equation}
h(t) = \int_{-\infty}^{+\infty} h e^{-i\Omega t} {d\Omega\over 2\pi}\;,
\label{hDef}
\end{equation}
and $x_{\rm BA}$ is the influence of the radiation-pressure back action. 
Notice our notation: $x$, $x_{\rm BA}$ and $h$ are the $\Omega$-dependent
Fourier transforms of $x(t)$, $x_{\rm BA}(t)$ and  $h(t)$. 

Elsewhere \cite{TestMassQM} we discuss the fact that $x_o$ and
$p_o$ influence the interferometer output only near zero frequency $\Omega \sim
0$, and their influence is thus removed when the output data are filtered.  For
this reason, we ignore them and rewrite $x(t)$ as
\begin{equation}
x(t) = \int_{-\infty}^{+\infty} (Lh + x_{\rm
BA})e^{-i\Omega t} {d\Omega\over 2\pi}\;.
\label{xhBA1}
\end{equation}

\subsection{Output field expressed in terms of input}

Because we have idealized the beam splitter as infinitesimally thin, 
the input field
emerging from it and traveling toward the arm cavities has the coherent laser
light in the same $\cos\omega_o t$ quadrature as the dark-port field amplitude
$a_1$.  We further idealize the distances between the beam splitter and the
arm-cavity input mirrors as integral multiples of the carrier wavelength
$\lambda_o = 2\pi c/\omega_o$ and as small compared to $2\pi c /\gamma \sim 
300{\rm m}$.  (These idealizations could easily be relaxed without change in
the ultimate results.)  

Relying on these idealizations, we show
in Appendix \ref{app:interferometer} that the annihilation operators
$b_j$ for the beam splitter's output quadrature fields 
$E_j(t)$ are related to the input
annihilation operators $a_j$ and the gravitational-wave signal $h$ by
the linear relations
\begin{eqnarray}
b_1 &=& \Delta b_1 = a_1 e^{2i\beta}\;, \nonumber\\
b_2 &=& \Delta b_2 + \sqrt{2{\cal K}} {h\over h_{\rm SQL}}e^{i\beta}\;, \quad
\Delta b_2 = (a_2 - {\cal K} a_1) e^{2i\beta}\;.
\label{bjFromaj}
\end{eqnarray}
Here and below, for any operator $A$, $\Delta A \equiv A- \langle A\rangle$. 
This input-output equation and the quantities appearing in it require
explanation.  

The quantities
$\Delta b_j$ are the noise-producing parts of $b_j$, which remain 
when the gravitational-wave signal is turned off.  The $a_j$ impinge on the arm
cavities at a frequency $\omega_o + \Omega$ that is off resonance, so they
acquire the phase shift $2\beta$ upon emerging, where
\begin{equation}
\beta \equiv \arctan(\Omega/\gamma)\;.
\label{betaDef}
\end{equation}
If the test masses were unable to move, then $\Delta b_j$ would just be $a_j
e^{2i\beta}$; however, the fluctuating light pressure produces 
the test-mass motion
$x_{\rm BA}$, thereby inducing a phase shift in the light inside the cavity,
which shows up in the emerging light as the term $-{\cal K}a_1$ in $b_2$.  
(cf.\ Appendix \ref{app:interferometer}).  The quantity 
\begin{equation}
{\cal K} \equiv {(I_o/I_{\rm SQL})
2\gamma^4 \over \Omega^2(\gamma^2 + \Omega^2)}
\label{KDef}
\end{equation}
is the coupling constant by which this radiation-pressure
back-action converts input $a_1$ into
output $\Delta b_2$.  In this coupling constant,
$I_{\rm SQL}$ is the input laser power required, in a conventional
interferometer (Sec.\ \ref{sec:ConventionalInterferometer}), 
to reach the standard quantum limit:
\begin{equation}
I_{\rm SQL} = {mL^2 \gamma^4 \over 4 \omega_o} \simeq 1.0 \times 10^4
{\rm W}\; .
\label{ISQLDef}
\end{equation}

In Eq.\ (\ref{bjFromaj}), the gravitational-wave signal shows up as the
classical piece $\sqrt{2{\cal K}} h/ h_{\rm SQL}$ of $b_2$.  Here, as
we shall see below,
\begin{equation}
h_{\rm SQL} \equiv \sqrt{8\hbar\over m\Omega^2 L^2}
\simeq 2\times10^{-24} {\gamma\over\Omega} {\rm Hz}^{-1/2} 
\label{hSQLDef}
\end{equation}
is the standard quantum limit for the square root
of the single-sided spectral density of $h(t)$, $\sqrt{S_h}$. 

\section{Conventional Interferometer}
\label{sec:ConventionalInterferometer}

In an (idealized) conventional interferometer, the beam-splitter's output
quadrature field $E_2(t)$ is measured by means of conventional 
photodetection.\footnote{\label{foot:Photodetect}
Here and throughout this paper we regard 
some particular
quadrature $E_\zeta(t)$
as being measured directly.  This corresponds 
to superposing on $E_\zeta(t)$ carrier light with the same
quadrature phase as $E_\zeta$ and then performing 
direct
photodetection, which
produces a
photocurrent whose time variations are proportional to $E_\zeta(t)$.
For a conventional interferometer the carrier light in the desired quadrature,
that of $E_2(t)$, can be produced by operating
with the dark port biased slightly away from the precise dark fringe.
In future research it 
might be necessary to modify the QND designs described in
this paper so as to accommodate the modulations that are actually used
in the detection process; see Sec.\ \ref{sec:Conclusions} and especially
Footnote \ref{fn:modulation}.
}  
The Fourier transform of this measured quadrature is proportional to the
field amplitude 
$b_2 = \Delta b_2 + \sqrt{2{\cal K}} 
(h/h_{\rm SQL})e^{i\beta}$; 
cf.\ Eqs.\
(\ref{E1E2Def}) and (\ref{bjFromaj}).  
Correspondingly, we can think of $b_2 = b_2(\Omega)$ as the
quantity measured, and when we compute, from the output, the Fourier transform
$h=h(\Omega)$ of the gravitational-wave signal, the noise in that computation
will be
\begin{equation}
h_n(\Omega) = {h_{\rm SQL}\over\sqrt{2{\cal K}}} \Delta b_2 e^{-i\beta}\;.
\label{hnb2'}
\end{equation}

This noise is an operator for the Fourier transform of a random
process, and the corresponding single-sided spectral density $S_h(f)$
associated with this noise is given by the standard formula 
\cite{QuantumMeasurement,CavesSchumaker,SchumakerCaves}
\begin{equation} 
{1\over2}2\pi\delta(\Omega-\Omega')S_h(f) = \langle {\rm in}| h_n(\Omega)
h_n^{\dag}(\Omega')|{\rm in}\rangle_{\rm sym}\;.
\label{Shhn}
\end{equation}
Here $f = \Omega/2\pi$ is frequency, 
$|{\rm in}\rangle$ is the quantum state of the input light field (the
field operators $a_1$ and $a_2$), and the subscript ``sym'' means ``symmetrize
the operators whose expectation value is being computed'', i.e., replace 
$h_n(\Omega) h_n^{\dag}(\Omega')$ by ${1\over2}\left(
h_n(\Omega) h_n^{\dag}(\Omega') + h_n^{\dag}(\Omega') h_n(\Omega)\right)$.
Note that when Eq.\ (\ref{hnb2'}) for $h_n$ is inserted into
Eq.\ (\ref{Shhn}), the phase factor $e^{-i
\beta}$ cancels, i.e.\ it has no influence on the noise $S_h$.  This allows us
to replace Eq.\ (\ref{hnb2'}) by 
\begin{equation}
h_n(\Omega) = {h_{\rm SQL}\over\sqrt{2{\cal K}}} \Delta b_2\;.
\label{hnb2}
\end{equation}

For a conventional interferometer, the dark-port input is in its vacuum state,
which we denote by
\begin{equation}
|{\rm in}\rangle = |0_a\rangle\;.
\label{inConventional}
\end{equation}
For this vacuum input, the standard relations
$a_+|{\rm 0_a}\rangle = a_-|{\rm 0_a}\rangle = 0$, together with 
Eqs.\ (\ref{a12Def}) and (\ref{a12Commutator}), imply 
\cite{CavesSchumaker,SchumakerCaves}
\begin{equation}
\langle 0_a|a_j {a}_{k'}^{\dag}|0_a\rangle_{\rm sym} = {1\over 2}
2\pi\delta(\Omega-\Omega')\delta_{jk}\;.
\label{Vacuumajak}
\end{equation}
Comparing this relation with Eq.\ (\ref{Shhn}) and its generalization to
multiple random processes, 
we see that (when $|{\rm in}\rangle = |0_a\rangle$)
$a_1(\Omega)$ and $a_2(\Omega)$
can be regarded as the Fourier transforms of
classical random processes with single-sided spectral
densities and cross-spectral density given by \cite{TestMassQM}
\begin{equation}
S_{a_1}(f) = S_{a_2}(f) = 1\;, \quad S_{a_1 a_2}(f) = 0\;.
\label{Sa1a2Vacuum}
\end{equation}

Combining Eqs.\ (\ref{bjFromaj}) and (\ref{hnb2})--(\ref{Vacuumajak})
[or, equally well, (\ref{bjFromaj}), (\ref{hnb2}), and (\ref{Sa1a2Vacuum})],
we obtain for the noise spectral density of the conventional interferometer
\begin{equation} S_h = {h^2_{\rm SQL}\over 2}\left( {1\over {\cal K}} + {\cal
K}\right)\;.
\label{ShConventional}
\end{equation}
This spectral density is limited, at all frequencies $\Omega$, by the standard 
quantum limit
\begin{equation}
S_h \ge h_{\rm SQL}^2 = {8\hbar\over m\Omega^2L^2}\;.
\label{ShSQL}
\end{equation}

\begin{figure}
\epsfxsize=3.2in\epsfbox{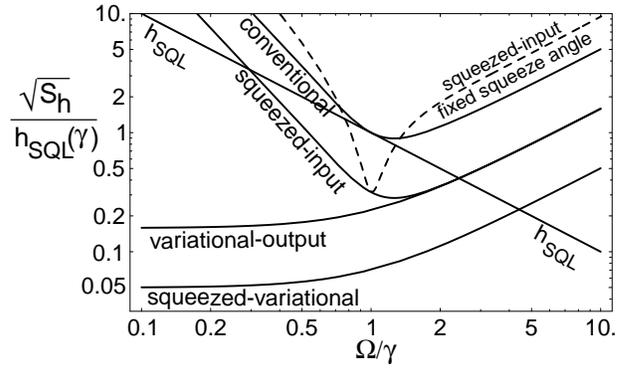}
\caption{The square root of the spectral density $\protect\sqrt{S_h}$ of the
gravitational-wave noise for several interferometer designs, 
as a function of angular frequency $\Omega$, with {\it optical losses
assumed negligible};  
$\protect\sqrt{S_h}$ is measured in units of the standard
quantum limit at frequency $\Omega=\gamma$, and $\Omega$ is measured in units
of $\gamma$.  The noise curves shown are: (i) the standard quantum limit
itself, $h_{\rm SQL}(\Omega)$ [Eq.\ (\protect\ref{hSQLDef})]; (ii) the noise
for a {\it conventional} 
interferometer with laser power $I_o = I_{\rm SQL}$ [Eq.\
(\protect\ref{ShConventionalOptimized})]; (iii) the noise for a 
{\it squeezed-input}
interferometer with $I_o = I_{\rm SQL}$, squeeze factor $e^{-2R} = 0.1$, and
(a) optimized FD squeeze angle $\lambda = - \Phi(\Omega)$ [Eq.\ 
(\protect\ref{ShSqueezedInputOptimized}); solid curve], (b)
optimized frequency-independent squeeze angle 
[Eq.\ (\protect\ref{ShSIFixedAngle}); dashed curve];
(iv) the noise for a {\it
variational-output} interferometer with $I_o = 10 I_{\rm SQL}$ and optimized
frequency-dependent homodyne phase $\zeta = \Phi(\Omega)$ [Eq.\
(\protect\ref{ShVariationalOutputOptimized})]; and (v) the noise for a {\it
squeezed-variational} interferometer with $I_o = 10 I_{\rm SQL}$, input squeeze
factor $e^{-2R}=0.1$, and optimized input squeeze angle $\lambda=\pi/2$ and
output homodyne phase $\zeta = \Phi(\Omega)$ [Eq.\ 
(\protect\ref{ShSqueezedVariationalOptimized})]. 
\label{fig:Fig4}
}
\end{figure}

Recall that $\cal K$ is a function of frequency $\Omega$ and is 
proportional to the input laser power $I_o$ [Eq.\ (\ref{KDef})]. 
In our conventional interferometer, we
adjust the laser power to $I_o = I_{\rm SQL}$
[Eq.\ (\ref{ISQLDef})],
thereby making ${\cal K}(\Omega = \gamma) = 1$, which minimizes 
$S_h$ at the interferometer's optimal frequency $\Omega = \gamma$.  
The noise spectral density then becomes [cf.\ Eqs.\ (\ref{ShConventional}) and
(\ref{KDef})]
\begin{equation}
S_h = {4\hbar\over m L^2\Omega^2}
\left[ {2\gamma^4\over\Omega^2(\gamma^2+\Omega^2)} + 
{\Omega^2(\gamma^2 + \Omega^2)\over2\gamma^4}\right]\;.
\label{ShConventionalOptimized}
\end{equation}
This optimized conventional noise is shown as a curve
in Fig.\ \ref{fig:Fig4},
along with the standard quantum limit $h_{\rm SQL}$ and the noise curves for
several QND interferometers to be discussed below.  This conventional noise
curve is currently a tentative goal for LIGO-II, when operating without signal
recycling \cite{WhitePaper}.

\section{Strategies to Beat the SQL, and Their Lossless Performance}
\label{sec:LosslessPerformance}

\subsection{Motivation: Ponderomotive squeezing}

The interferometer's input-output relations $\Delta b_1 = a_1 e^{2i\beta}$, 
$\Delta b_2 = (a_2-{\cal K}a_1)e^{2i\beta}$ can be regarded as consisting of 
the uninteresting phase shift $e^{2i\beta}$, and a
rotation in the $\{a_1,a_2\}$ plane (i.e., $\{\cos\omega_o t, \sin\omega_o t\}$
plane), followed by a squeeze:
\begin{equation}
b_j = S^{\dag}(r,\phi) R^{\dag}(-\theta) a_j e^{2i\beta} R(-\theta)
S(r,\phi)\;.
\label{bjSqueezeaj}
\end{equation}  
Here $R(-\theta)$ is the rotation operator and $S(r,\phi)$ the
squeeze operator for two-photon quantum optics; see Appendix \ref{app:2photon}
for a very brief summary, and Refs.\ \cite{CavesSchumaker,SchumakerCaves} for
extensive detail.  The rotation angle $\theta$, squeeze angle $\phi$ and
squeeze factor $r$ are given by
\begin{equation}
\theta = \arctan({\cal K}/2), \;\; \phi = {1\over2} {\rm arccot}
({\cal K}/2),
\;\; r = {\rm arcshinh}({\cal K}/2).
\label{RotSqueezeParams}
\end{equation}
Note that, because the coupling constant $\cal K$ depends on frequency
$\Omega$ [Eq.\ (\ref{KDef})], the rotation angle, squeeze angle, and squeeze
factor are frequency dependent.  This frequency dependence will have
major consequences for the QND interferometer designs discussed below.

The rotate-and-squeeze transformation 
(\ref{bjSqueezeaj}) 
for the two-photon amplitudes implies corresponding rotate-and-squeeze
relations for the one-photon creation and annihilation operators
\begin{equation}
b_\pm =  
S^{\dag}(r,\phi) R^{\dag}(-\theta) a_\pm e^{\pm 2i\beta} R(-\theta)
S(r,\phi)\;.
\label{bpmSqueezeapm}
\end{equation}  
Denote by $|0_{a_+}\rangle$ the vacuum for the 
{\it in} mode at
frequency $\omega_o+\Omega$, by $|0_{a_-}\rangle$ that for the {\it in} mode
at
$\omega_o-\Omega$, and by $|0_{a_\pm}\rangle$ the vacuum for one or the other of
these modes; and denote similarly the vacuua for the 
{\it out} modes,
$|0_{b\pm}\rangle$.  Then $|0_{a_\pm}\rangle$ is the 
state annihilated by $a_\pm$
and $|0_{b_\pm}\rangle$ is that annihilated by $b_\pm$.
Correspondingly, the rotate-squeeze relation (\ref{bpmSqueezeapm}) implies that
\begin{equation} b_\pm |0_{b_\pm}\rangle = S^{\dag}R^{\dag}a_\pm e^{\pm2i\beta}
RS|0_{b_\pm}\rangle = 0\;,
\label{b0manipulate1}
\end{equation}
where the parameters of the squeeze and rotation operators are those given in
Eqs.\ (\ref{RotSqueezeParams}) and (\ref{bpmSqueezeapm}).  
This equation implies
that $e^{\pm2i\beta} RS|0_{b_\pm}\rangle$ is annihilated by $a_\pm$ and 
therefore is the {\it in} vacuum $|0_{a_\pm}\rangle$ for the {\it in} mode
$\omega_o\pm\Omega$:  
\begin{equation} e^{\pm2i\beta} RS|0_{b_\pm}\rangle = |0_{a_\pm}\rangle\;.
\label{b0manipulate2}
\end{equation}
Applying $R^{\dag}$ 
and noting that $R^{\dag} |0_{a_\pm}\rangle =
|0_{a_\pm}\rangle$ (the vacuum is rotation invariant), we obtain 
\begin{equation}
|0_{a_\pm}\rangle = e^{\pm2i \beta}S(r,\phi)|0_{b_\pm}\rangle\;.
\label{0a0b}
\end{equation}
Thus, {\sl the} in {\sl vacuum is equal to a squeezed} out {\sl
vacuum, aside from an
uninteresting, frequency-dependent phase shift}.  The meaning of
this statement in the context of a conventional interferometer is the 
following:

For a conventional interferometer, the {\it in} state is 
\begin{equation}
|{\rm in}\rangle = |0_{a_\pm}\rangle = e^{\pm2i\beta}
S(r,\phi)|0_{b_\pm}\rangle\;;
\label{in0b}
\end{equation}
and because we are using the Heisenberg Picture where the state does not
evolve, the light emerges from the interferometer in this state.
However, in passing through the interferometer, the light's quadrature
amplitudes evolve from $a_j$ to $b_j$.
Correspondingly, at the output we should discuss the properties of the
unchanged state in terms of a basis built from the {\it out} vacuum
$|0_{b_\pm}\rangle$.  Equation (\ref{0a0b}) says that in this {\it out}
language, the light has been squeezed at
the angle $\phi$ and squeeze-factor $r$
given by Eq.\ (\ref{RotSqueezeParams}).  
This squeezing is produced by the back-action force of fluctuating
radiation pressure on the test masses.  That back action has the character
of a ponderomotive nonlinearity 
first recognized by Braginsky and Manukin 
\cite{BraginskySqueeze}.\footnote{Recently 
it has been recognized that this ponderomotive nonlinearity
acting on a movable mirror in a Fabry-Perot resonator may provide a practical
method for generating  bright squeezed light \cite{PSqueeze}.}  
The correlations inherent in this squeezing form the foundation for the QND
interferometers discussed below.

One can also deduce this ponderomotive squeezing from the in-out
relations $\Delta b_1 = a_1 e^{2i\beta}$, $\Delta b_2 = (a_2 - {\cal K}
a_1)e^{2i\beta}$ [Eq.\ (\ref{bjFromaj})], the expressions 
\begin{eqnarray}
{1\over 2} 2 \pi \delta(\Omega-\Omega')S_{b_j}(f) &=& \langle {\rm in} | \Delta
b_j \Delta {b_{j'}}^{\dag}|{\rm in}\rangle_{\rm sym}\;, \nonumber\\
{1\over 2} 2 \pi \delta(\Omega-\Omega')S_{b_1 b_2}(f) &=& 
\langle {\rm in} | {1\over2}(\Delta b_1 \Delta {b_{2'}}^{\dag} + \Delta b_1^{\rm
\dag} \Delta b_{2'})|{\rm in}\rangle_{\rm sym} \nonumber\\
\label{Sb1b2Def}
\end{eqnarray}
for the spectral densities and cross spectral densities of $b_1$ and $b_2$,
and the spectral densities 
$S_{a_1}= S_{a_2}= 1$, $S_{a_1 a_2} = 0$ [Eqs.\ (\ref{Sa1a2Vacuum})].  
These imply that for a conventional interferometer 
\begin{equation}
S_{b_1} = 1\;, \quad S_{b_2} = 1+{\cal K}^2\;, \quad
S_{b_1b_2} = - {\cal K}\;.
\label{Sb1b2Vacuum}
\end{equation}
Rotating $\Delta b_j$ through the angle 
$\phi = {1\over2}{\rm arccot}({\cal K}/2)$
to obtain
\begin{equation}
b'_{1} = b_1 \cos\phi + b_2 \sin\phi\;, \quad b'_{2} = b_2 \cos\phi - b_1
\sin\phi\;,
\label{bjRotated}
\end{equation}
and using Eqs.\ (\ref{Sb1b2Def}) and (\ref{Sb1b2Vacuum}), we obtain
\begin{eqnarray}
S_{b'_{1}} &=& e^{-2r} = \left(\sqrt{1+({\cal K}/2)^2} - {\cal K}/2\right)^2
\simeq 1/{\cal K} \hbox{ if }{\cal K}\gg1\;, 
\nonumber\\
S_{b'_{2}} &=& e^{+2r} = \left(\sqrt{1+({\cal K}/2)^2} + {\cal K}/2\right)^2\;, 
\quad
S_{b'_{1} b'_{2}} = 0\;, \nonumber\\
\label{Sb1'b2'}
\end{eqnarray}
which represents a squeezing of the input vacuum noise in the manner 
described formally by Eqs.\ (\ref{in0b}) and (\ref{RotSqueezeParams}). 

\begin{figure}
\epsfxsize=3.2in\epsfbox{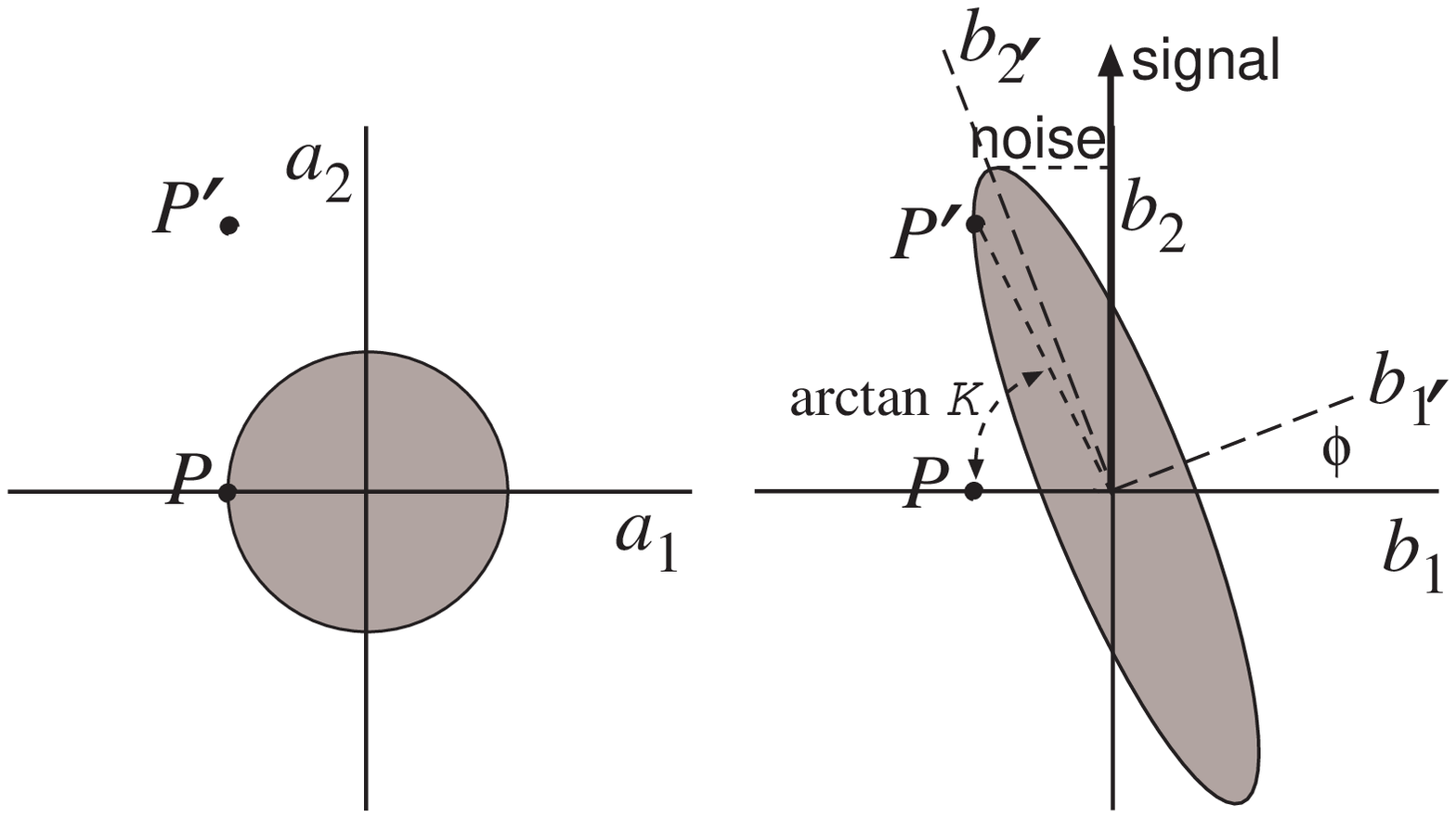}
\caption{
Noise ellipses for a conventional interferometer. {\it Left:} Noise
for vacuum that enters the interferometer's dark port.  {\it Right:} Noise for
ponderomotively squeezed vacuum that exits at the dark port along with the
gravitational-wave signal; 
the ponderomotive squeeze has moved the point $P$ to
the new point $P'$ [$b_1=a_1$, $b_2 = a_2 - {\cal K}a_1$, Eqs.\ 
(\protect\ref{bjFromaj})].  
These noise ellipses have dimensions and shapes
described by the noise spectral densities (\protect\ref{Sa1a2Vacuum}), 
(\protect\ref{Sb1b2Vacuum}) and (\protect\ref{Sb1'b2'}), and by the 
squeeze equations (\protect\ref{in0b}) and
(\protect\ref{RotSqueezeParams}).  The minor radius of the output noise
ellipse is $\protect\sqrt{S_{b'_{1}}} = e^{-r}$, and its major radius is 
$\protect\sqrt{S_{b'_{2}}} = e^{+r}$, 
where $r$ is the squeeze factor; cf.\ Eqs.\ 
(\protect\ref{RotSqueezeParams}) and (\protect\ref{Sb1'b2'}).
The conventional interferometer measures
$b_2$, which contains the indicated noise [cf.\ Eq.\ (\protect\ref{hnb2})]
and the indicated signal [$\delta b_2 = \protect\sqrt{2{\cal K}} h/h_{\rm
SQL}$; cf.\ Eq.\ (\protect\ref{bjFromaj})].   For a detailed discussion of
noise ellipses in 2-photon quantum optics see, e.g., Ref.\ 
\protect\cite{CavesSchumaker}.
\label{fig:Fig5}
}
\end{figure}

This ponderomotive squeezing is depicted by the noise ellipse of 
Fig.\ \ref{fig:Fig5}.  For a conventional interferometer ($b_2$
measured via 
photodetection\footnote{See footnote \ref{foot:Photodetect}.}), 
the signal is the arrow along the $b_2$ axis, and
the square root of the noise spectral density $S_{b_2}$ is 
the projection of the noise ellipse onto the $b_2$ axis.  For a detailed
discussion of this type of graphical representation of noise in two-photon
quantum optics see, e.g., Ref.\ \cite{CavesSchumaker}.

\subsection{Squeezed-Input Interferometer}
\label{sec:SqueezedInputInterferometer}

Interferometer designs that can beat the SQL (\ref{ShSQL}) are sometimes called
``QND interferometers''.  Unruh \cite{Unruh} has devised a QND interferometer 
design based on (i) putting the input electromagnetic 
fluctuations at the dark port
($a_1$ and $a_2$) into a squeezed state, and (ii) using standard photodetection
to measure the interferometer's output field.  We shall call this a
{\it squeezed-input interferometer}.  The squeezing of the input has been
envisioned as achieved using nonlinear crystals
\cite{Xaio,Grangier},
but one might also use
ponderomotive squeezing.

The squeezed-input interferometer is identical to the conventional
interferometer of Sec.\ \ref{sec:ConventionalInterferometer}, except
for the choice of the {\it in} 
state $|{\rm in}\rangle$ for the dark-port field. 
Whereas a conventional interferometer has $|{\rm in}\rangle = |0_a\rangle$, the
squeezed-input interferometer has
\begin{equation}
|{\rm in}\rangle = S(R,\lambda)|0_a\rangle\;,
\label{InSqeezed}
\end{equation}
where $R$ is the largest squeeze factor that the experimenters are able to
achieve 
($e^{-2R} \sim 0.1$ in the LIGO-III time frame), 
and $\lambda = \lambda(\Omega)$ is a squeeze angle
that depends on side-band frequency.  One adjusts
$\lambda(\Omega)$ so as to minimize the noise in the output quadrature
amplitude $b_2$, which (i) contains the gravitational-wave signal and (ii) is
measured by standard photodetection.

The gravitational-wave noise for such an interferometer is proportional to
\begin{equation}
\langle{\rm in}|h_n h_{n'}|{\rm in}\rangle = \langle 0_a | h_{ns}
h_{ns'}|0_a\rangle
\label{EVhn}
\end{equation}
[Eq.\ (\ref{Shhn})], where $h_{ns}$ is the {\it squeezed} gravitational-wave
noise operator
\begin{equation}
h_{ns} = S^{\dag}(R,\lambda) h_n S(R,\lambda)
\label{hns}
\end{equation}
and $h_n' \equiv h_n(\Omega')$. 
By inserting expression (\ref{hnb2'}) for $h_n$ into Eq.\ (\ref{hns}) and
then combining the interferometer's ponderomotive squeeze relation
$\Delta b_2 = (a_2 - {\cal K}a_1)e^{2i\beta}$ with the action of the squeeze
operator on $a_1$ and $a_2$ [Eq.\ (\ref{Squeezea12})], we obtain
\begin{eqnarray}
&& h_{ns} = - {h_{\rm SQL}\over\sqrt{2{\cal K}}} 
\sqrt{(1+{\cal K}^2)}\;e^{i\beta} 
\nonumber\\
&& \quad \times 
\left( a_1 \{\cosh R \cos\Phi - \sinh R \cos[\Phi - 2(\Phi+\lambda)]\} \right.
\nonumber\\
&&\quad\;
\left. - a_2 \{\cosh R \sin\Phi - \sinh R \sin[\Phi - 2(\Phi+\lambda)]\}\right)
\;,
\label{hnsExpand}
\end{eqnarray}
where
\begin{equation}
\Phi \equiv {\rm arccot}{\cal K}\;.
\label{PhiDef}
\end{equation}

We can read the spectral density of the gravitational-wave noise off of Eq.\
(\ref{hnsExpand}) by recalling that in the $|0_a\rangle$ vacuum state [which is
relevant because of Eq.\ (\ref{EVhn})], $a_1$
and $a_2$ can be regarded as random processes with spectral sensities $S_{a_1}
= S_{a_2} = 1$ and vanishing cross spectral density [Eqs.\
(\ref{Sa1a2Vacuum})]:  
\begin{equation}
S_h = {h_{\rm SQL}^2\over2}\left({1\over{\cal K}}+{\cal K}\right) \Big(\cosh2R
- \cos[2(\lambda+\Phi)]\sinh2R\Big).
\label{SqueezeSb2New}
\end{equation}
It is straightforward to verify that this noise 
is minimized by making it
proportional to $\cosh 2R - \sinh 2R = e^{-2R}$, which is achieved by 
choosing for the input squeeze angle
\begin{equation}
\lambda(\Omega) = -\Phi(\Omega) \equiv - {\rm arccot}{\cal K}(\Omega)\;.
\label{lambdaOpt}
\end{equation}
The result is
\begin{equation}
S_h = {h_{\rm SQL}^2\over2}\left({1\over{\cal K}} + {\cal K}\right)e^{-2R}\;.
\label{ShSqueezedInput}
\end{equation}

This says that {\it the squeezed-input interferometer has the same noise
spectral density
as the
conventional interferometer, except for an overall reduction by} $e^{-2R}$,
where $R$ is the squeeze factor for the dark-port input field (a result
deduced by Unruh \cite{Unruh} and later confirmed 
by Jaekel and Reynaud \cite{JaekelReynaud} using a different method);
see Fig.\ \ref{fig:Fig4}.
This result implies that the squeezed-input interferometer can beat the 
amplitude SQL
by a factor $e^{-R}$. 

When the laser power $I_o$ of the squeezed-input interferometer is optimized
for detection at the frequency $\Omega=\gamma$ ($I_o = I_{\rm SQL}$ as for a
conventional interferometer), the noise spectrum becomes
\begin{equation}
S_h = {4\hbar\over mL^2\Omega^2} 
\left[ {2\gamma^4\over\Omega^2(\gamma^2+\Omega^2)} +
{\Omega^2(\gamma^2 + \Omega^2)\over2\gamma^4}\right]e^{-2R}\;.
\label{ShSqueezedInputOptimized}
\end{equation}
This optimized noise is shown
in Fig.\ \ref{fig:Fig4}
for $e^{-2R} = 0.1$,
along with the noise spectra for other
optimized interferometer designs.

In previous discussions of this squeezed-input scheme
\cite{Unruh,JaekelReynaud,PCW}, no attention has been paid to the 
practical problem of how to produce the necessary frequency dependence
\begin{equation}
\lambda(\Omega) = - \Phi(\Omega) = -{\rm arccot}{2\gamma^4\over \Omega^2(\gamma^2 +
\Omega^2)}
\label{lambdaOptOmega}
\end{equation}
of the squeeze angle.  In Sec.\ \ref{sec:FDSqueeze}, we shall show
that this $\lambda(\Omega)$
can be achieved by squeezing at a frequency-independent squeeze 
angle 
(using, e.g., a nonlinear crystal for which the squeeze angle will be 
essentially
frequency-independent because the gravity-wave bandwidth, $<1000$ Hz, is so
small
compared to usual optical bandwidths 
) and then
filtering through two Fabry-Perot cavities.  This squeezing and filtering must
be performed before injection into the interferometer's dark port; 
see Fig.\ 
\ref{fig:Fig1} for a schematic diagram.

\begin{figure}
\epsfxsize=3.2in\epsfbox{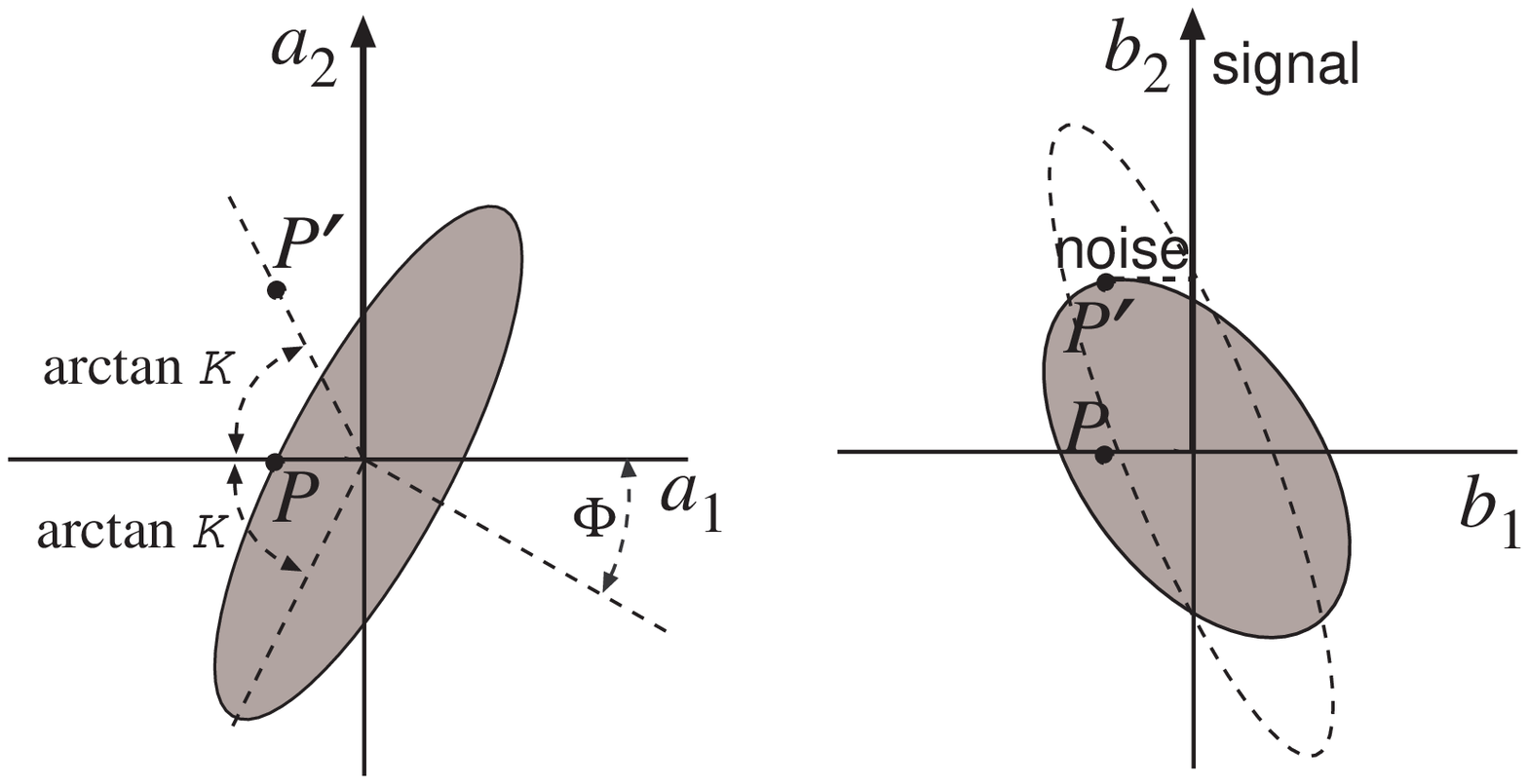}
\caption{Noise ellipses for a squeezed-input interferometer. {\it Left:} Noise
for squeezed vacuum that enters the interferometer's dark port.  The field
is squeezed at the angle $\lambda = - \Phi$.  {\it Right:} Noise for
the field that exits at the dark port along with the
gravitational-wave signal.  This output field results from the interferometer's
ponderomotive
squeezing of the input field 
[e.g., point $P$ goes to point $P'$ in accord with
$b_1 = a_1$, $b_2 = a_2 - {\cal K}a_1$; Eqs. (\protect\ref{bjFromaj})]. 
If the input field had been vacuum as in a
conventional interferometer [Fig.\ \protect\ref{fig:Fig5}], then
the output would have been squeezed in the manner of the dashed ellipse.  The
two squeezes (input and ponderomotive) result in the shaded ellipse, whose
projection along the axis measured by the photodetector ($b_2$ axis) has been
minimized by the choice of squeeze angle, $\lambda = - \Phi$. 
\label{fig:Fig6}
}
\end{figure}

The signal and noise for this squeezed-input interferometer are depicted in
Fig.\ \ref{fig:Fig6}. 

We comment, in passing, on two other variants of a squeezed-input
interferometer:  

(i) If, for some reason, the filter cavities cannot be implemented
successfully, one can still inject squeezed vacuum at the dark port with
a frequency-independent phase that is optimized for the lowest point in the
noise curve, $\Omega=\gamma$; i.e.\ (with the input
power optimized to $I_o = I_{\rm SQL}$):
\begin{equation}
\lambda = - \Phi(\gamma) = - \pi/4\;;
\label{lambdagamma}
\end{equation}
cf.\ \ Eq.\ (\ref{lambdaOptOmega}).  In this case the noise spectrum is
\begin{eqnarray}
&&
S_h = {h_{\rm SQL}^2\over2}\left({1\over{\cal K}} + {\cal K}\right)
\nonumber\\
&&\quad \times
\left[ (\cosh R \cos\Phi - \sinh R \sin\Phi )^2 \right.
\nonumber\\
&&\quad\;
\left. + (\cosh R \sin\Phi - \sinh R \cos\Phi)^2\right]
\;;
\label{ShSIFixedAngle}
\end{eqnarray}
[Eq.\ (\ref{SqueezeSb2New}), translated into gravitational-wave
noise via Eq.\ (\ref{hnb2})].
This noise spectrum is shown as a dashed curve in Fig.\ 
\ref{fig:Fig4}, for $e^{-2R} = 0.1$.  The SQL is beat
by the same factor $\mu = \sqrt{e^{-2R}} \simeq 0.32$ as in the case of
a fully optimized squeezed-input interferometer, but the frequency band over
which the SQL is beat is significantly smaller than in the optimized case, and
the noise is worse than for a conventional interferometer outside that band.  

(ii) Caves \cite{Caves}, in a paper that preceeded Unruh's and
formed a foundation for Unruh's ideas, proposed a squeezed-input interferometer
with the  squeeze angle set to $\lambda = \pi/2$ independent of frequency.  
In this case, Eq.\ (\ref{SqueezeSb2New}), translated into gravitational-wave
noise via Eq.\ (\ref{hnb2}), says that
\begin{equation}
S_h = {h_{\rm SQL}^2\over2}\left({1\over{e^{2R}\cal K}} + e^{2R}{\cal K}\right)
\;.
\label{ShCaves}
\end{equation}
Since $\cal K$ is proportional to the input laser power $I_o$, {Caves'
interferometer produces the same noise spectral density as a conventional
interferometer [Eq.\ (\ref{ShConventional})] but with an input power that is
reduced by a factor $e^{-2R}$.}  This is a well-known result.

\subsection{Variational-output interferometer}
\label{sec:VariationalOutput}

Vyatchanin, Matsko and Zubova \cite{VMZ,VM1,VM2} have devised a QND interferometer
design based on (i) leaving the dark-port input field in its vacuum state,
$|{\rm in}\rangle = |0_a\rangle$, 
and (ii) changing the output measurement from standard
photodetection (measurement of $b_2$) to homodyne detection at an appropriate,
frequency-dependent (FD) homodyne phase $\zeta(\Omega)$ -- i.e., 
measurement of
\begin{equation}
b_\zeta = b_1 \cos\zeta + b_2 \sin\zeta\;.
\label{bzetaDef}
\end{equation}
In their explorations of this idea, Vyatchanin, Matsko and Zubova  \cite{VMZ,VM1,VM2}
did not identify any practical scheme for achieving such a FD homodyne
measurement, so they approximated it by homodyne detection with a homodyne
phase that depends on time rather than frequency
--- a technique that they call a ``quantum
variational measurement''.  

In Sec.\ \ref{sec:FDHomodyne} below, we show that the optimized 
FD homodyne measurement can, in fact, be achieved by filtering the
interferometer output through two Fabry-Perot cavities and then performing
standard, balanced homodyne detection at a frequency-independent homodyne
phase;
see Fig.\ \ref{fig:Fig2} for a schematic diagram.  
We shall call such an scheme a {\it variational-output interferometer}.
The word ``variational'' refers to 
(i) the fact that the measurement entails monitoring a
frequency-varying quadrature of the output field, as well as 
(ii) the fact that the goal is to measure
variations of the classical force acting on the interferometer's test mass
(the original Vyatchanin-Matsko-Zubova motivation for the word). 

The monitored FD amplitude 
$b_\zeta$ [Eq.\ (\ref{bzetaDef})] can be expressed in
terms of the interferometer's dark-port input amplitudes $a_1$, $a_2$ and the
Fourier transform of the gravitational-wave field $h$ as 
\begin{equation}
b_\zeta = \sin\zeta \left( \sqrt{2 {\cal K}} {h\over h_{\rm SQL}} e^{i\beta}
+ [a_2 + (\cot\zeta - {\cal K})a_1] e^{2i\beta} \right)\;;
\label{bzetaOfa1a2h}
\end{equation}
cf.\ Eqs.\ (\ref{bjFromaj}) and (\ref{bzetaDef}).  Correspondingly, the
operator describing the Fourier transform of the interferometer's 
gravitational-wave noise is
\begin{equation}
h_n(\Omega) = {h_{\rm SQL}\over\sqrt{2{\cal K}}}e^{i\beta}
[a_2 + a_1(\cot\zeta - {\cal
K})]\;;
\label{hnb2Variational}
\end{equation}
cf.\ Eq.\ (\ref{hnb2}).

\begin{figure}
\epsfxsize=3.2in\epsfbox{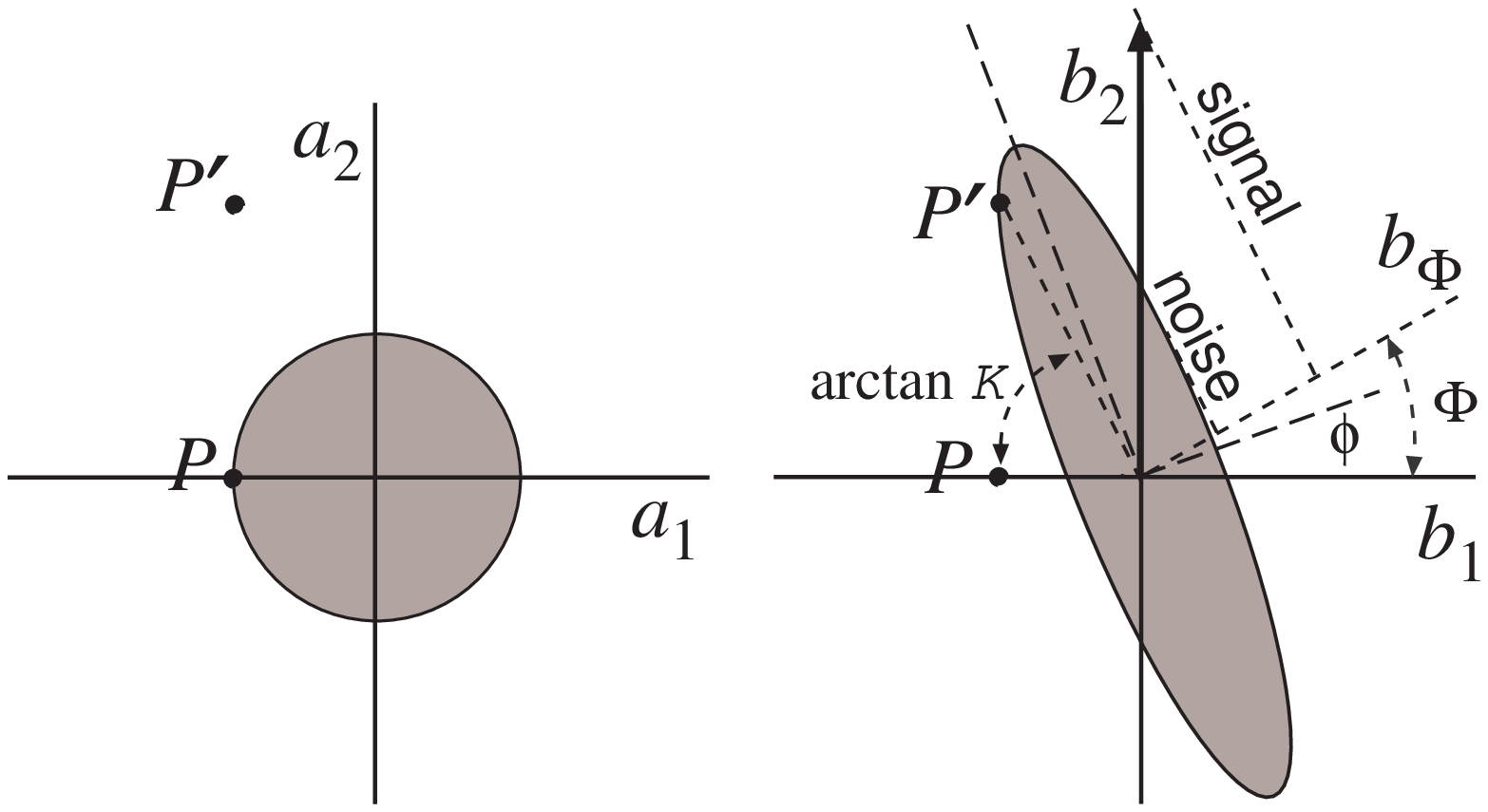}
\caption{Noise ellipses for a variational-output interferometer. 
{\it Left:} Noise for the ordinary vacuum that enters the interferometer's 
dark port.  {\it Right:} Noise for
the field that exits at the dark port along with the
gravitational-wave signal.  These noise ellipses are the same as for
a conventional interferometer, Fig.\ \protect\ref{fig:Fig5}, 
but here the quantity measured is the quadrature amplitude $b_\Phi$
with frequency
dependent phase $\Phi \equiv {\rm arccot}{\cal K}$. 
It is informative to compare the measured phase $\Phi$ with the
angle of ponderomotive squeeze $\phi = {1\over2}{\rm arccot}({\cal K}/2)$.  
They
are related by $\tan\Phi = {1\over2}\tan2\phi = \tan\phi/(1-\tan^2\phi)$, so
$\Phi$ is always larger than $\phi$; but
for large ${\cal K}$ (strong beating of the SQL), they
become small and nearly equal. 
\label{fig:Fig7}
}
\end{figure}

The radiation-pressure-induced back action of the measurement on the
interferometer's test masses is embodied in the $-{\cal K}a_1$ term of this
equation; cf.\ Eq.\ (\ref{bjFromaj}) and subsequent discussion.  It should be
evident that {\it by choosing}
\begin{equation}
\zeta = \Phi \equiv {\rm arccot}{\cal K}\;,
\label{zetaOptVariational}
\end{equation}
{\it we can completely remove the back-action noise from the measured 
interferometer output}; cf.\ Fig.\ \ref{fig:Fig7}.  
This optimal choice of the FD homodyne phase, together
with the fact that the input state is vacuum, $|{\rm in}\rangle = |0_a\rangle$,
leads to the gravitational-wave noise 
\begin{equation}
S_h = {h^2_{\rm SQL}\over 2} {1\over {\cal K}} = 
{1\over I_o/I_{\rm SQL}} \left({4\hbar\over mL^2\Omega^2}\right)
{\Omega^2 (\gamma^2 + \Omega^2)\over 2\gamma^4}\;. 
\label{ShVariationalOutputOptimized}
\end{equation}
cf.\ Eqs.\ (\ref{Shhn}) and (\ref{Vacuumajak}).  

This noise for an optimized variational-output interferometer is entirely due
to shot noise of the measured light, and continues to improve 
$\propto 1/I_o$ even
when the input light power $I_o$ exceeds $I_{\rm SQL}$.  Figure 
\ref{fig:Fig4} 
shows this noise, along with the noise spectra for other
optimized interferometer designs.

It is interesting that the optimal frequency-dependent homodyne phase $\Phi$
for this variational-output interferometer is the same, aside from sign, as the
optimal frequency-dependent squeeze angle for the squeezed-input
interferometer; cf.\ Eq.\ (\ref{lambdaOpt}). 

\subsection{Comparison of squeezed-input and variational-output
interferometers}

The squeezed-input and variational-output interferometers described above are
rather idealized, most especially because they assume perfect, lossless optics.
When we relax that assumption in Sec.\ \ref{sec:LossyInterferometer} below,
we shall see that, for realistic squeeze factors $e^{-2R}$ and losses
$\epsilon_*$, the two interferometers have essentially the same performance,
but the variational-output intefermometer requires $\sim 10$ times higher 
input power $I_o$.
In this section we shall seek insight into the physics of these interferometers
by comparing them in the idealized, lossless limit.

Various comparisons are possible.  
The noise curves in Fig.\ \ref{fig:Fig4} illustrate one comparison:
When the FD homodyne angle has been
optimized, a 
lossless
variational-output interferometer reduces shot noise below the
SQL and completely removes back-action noise; by contrast, when the FD squeeze
angle has been optimized, a 
squeezed-input interferometer reduces shot noise 
and reduces but does not remove back-action noise; cf.\ Eqs.\ 
(\ref{ShVariationalOutputOptimized}) and (\ref{ShSqueezedInput}).  

In variational-output interferometers, after optimizing the FD homodyne angle,
the experimenter has further control of just one input/output
parameter: the laser intensity or equivalently 
$I_o/I_{\rm SQL} = {\cal K}(\Omega=\gamma)$.  When $I_o/I_{\rm SQL}$ 
is increased, the shot
noise decreases; independent of its value, the back-action noise has already
been removed completely; cf.\ Eq.\ (\ref{ShVariationalOutputOptimized}).
By contrast, in squeezed-input interferometers, after optimizing the FD squeeze
phase, the experimenter has control of two parameters: $I_o/I_{\rm SQL}$, which
moves the minimum of the noise curve back and forth in frequency but does not
lower its minimum \cite{Caves}, and the squeeze factor $R$, which reduces the
noise by $e^{-2R}$; cf.\ Eq.\ (\ref{ShSqueezedInput}).  

Present technology
requires that $R$ be approximately constant over the LIGO frequency band.
However, in the same spirit as our assumption
that the FD homodyne phase can be optimized at all frequencies, it is
instructive to ask what can be achieved with an unconstrained,
frequency-dependent (FD) squeeze factor $R(\Omega)$, when coupled to an
unconstrained FD squeeze angle $\lambda(\Omega)$.

One instructive choice is $\lambda = -\hbox{arccot}{\cal K}$ as in our previous,
optimized interferometer [Eq.\ (\ref{lambdaOpt})], and 
$e^{-2R} = 1/(1+ {\cal K}^2)$.
In this case, the squeezed-input interferometer has precisely the same noise
spectrum as the lossless variational-output interferometer 
\begin{equation}
S_h = {h_{\rm SQL}^2 \over 2{\cal K}}\;;
\label{Example1}
\end{equation}
[Eq.\ (\ref{ShVariationalOutputOptimized})], and achieves it with 
precisely the same laser power.

Another instructive choice is an input squeeze that is inverse to
the interferometer's ponderomotive squeeze (a configuration we shall call
``inversely input squeezed'' or IIS):  Let the dark-port input
field before squeezing be described by annihilation operators $c_\pm$, so
\begin{equation}
c_\pm |{\rm in}\rangle = 0\;,
\label{PreSqueezeIn}
\end{equation}
i.e.\ the pre-squeeze field is vacuum.  Then,
denoting by $c_1$, $c_2$ the quadrature amplitudes of this pre-squeeze field,
the IIS input squeezing is
\begin{equation}
a_1 = c_1\;, \quad a_2 = c_2 + {\cal K} c_1\;,
\label{ReciprocityInput}
\end{equation}
where ${\cal K}(\Omega)$ is the interferometer's frequency-dependent
coupling constant (\ref{KDef}).  The interferometer's ponderomotively squeezed
output noise is then
\begin{equation}
\Delta b_1 = a_1 e^{2i\beta} = c_1 e^{2i\beta}\;, \quad 
\Delta b_2 = (a_2 - {\cal K}a_1)e^{2i\beta} = c_1 e^{2i\beta}
\label{ReciprocityOutput} 
\end{equation}
[cf.\ Eq.\ (\ref{bjFromaj})], i.e., the noise of the output light is that
of the vacuum with a phase
shift, but since the vacuum state is insensitive to phase, it is actually 
just the noise of the vacuum.  

\begin{figure}
\epsfxsize=3.2in\epsfbox{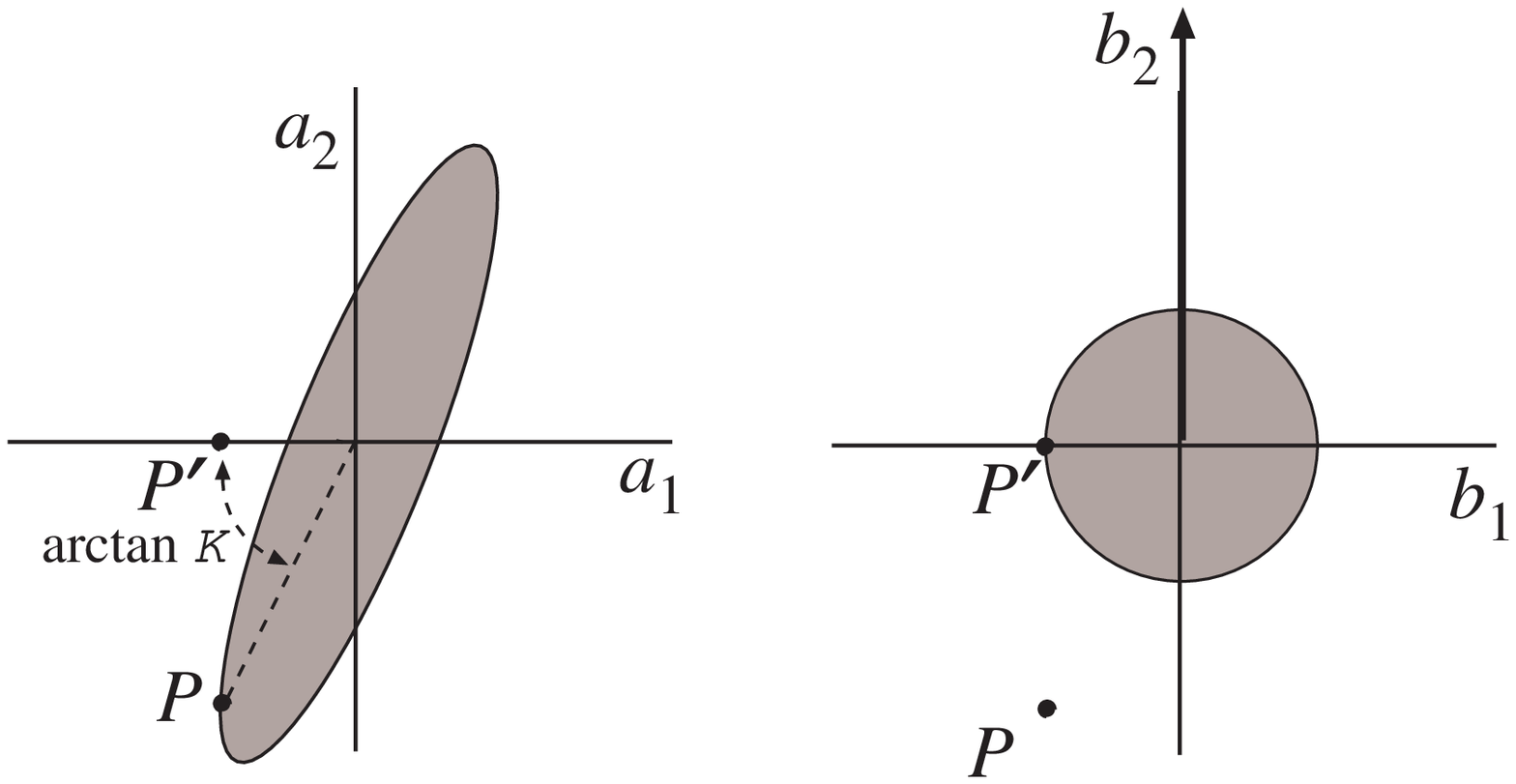}  
\caption{Noise ellipses for a squeezed-input interferometer whose input
squeeze is inverse to the interferometer's ponderomotive squeeze (``IIS
interferomter''). 
\label{fig:Fig8}
}
\end{figure}

Figure \ref{fig:Fig8} illustrates
this:  The IIS input light is squeezed in a manner that gets perfectly undone
by the ponderomotive squeeze, so the output light has no squeeze at all.
The fact that the input squeeze is inverse to the ponderomotive squeeze 
shows up in this diagram as an input noise ellipse that is the same as
the output ellipse of the ponderomotively squeezed vacuum, Fig.\
\protect\ref{fig:Fig5}, except for a reflection in the
horizontal axis.  

Because the output of the IIS interferometer is $b_2$ (ordinary photodetection)
and the output light's state is the ordinary vacuum, its
gravitational-wave noise is
\begin{equation}
S_h = {h_{\rm SQL}^2 \over 2{\cal K}}\;;
\label{ReciprocalNoise}
\end{equation}
cf.\ Eqs.\ (\ref{hnb2}), (\ref{Shhn}) and (\ref{Sa1a2Vacuum}) 
(with $a_j$ replaced by
$b_j$).  
Notice that this is identically the same noise spectral density 
as for our previous example [Eq.\ (\ref{Example1})] and as for
a variational-output interferometer, and it is achieved in all three
cases with the same light power.  

The fact that our two squeezed-input examples produce the same noise spectrum
using different squeeze angles and squeeze factors should not be surprising.
The noise spectrum is a single function of $\Omega$ and it is being
shaped jointly by the two squeeze functions $\lambda(\Omega)$ and $R(\Omega)$.

The fact that the IIS interferometer and the variational output interferometer
produce the same noise spectra results from a {\it reciprocity} between
the IIS 
and the variational-output configurations:  The IIS interferometer has its
input squeezed at the angle $-\Phi = -{\rm arccot} {\cal
K}$ and it 
has vacuum-noise output, whereas the variational-output interferometer
has vacuum-noise input and is measured at the homodyne angle $+\Phi = + {\rm
arccot}
{\cal K}$.  

Note that
the IIS interferometer has a different input squeeze
angle [$\lambda = -{1\over2} {\rm arccot}({\cal K}/2)$; cf.\ Eq.\ 
(\ref{RotSqueezeParams})] from that of the {\it angle-optimized} squeezed-input
interferometer of Sec.\ \ref{sec:SqueezedInputInterferometer} 
[$\lambda = - {\rm arccot} 
{\cal K}$; cf.\ Eq.\ (\ref{lambdaOpt})].  This difference shows clearly
in the noise ellipses of Fig.\ \ref{fig:Fig8} 
(the IIS interferometer)
and Fig.\ \ref{fig:Fig6} (the angle-optimized interferomter).  
Meoreover,
this difference implies that by optimizing the IIS interferometer's squeeze
angle (changing it to $\lambda = - {\rm arccot}{\cal K}$), while keeping its
squeeze factor unchanged [$R = {\rm arcshinh}({\cal K}/2)$; cf.\ 
Eq.\ (\ref{RotSqueezeParams})], we can improve its noise performanance
slightly. 
The improvement is from 
(\ref{ReciprocalNoise}) to
\begin{equation}
S_h = {h^2_{\rm SQL}\over2{\cal K}} \left[{1+{\cal K}^2 \over
1+ {1\over2}\left({\cal K}^2 + {\cal K}\sqrt{{\cal K}^2 + 4}\right) }
\right]
\label{ReciprocalNoiseOptimize}
\end{equation}
[which can be derived by 
setting $\lambda = -\Phi = - {\rm arccot}{\cal K}$ and $R= {\rm arcshinh}({\cal
K}/2)$ in Eq.\ (\ref{SqueezeSb2New}), or by 
inserting $R = {\rm arcshinh}({\cal K}/2)$
into Eq.\ (\ref{ShSqueezedInput})---note that 
(\ref{ShSqueezedInput})
is valid for any angle-optimized,
squeezed-input interferometer but not for the IIS interferometer].  
The improvement factor in square brackets is quite modest; it
lies between 0.889 and unity.

We reiterate, however, that the above comparison of interferometer
designs is of pedagogical interest only.  In
the real world, the noise of a QND interferometer is strongly influenced 
by losses, which we consider in Sec.\ \ref{sec:LossyInterferometer} below.

\subsection{Squeezed-variational interferometer}

The squeezed-input and variational-output techniques are complementary.  By
combining them, one can beat the SQL more strongly than using either one alone.
We call an interferometer that uses the two techiques simultaneously a {\it
squeezed-variational interferometer}.

The dark-port input of such an interferometer is squeezed by the maximum
achievable squeeze factor $R$ at a (possibly frequency dependent)
squeeze angle 
$\lambda(\Omega)$, so 
\begin{equation}
|{\rm in}\rangle = S(R,\lambda)|0_a\rangle\;.
\label{InSqeezed'}
\end{equation}
The dark-port output is subjected to FD homodyne detection with (possibly
frequency dependent) homodyne angle 
$\zeta(\Omega)$; i.e., the measured quantity is the 
same output quadrature as for a variational-output interferometer,
$b_\zeta$ [Eq.\ (\ref{bzetaOfa1a2h})], so the gravitational-wave
noise operator is also the same 
\begin{equation}
h_n(\Omega) = {h_{\rm SQL}\over\sqrt{2{\cal K}}}\;e^{i\beta}\;
[a_2 + a_1(\cot\zeta - {\cal
K})]e^{i\beta}
\label{hnb2Variational'}
\end{equation}
[Eq.\ (\ref{hnb2Variational})]. 

As for a squeezed-input interferometer,
the gravitational-wave noise is proportional to 
\begin{equation}
\langle{\rm in}|h_n h_{n'}|{\rm in}\rangle = \langle 0_a | h_{ns}
h_{ns'}|0_a\rangle
\label{EVhn'}
\end{equation}
[Eq.\ (\ref{Shhn})], where $h_{ns}$ is the squeezed gravitational-wave
noise operator
\begin{equation}
h_{ns} = S^{\dag}(R,\lambda) h_n S(R,\lambda)\;.
\label{hns'}
\end{equation}
By inserting expression (\ref{hnb2Variational'}) for $h_n$ into Eq.\   
(\ref{hns'}) and invoking the action of the squeeze operator on $a_1$ and $a_2$
[Eq.\ (\ref{Squeezea12})], we obtain
\begin{eqnarray}
&\null&
h_{ns} = - {h_{\rm SQL}\over\sqrt{2{\cal K}}}
\sqrt{1+\tilde{\cal K}^2}\; e^{i\beta}
\nonumber\\
&& \quad \times
\left( a_1 \{\cosh R \cos\tilde\Phi - \sinh R \cos[\tilde\Phi - 
2(\tilde\Phi+\lambda)]\} \right.
\nonumber\\
&&\quad\;
\left. - a_2 \{\cosh R \sin\tilde\Phi - \sinh R \sin[\tilde\Phi - 
2(\tilde\Phi+\lambda)]\}\right)
\;,
\label{hns''}
\end{eqnarray}
where
\begin{equation}
\tilde{\cal K} = {\cal K} - \cot\zeta\;,\quad
\tilde\Phi = {\rm arccot \tilde{\cal K}}
\;.
\label{tildeKtildePhiDef}
\end{equation}

As for a squeezed-input interferometer [see passage following Eq.\  
(\ref{PhiDef})], we can read the gravitational-wave spectral density
off of Eq.\ (\ref{hns''}) by regarding $a_1$ and $a_2$ as random processes
with unit spectral densities and vanishing cross spectral density.  The result
is
\begin{eqnarray}
S_h &=& {h_{\rm SQL}^2\over2{\cal K}} (1+\tilde{\cal K}^2)
\nonumber\\
&&\times \left\{e^{-2R} + 
\sinh2R [1-\cos2(\tilde\Phi+\lambda)]\right\}\;.
\label{ShSqueezedVariational}
\end{eqnarray}

This noise is minimized by setting the input squeeze angle $\lambda$ and
output homodyne phase $\zeta$ to
\begin{equation}
\lambda = \pi/2\;, \quad \zeta = \Phi = {\rm arccot}{\cal K}\;,
\label{lambdazetaOpt}
\end{equation}
which produces $\tilde{\cal K} = 0$ and $\lambda=\tilde\Phi = \pi/2$, so
\begin{eqnarray}
S_h &=& {h_{\rm SQL}^2\over2{\cal K}} e^{-2R} \nonumber\\
&=&
{e^{-2R}\over I_o/I_{\rm SQL}} \left({4\hbar\over mL^2\Omega^2}\right)
{\Omega^2 (\gamma^2 + \Omega^2)\over 2\gamma^4}\;;
\label{ShSqueezedVariationalOptimized}
\end{eqnarray}
see Fig.\ \ref{fig:Fig4}
 
\begin{figure}
\epsfxsize=3.2in\epsfbox{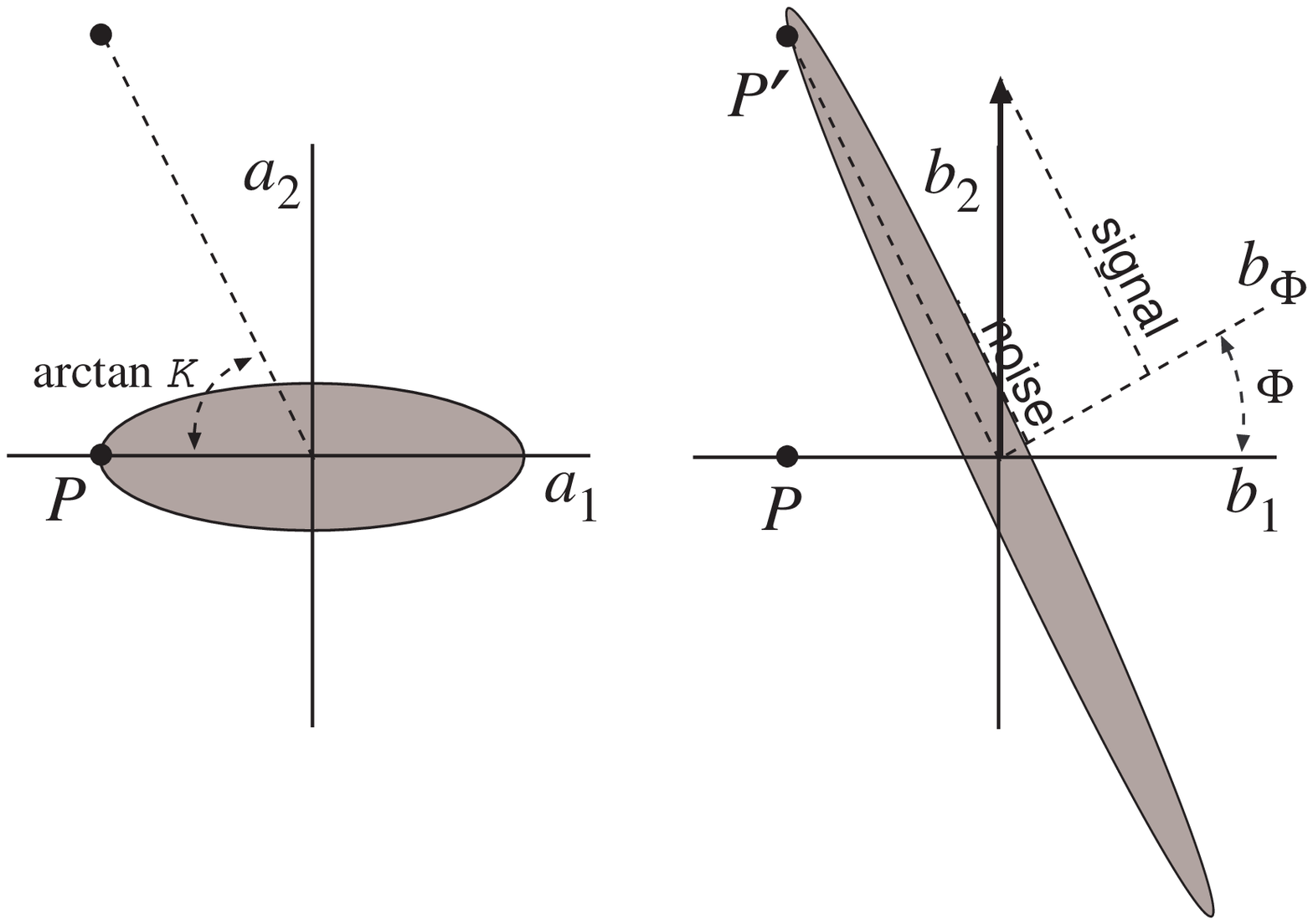}
\caption{Noise ellipses for a squeezed-variational interferometer.
{\it Left:} Noise for the squeezed vacuum that enters the interferometer's
dark port.  {\it Right:} Noise for
the field that exits at the dark port along with the
gravitational-wave signal.  
\label{fig:Fig9}
}
\end{figure}

Equation (\ref{lambdazetaOpt}) says
that, to optimize the (lossless) squeezed-variational interferometer, one
should squeeze the dark-port input field at the frequency-independent squeeze
angle $\zeta = \pi/2$ (which ends up squeezing the interferometer's shot
noise), and measure the output field at the same FD homodyne phase $\zeta =
\Phi$ as for a variational-output interferometer; see Fig.\ 
\ref{fig:Fig9}.  Doing so produces an output, Eq.\ 
(\ref{ShSqueezedVariationalOptimized}), in which the radiation-pressure-induced
back-action noise has been completely removed, and the shot noise has been
reduced by the input squeeze factor $e^{-2R}$.

Because the optimal input squeeze angle is frequency independent,
the squeezed variational interferometer needs no filter cavities on the input.
However, they are needed on the output to enable FD homodyne detection; see
Fig.\ \ref{fig:Fig2} for a schematic diagram. 

\section{FD Homodyne Detection and Squeezing}
\label{sec:FDHomodyne}

Each of the QND schemes discussed above requires homodyne detection with a
frequency-dependent phase (FD homodyne detection) and/or input squeezed vacuum
with a frequency-dependent squeeze angle (FD squeezed vacuum).  In this section
we sketch how such FD homodyne detection and squeezing can be achieved.

\subsection{General Method for FD Homodyne Detection}

The goal of FD Homodyne Detection is to measure the electric-field quadrature
\begin{equation}
E_\zeta(t) = \sqrt{4\pi\hbar\omega_o\over{\cal A}c}\int_0^{\infty} \left(b_\zeta
e^{-i\Omega t} + b_\zeta^{\dag}e^{+i\Omega t}\right) {d\Omega\over2\pi}\;,
\label{EzetaDef}
\end{equation}
for which the quadrature amplitude is
\begin{equation}
b_\zeta = b_1\cos\zeta + b_2\sin\zeta\;, \quad \zeta = \zeta(\Omega)\;;
\label{bzetaDef'}
\end{equation}
cf.\ Eqs.\ (\ref{E1E2Def}) and (\ref{bzetaDef}).  If $\zeta$ were frequency
independent, the measurement could be made by conventional balanced homodyne 
detection, with homodyne phase $\zeta$.  In this subsection we shall show that,
{\it when $\zeta$ depends on frequency, 
the measurement can be achieved in two steps:
first send the light through an appropriate filter (assumed to be lossless),
and then perform conventional balanced homodyne detection.}

The filter puts onto the light a phase shift $\alpha$ that
depends on frequency.  Let the phase shift be $\alpha_+$ for light frequency
$\omega_o+\Omega$, and $\alpha_-$ for $\omega_0-\Omega$.  The input to the
filter has amplitudes (annihilation operators) $b_\pm$ at these two sidebands,
and the filter output has amplitudes (denoted by a tilde)
\begin{equation}
\tilde b_\pm = b_\pm e^{i\alpha_\pm}\;.
\label{tildebpm}
\end{equation}
The corresponding quadrature amplitudes are 
\begin{equation}
b_1 = {b_+ + b_-^{\dag} \over \sqrt2}\;, \quad
b_2 = {b_+ - b_-^{\dag} \over \sqrt2 i}
\label{b12Def}
\end{equation}
at the input [Eqs.\ (\ref{a12Def})], 
and the analogous expression with tildes at the output.
Combining Eqs. (\ref{b12Def}) with and without tildes, and Eq.\ 
(\ref{tildebpm}), we obtain for the output quadrature amplitudes in terms of
the input 
\begin{eqnarray}
\tilde b_1 &=& e^{i\alpha_{\rm m}} ( b_1 \cos\alpha_{\rm p} - b_2
\sin\alpha_{\rm p})\;,\nonumber\\
\tilde b_2 &=& e^{i\alpha_{\rm m}}  (b_2 \cos\alpha_{\rm p} + b_2
\sin\alpha_{\rm p})\;.
\end{eqnarray}
Here
\begin{equation}
\alpha_{\rm m} = {1\over2} (\alpha_+ - \alpha_-)\;, \quad
\alpha_{\rm p} = {1\over2} (\alpha_+ + \alpha_-)\;.
\label{alphamp}
\end{equation}

The light with the output amplitudes $\tilde b_1$, $\tilde b_2$
is then subjected to conventional balanced
homodyne detection with frequency-independent homodyne angle $\theta$, 
which measures an electric-field quadrature with amplitude
\begin{eqnarray}
\tilde b_\theta &=& \tilde b_1\cos\theta + \tilde b_2 \sin\theta
\nonumber\\
&=& e^{i\alpha_{\rm m}} [ b_1\cos(\theta - \alpha_{\rm p}) + b_2\sin(\theta -
\alpha_{\rm p})]\;.
\label{btheta}
\end{eqnarray}

If we adjust the filter and the constant homodyne phase so that
\begin{equation}
\theta - \alpha_{\rm p} \equiv \theta - {1\over2}(\alpha_+ + \alpha_-) =
\zeta(\Omega)\;,
\label{FilterParameterszeta}
\end{equation}
then, aside from 
the frequency-dependent phase shift $\alpha_{\rm m}$, the output
quadrature amplitude will be equal to our desired FD amplitude:
\begin{equation}
\tilde b_\theta = e^{i\alpha_{\rm m}} b_\zeta\;.
\label{bthetabzeta}
\end{equation}
The phase shift $\alpha_{\rm m}(\Omega)$ is actually unimportant; it can be
removed from the signal in the data analysis (as can be the phase shift
$\beta(\Omega)$ produced by the interferometer's arm cavities).

To recapituate: {\it FD homodyne detection with homodyne phase 
$\zeta(\Omega)$ can be achieved by filtering 
and conventional homodyne
detection, with the filter's phase shifts
$\alpha_\pm$ (at $\omega = \omega_o \pm \Omega$) and
the constant homodyne phase $\theta$ adjusted to satisfy Eq.\ 
(\ref{FilterParameterszeta}).}

\subsection{Realization of the filter}

The desired FD homodyne phase is 
\begin{eqnarray}
\zeta = \Phi(\Omega) = {\rm arccot}{\cal K} &=& {\rm
arccot}\left({\Lambda^4\over\Omega^2(\gamma^2+\Omega^2)} \right) \nonumber\\
&=& {\rm
arctan}\left(\Omega^2(\gamma^2+\Omega^2)\over\Lambda^4\right)\;,
\label{PhiDesired}
\end{eqnarray}
where 
\begin{equation}
\Lambda^4 = (I_o/I_{\rm SQL})2\gamma^4
\label{LambdaDef}
\end{equation}
[cf.\ Eqs.\ (\ref{KDef}) and (\ref{PhiDef})].  Recall that $\gamma \simeq
2\pi\times 100$Hz is the optimal frequency of operation of the interferometer,
and to beat the SQL by a moderate amount will require $I_o/I_{\rm SQL} \sim 10$
so $\Lambda^4 \sim 20 \gamma^4$, i.e., $\Lambda \sim 2 \gamma$. 

In Appendix \ref{app:filters} we show that this desired FD phase
can be achieved
by 
filtering the light with 
two successive lossless Fabry-Perot filter
cavities, followed by conventional homodyne detection at homodyne angle
\begin{equation}
\theta = \pi/2
\label{theta}
\end{equation} 
[i.e., homodyne measurement of $\tilde b_2$ at the filter
output; cf.\ Eq.\ (\ref{btheta})].\footnote{The fact that only 
two cavities are needed to produce the desired FD
homodyne phase (\ref{PhiDesired}) is a result of the simple quadratic
form of $\tan\Phi(\Omega^2)$.  If the desired phase were significantly more
complicated, a larger number of filter cavities would be needed; cf.\
Eq.\ (\ref{Requirement}) and associated analysis.  It
would be interesting to explore what range of FD homodyne phases
can be achieved, with what accuracy, using what number of cavities.
}
The two filter cavities (denoted I and
II) produce phase shifts $\alpha_{{\rm I}\pm}$ and $\alpha_{\rm II}\pm$ on the
$\omega_o\pm\Omega$ side bands, so upon emerging from the second cavity, the
net phase shifts are
\begin{equation}
\alpha_\pm = \alpha_{{\rm I}\pm} + \alpha_{{\rm II}\pm}\;.
\label{alphapmIandII}
\end{equation}

Each cavity ($J=$I or II) is
characterized by two parameters: its decay rate (bandwidth)
$2\delta_J$ (with $J= \hbox{I or II}$), and its fractional
resonant-frequency offset from the light's carrier frequency $\omega_o$, 
\begin{equation}
\xi_J \equiv {\omega_o - \omega_{{\rm res}\,J} \over \delta_J}\;.
\label{xiJDef}
\end{equation}
Here $\omega_{{\rm res}\,J}$ is the resonant frequency of cavity $J$.  In terms
of these parameters, the phase shifts produced in the $\omega_o\pm\Omega$ side
bands by cavity $J$ are
\begin{equation}
\alpha_{J\pm} = {\rm arctan}(\xi_J \pm \Omega/\delta_J)\;.
\label{alphaJpm}
\end{equation}
The filters' parameters must be adjusted so that the net phase shift
(\ref{alphapmIandII}), together
with the final homodyne angle $\theta = \pi/2$, produce the desired FD phase,
Eqs.\ (\ref{FilterParameterszeta}) and (\ref{PhiDesired}).

In Appendix \ref{app:filters} we derive the following values for the filter
parameters $\xi_{\rm I}$, $\delta_{\rm I}$, $\xi_{\rm II}$ and $\delta_{\rm
II}$ as functions of the parameters $\Lambda$ and $\gamma$ that appear in the
desired FD homodyne phase.  Define the following four
functions of $\Lambda$ and $\gamma$
\begin{mathletters}
\label{FilterParameters}
\begin{eqnarray}
P &\equiv& {4\Lambda^4\over\gamma^4}\;, \quad
Q \equiv {1+\sqrt{1+P^2}\over2}\;, 
\label{PQ} \\
A_+ &\equiv& {Q+\sqrt{Q}\over P}\;, \quad
A_- \equiv {Q-\sqrt{Q}\over P}\;.
\label{ApAm}
\end{eqnarray}
Then in terms of these functions, the filter parameters are
\begin{eqnarray}
\xi_{\rm I} &=& {1\over2A_+} + \sqrt{1+{1\over(2A_+)^2}} \;, 
\label{xiI} \\
\xi_{\rm II} &=& {1\over2A_-} - \sqrt{1+{1\over(2A_-)^2}} \;, 
\label{xiII} \\
{\delta_{\rm I}\over\gamma} &=& 
\sqrt{P\over 8 \xi_{\rm I} \sqrt{Q}} \;, 
\label{deltaI} \\
{\delta_{\rm II}\over\gamma} &=& 
\sqrt{P\over 8 (-\xi_{\rm II}) \sqrt{Q}} \;.
\label{deltaII}
\end{eqnarray}
\end{mathletters}

\begin{figure}
\epsfxsize=3.2in\epsfbox{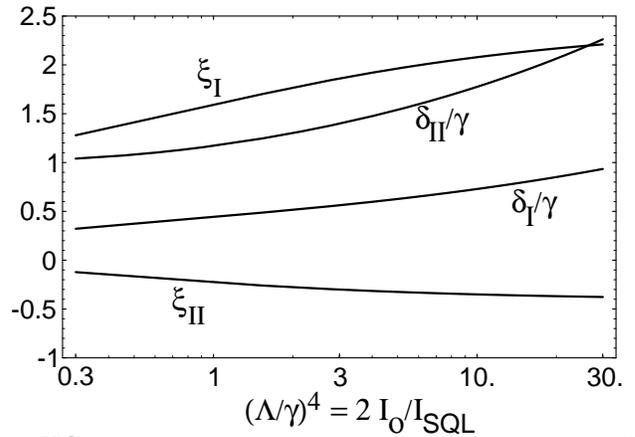}
\caption{The parameters characterizing the two Fabry-Perot cavities that
are used, together with conventional homodyne detection at phase $\theta =
\pi/2$, to produce FD homodyne detection at the desired frequency-dependent
phase (\protect\ref{PhiDesired}).  The quantities $\xi_{\rm I}$ and $\xi_{\rm
II}$ are the filters' fractional frequency offsets from the light's carrier
frequency (\protect\ref{xiJDef});  $\delta_{\rm I}/\gamma$ and $\delta_{\rm
II}/\gamma$ are the filters' half bandwidths in units of the half-bandwidth of
the interferometer's identical arm cavities.  The functional forms of these
parameters are Eqs.\ (\protect\ref{FilterParameters}).
\label{fig:Fig10}
}
\end{figure}

Note that, when the cavity half-bandwidths $\delta_J$ are expressed in terms of
the half-bandwidth $\gamma$ of the interferometer's arm cavities, as in Eqs.\ 
(\ref{deltaI}) and (\ref{deltaII}), then the filter parameters depend on only
one characteristic of the desired FD homodyne phase: the quantity
$(\Lambda/\gamma)^4 = 2I_o/I_{\rm SQL}$.
Figure \ref{fig:Fig10} depicts the filter parameters as functions 
of this quantity.

As Fig.\ \ref{fig:Fig10} shows, the half-bandwidths of the two
filter cavities are within a factor $\sim 2$ of that of the interferometer's
arm cavities.  This is so for the entire range of laser 
powers, $I_o/I_{\rm SQL}$, that
are likely to be used in QND interferometers, at least in the early years
(e.g., LIGO-III; ca.\ 2008--2010).  Moreover, the filter cavities'
fractional frequency offsets
$\xi_J$ are of order unity ($-0.5 < \xi_J \alt 2$).  Thus, the desired
properties of the filter cavities are not much different from those of the
interferometer's arm cavities.

In Sec.\ \ref{sec:LossyInterferometer} 
below, we shall see that the most serious limitation on the
sensitivities of variational-output and squeezed-variational interferometers
is optical loss
in the filter cavities.  To minimize losses,
the cavities should be very long (so the cavities' stored light encounters 
the mirrors a minimum number of times).  This suggests placing the filter
cavities in the interferometer's 4km-long arms, alongside the interferometer's
arm cavities. 

\subsection{Squeezing with frequency-dependent squeeze angle}
\label{sec:FDSqueeze}

Just as the variational-output and squeezed-variational interferometers require
homodyne detection at a FD phase, so a squeezed-input interferometer requires
squeezing at a FD angle $\lambda(\Omega)$. 

The nonlinear-optics techniques currently used for squeezing will
produce a squeeze angle that is nearly constant over the very
narrow frequency band of gravitational-wave interferometers, 
$|\omega - \omega_o| \alt \hbox{(a few)}\times \gamma \sim 10^{-12}\omega_o$.
What we need is a way to change the squeeze angle from its constant 
nonlinear-optics-induced value to the desired frequency-dependent value, 
$\lambda = - \Phi(\Omega)$ 
[Eq.\ (\ref{lambdaOptOmega})].

Just as FD homodyne detection can be achieved by sending the light field through
appropriate filters followed by a frequency-independent homodyne device, 
so also FD squeezing can be achieved by squeezing the input field in the
standard frequency-independent way, and then sending it through appropriate
filters.  Moreover, since the necessary squeeze angle has the same frequency
dependence $-\Phi(\Omega)$ as the homodyne phase (aside from sign), the
filters needed in FD squeezing are 
nearly the same as those needed in
FD homodyne detection:  The filtering can be achieved by sending the squeezed
input field through two Fabry-Perot cavities before injecting it into the
interferometer, and the cavity parameters are given by 
formulae analogous to Eqs.\ (\ref{FilterParameters}): 
\begin{mathletters}
\label{FilterParametersSqueeze}
\begin{eqnarray}
Q &\equiv& {1 + \sqrt{65}\over 2}\;, \quad A_\pm \equiv - {Q \pm\sqrt
Q \over 8}\;, 
\label{QApmnew} \\
\xi_{\rm I} &=& {1\over2A_+} - \sqrt{1+{1\over(2A_+)^2}} \;, 
\label{xiIsqueeze} \\
\xi_{\rm II} &=& {1\over2A_-} + \sqrt{1+{1\over(2A_-)^2}} \;. 
\label{xiIIsqueeze} \\ 
{\delta_{\rm I}\over\gamma} &=&
\sqrt{1\over (-\xi_{\rm I}) \sqrt{Q}} \;,
\label{deltaIsqueeze} \\
{\delta_{\rm II}\over\gamma} &=&
\sqrt{1\over \xi_{\rm II} \sqrt{Q}} \;.
\label{deltaIIsqueeze}
\end{eqnarray}
\end{mathletters}
The details of the calculations are essentially the same as Appendix C,
but with Eq.\ (\ref{FilterRequirement}) 
changed into the following expression for the 
initial frequency-independent squeeze angle $\theta$ and the
cavities' frequency-dependent phase shifts $\alpha_{J\pm}$: 
\begin{eqnarray}
\tan\Phi(\Omega) &\equiv& - {\Omega^2(\gamma^2+\Omega^2)\over2\gamma^4} 
\nonumber\\
&=& \tan\left(\theta - {\alpha_{I+} + \alpha_{I-} + \alpha_{II+} +
\alpha_{II-}\over2}\right)\;.
\label{SqueezeFiltFun}
\end{eqnarray}

\section{Influence of Optical Losses on QND Interferometers} 
\label{sec:LossyInterferometer}

\subsection{The role of losses}

It is well known that, when one is working with squeezed light, any source of
optical loss (whether fundamentally irreversible or not)
can debilitate the light's squeezed state.  This is
because, wherever the squeezed light can leave one's optical system, 
vacuum field can (and must) enter by the inverse route; and the entering vacuum 
field will generally be unsqueezed \cite{LesHouches}. 

All of the QND interferometers discussed in this paper rely on
squeezed-light correlations in order to beat the SQL ---
with the squeezing always produced ponderomotively inside the interferometer
and, in some designs, also present in the dark-port input field.  Thus,
optical loss is a serious issue for all the QND interferometers.

In this section we shall study the influence of optical losses on the 
optimized sensitivities of 
our three types of QND interferometers.

\subsection{Sources of optical loss}
\label{sec:DissipationDescription}

The sources of optical loss in our interferometers are the following:
\begin{enumerate}
\item[$\bullet$] For light inside the interferometer's arm cavities and 
inside the Fabry-Perot
filter cavities: scattering and absorption on the mirrors and finite
transmissivity through the end mirrors.  We shall discuss these quantitatively
at the end of the present subsection.
(In addition, wave front errors and birefringence produced in the arm
cavities and filters, e.g.\ via
power-dependent changes in the shapes and optical properties of the mirrors,
will produce mode miss-matching and thence losses in subsequent elements 
of the output optical train.)

\item[$\bullet$] For squeezed vacuum being injected into the interferometer:
fractional photon losses $\epsilon_{\rm circ}$ in the 
{\it circulator}\footnote{ \label{foot:Circulator}
The circulator is 
a four-port optical device that separates spatially the injected input
and the returning output from the interferometer; see Fig.\ \ref{fig:Fig1}. 
It can be implemented 
via a Faraday rotator in
conjuction with two linear polarizers.  
}
used to do the injection, in the beam splitter $\epsilon_{\rm bs}$, and in 
mode-matching into the interferometer $\epsilon_{\rm mm}$. 
\item[$\bullet$] For the signal light traveling out of the interferometer:
In addition to losses in the arm cavities and filter cavities, also 
fractional photon losses in
the beam splitter $\epsilon_{\rm bs}$, in the circulator 
$\epsilon_{\rm circ}$, in mode-matching into each of the filter
cavities $\epsilon_{\rm mm}$, in mode-matching with the local-oscillator light
used in the homodyne detection $\epsilon_{\rm lo}$, and in the photodiode
inefficiency $\epsilon_{\rm pd}$.
\end{enumerate}

It is essential to pursue R\&D with the aim of driving these 
fractional 
photon losses
down to 
\begin{equation}
\epsilon_{\rm circ} \sim \epsilon_{\rm bs} \sim \epsilon_{\rm mm} \sim 
\epsilon_{\rm lo} \sim \epsilon_{\rm pd}\sim 0.001\;.
\label{epsilonValues}
\end{equation}
These loss levels are certainly daunting. However, it is well to keep
in mind that attaining the absolute lowest loss levels will likely be an
essential component of any advanced interferometer that attempts to challenge
and surpass the SQL. In the current case, discussions with Stan Whitcomb and
the laboratory experience of one of the authors (HJK) lead us to suggest that
it may be technically plausible to achieve the levels of Eq. 
(\ref{epsilonValues}) in the LIGO-III
time frame, though a vigorous research effort will be needed to
determine the actual feasibility.

These loss levels are certainly daunting, but
based on discussions with Stan Whitcomb and on the laboratory experience of 
one of the authors (HJK), 
it seems technically plausible that they can be achieved in the
LIGO-III time frame. 

The arm cavities are a dangerous source of losses because the light bounces
back and forth in them so many times.
We denote by $\cal L$
the probability that a photon in an arm cavity gets lost during one round-trip
through the cavity, due to scattering and absorption in each of the two mirrors
and transmission through the end mirror.  
With much R\&D, by the LIGO-III time frame this {\it loss coefficient}
per round trip may be as low as
\begin{equation}
{\cal L} \sim 20\times 10^{-6}\;. 
\label{calLDef}
\end{equation}
A fraction 
\begin{equation}
\epsilon \equiv {2{\cal L}\over T} = 
{{\cal L}\over 2\gamma L/c} \simeq 0.0012 
\label{epsilonDef}
\end{equation}
of the carrier photons that impinge on each arm cavity gets lost 
in the cavity
[cf.\ Eq.\ (\ref{gj1}) on resonance so
${\cal E} = \epsilon$].  
(Note the absence of any subscript on this particular $\epsilon$.)
For side-band light the net fractional loss [denoted ${\cal 
E}(\Omega)$;
Eq.\ (\ref{calEDef}) below] is also of order $\epsilon$.

Each filter cavity, $J = {\rm I}$ or II, has an analogous loss coefficient
${\cal L}_J \simeq {\cal L}$ and
fractional loss of resonant photons
\begin{equation}
\epsilon_J \equiv {2{\cal L}_J \over T_J} \simeq 
{{\cal L}\over 2\delta_J L_J/c}\;.
\label{epsilonJDef}
\end{equation}
Because (as we shall see), the filter cavities' losses place severe limits
on the interferometer sensitivity, we shall minimize their net fractional
loss in our numerical
estimates by making the filter cavities as long as possible: 
$L_J = L = 4$km.  Then the ratio of Eqs.\ (\ref{epsilonJDef}) and
(\ref{epsilonDef}) gives
\begin{equation}
\epsilon_J = \epsilon 
(\gamma/\delta_J) \sim (0.5 \hbox{ to } 2) \epsilon\;.
\label{epsilonJepsilon}
\end{equation}

\subsection{Input-output relation for lossy interferometer}
\label{sec:InOutLossy}

We show in Appendix \ref{app:interferometer} that, 
accurate to first order in the arm-cavity losses
(and ignoring beam-splitter losses which we shall deal with separately below),
the relation between the input to the
interferometer's beam splitter (field amplitudes $a_j$)
and the output from the beam splitter (field amplitudes $b_j$) takes 
the following form 
\begin{equation}
b_1 = \Delta b_1\;, \quad b_2 = \Delta b_2 +
\sqrt{2{\cal K}_*} {h\over h_{\rm SQL}}e^{i\beta_*}
\label{bjFromajLossy}
\end{equation}
[cf.\ 
the last sentence of App.\ B; also
the lossless input-output relation (\ref{bjFromaj}) and Fig.\
\ref{fig:Fig3}].  
Here, accurate to first order in $\epsilon$, 
\begin{equation}
\beta_* \equiv {\rm arctan}\left({\Omega/\gamma \over 1+\epsilon/2}\right)
= \beta - {\epsilon/2 \over \Omega/\gamma + \gamma/\Omega}
\label{betaDefLossy}
\end{equation}
is the loss-modified\footnote{
The loss modification, i.e.\ the difference between $\beta_*$ and
$\beta$, turns out to influence the gravitational-wave noise only at second order in
$\epsilon$ and thus is unimportant; see Footnote \ref{betastarnote} below.
}
phase $\beta$ [Eq.\ (\ref{betaDef})], 
and the coupling coefficient is reduced slightly by the 
losses:\footnote{As is
discussed in Footnote \ref{fn:CarrierAttenuation}, in 
Eq.\ (\ref{KDefLossy}) for ${\cal K}_*$, strictly
speaking, $I_o$ is not the input power to the interferometer, but rather is
the input power reduced by the losses that occur 
in the input optics, beamsplitter, and arm cavities.  We ignore this delicacy
since its only effect in our final formulas is a slight renormalization of
$I_o$.}
\begin{equation}
{\cal K}_* \equiv {(I_o/I_{\rm SQL})
2\gamma^4 \over \Omega^2[\gamma^2(1+\epsilon/2)^2 + \Omega^2]}
= {\cal K}\left(1-{1\over2}{\cal E}\right)\;
\label{KDefLossy}
\end{equation}
[cf.\ Eq.\  (\ref{KDef})], where
\begin{equation}
{\cal E} = {2\gamma^2 \over \gamma^2 + \Omega^2} \epsilon = 
{2\epsilon\over 1+(\Omega/\gamma)^2}\;
\label{calEDef}
\end{equation}
is the net fractional loss of sideband photons in the arm cavities
[cf.\ Eq.\ (\ref{gj1})].   
Accurate to first order in the losses, 
the output quadrature noise operators in Eq.\ (\ref{bjFromajLossy}) have the
form
\begin{eqnarray}
\Delta b_1 &=& a_1 e^{2i\beta}\left(1-{1\over2}{\cal E}\right) + \sqrt{\cal
E}e^{i\beta}n_1\;, \nonumber\\
\Delta b_2 &=& a_2 e^{2i\beta}\left(1-{1\over2}{\cal E}\right) + \sqrt{\cal
E}e^{i\beta}n_2 \nonumber\\
&&- {\cal K}_* (a_1 + \sqrt{\epsilon/2}\;n_1)e^{2i\beta_*}
\label{DeltabjLossy}
\end{eqnarray}
[cf.\ last sentence of App.\ B and cf.\ Eq.\ (\ref{bjFromaj})]. 
Here
$n_1$ and $n_2$ are the net 
quadrature field amplitudes that impinge on the interferometer's arm 
cavities at their various sites of optical loss.  We shall call $n_j$ the
quadrature amplitudes of the arm cavities' 
{\it loss-noise field}.  
They are complete analogs of the input and output fields' quadrature 
amplitudes $a_j$ and $b_j$:  they are related to the loss-noise field's
annihilation and creation operators $n_{\pm}$ and $n_\pm^{\dag}$ in the
standard way [analog of Eqs.\ (\ref{a12Def})], they have the standard
commutation relations [analog of Eqs.\ (\ref{a12Commutator})], and they
commute with the dark-port input field amplitudes $a_j$.

Equations (\ref{DeltabjLossy}) have a simple physical interpretation.  The
dark-port input field $a_j$ at frequency $\omega_o \pm\Omega$
gets 
attenuated
by a fractional amount ${\cal E}/2$ while in the interferometer
(corresponding to a photon-number fractional loss ${\cal E}$),
and the lost field gets replaced, in the output light, by a small bit 
of loss-noise field
$\sqrt{\cal E}n_j$.  The phase shift $\beta$
that the interferometer cavities put onto
the loss-noise field is half that put onto the dark-port input
field because of the different routes by which the $a_j$ and $n_j$ get into the
arm cavities.  

The radiation-pressure back-action force on the test mass 
is produced by a beating of the laser's carrier light against the in-phase
quadrature of the inside-cavity noise field 
$a_1 + \sqrt{\epsilon/2}\;n_1$.  
Thus,
it is $a_1 + \sqrt{\epsilon/2}\;n_1$ 
that appears in the output light's back-action
noise (last term of $\Delta b_2$).

\subsection{Noise from losses in the output optical train and the
homodyne filters}

The output quadrature operators $b_j$ get fed through an output
optical train including the beam splitter, circulator (if present), filter
cavities (if present in the output as opposed to the input), local-oscillator
mixer, and photodiode.  Losses in all these elements will modify the
$\Delta b_j$.  In analyzing these modifications, 
we shall not assume, initially, that the FD homodyne phase
is $\Phi(\Omega)$; rather, we shall give it an arbitrary value $\zeta(\Omega)$
(as we did in our lossless analysis, Sec.\ \ref{sec:VariationalOutput}), and
shall optimize $\zeta$ at the end.  The optimal $\zeta$ will turn out to be
affected negligibly by the losses; i.e., it will still be $\Phi(\Omega) \equiv
{\rm arccot}{\cal K}$.

\begin{figure}
\epsfxsize=3.45in\epsfbox{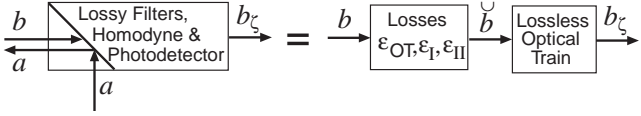}
\caption{
The output light $b$ is sent through a lossy output optical train, including a
beam splitter, circulator, cavity filters I and II, a mixer with local
oscillator light and a photodiode.  The
result (aside from an unimportant phase shift $\alpha_{\rm m}$) is the desired
measured quantity $b_{\rm \zeta(\Omega)}$.  This actual process, sketched on
the left side of the equality sign, is mathematically equivalent to the 
idealized process sketched on the right side:  The cavities' 
loss effects are 
introduced first, producing $\breve b$, which is then sent through 
an idealized, lossless optical train including the filters. 
\label{fig:Fig11}
}
\end{figure}

By analogy with the effects of arm-cavity
losses [factors $\cal E$ in Eqs.\ (\ref{DeltabjLossy})],
the effects of the optical-train losses 
on the output fields $b_j$ can
be computed in the manner sketched in Fig.\ \ref{fig:Fig11}:  The
process of sending the quadrature amplitudes $b_j$ through the optical train
is equivalent to (i) sending $b_j$ 
through a ``loss device'' to obtain loss-modified fields $\breve b_j$, and then
(ii) sending $\breve b_j$ through the lossless optical train.\footnote{Yanbei
Chen 
\protect\cite{Chen} has shown that it does not matter whether the losses are 
placed before or after the lossless train.}

Because the filter cavities 
have frequency offsets $\xi_J$
that make their losses different in the upper and lower
side bands,  
the influence of the losses 
is most simply
expressed in terms of the annihilation operators for the side bands $\breve
b_\pm$, rather
than in terms of the quadrature amplitudes $\breve
b_j$.  In terms of $b_\pm$,
the equation describing the influence of losses is identical to
that in the case of the arm cavities with fixed mirrors, Eqs.\ 
(\ref{DeltabjLossy}) with ${\cal K} =0$: 
\begin{equation}
\breve b_\pm = \left( 1- {1\over2}\sum_J {\cal E}_{J\pm}\right)b_\pm + \sum_J
\sqrt{{\cal E}_{J\pm}}\, n_{J\pm}\;,
\label{brevebpm}
\end{equation}
Here (i) the sum is over 
the two filter cavities 
$J= {\rm I}$ and II
(which must be treated specially) and over the rest of the output
optical train, denoted $J={\rm OT}$;
(ii) ${\cal E}_J$ is the net fractional loss of photons 
in element $J$; (iii) $n_{J\pm}$ is the annihilation operator for
the loss-noise field introduced by element $J$; (iv) for each filter I or II,
the analog of 
the phase factor $2\beta$ of Eq.\ (\ref{DeltabjLossy}) 
gets put onto the light in the subsequent lossless filter and thus is
absent here; and (v) we have absorbed a phase factor into the definition of
$n_{J\pm}$.  

The net fractional photon loss in a filter cavity must be identical 
to that
in an arm cavity, Eq.\ (\ref{calEDef}), if written in terms of the
cavity's half bandwidth ($\gamma$ for arm cavity, $\delta_J$ for filter cavity)
and the
difference between the field's frequency $\omega = \omega_o \pm\Omega$ and 
the cavity's resonant frequency $\omega_{\rm res}$
($\omega - \omega_{\rm res} = \pm\Omega$ for arm cavity; 
$\omega - \omega_{\rm res} = \xi_J \delta_J\pm\Omega$ for filter cavity).  
Therefore, Eq.\ (\ref{calEDef}) implies that
\begin{equation}
{\cal E}_{J\pm} = {2\epsilon_J\over1+(\xi_J\pm\Omega/\delta_J)^2} 
\quad \hbox{for }
J={\rm I},{\rm II}\;.
\end{equation}
For the remainder of the optical train, the net fractional photon loss ${\cal
E}_{\rm OT}$ is the sum of the contributions from the various elements
and is independent of frequency:
\begin{equation}
{\cal E}_{{\rm OT}\pm} =  {\cal E}_{\rm OT} 
= \epsilon_{\rm bs} + \epsilon_{\rm circ} + 2 \epsilon_{\rm
mm} + \epsilon_{\rm lo} + \epsilon_{\rm pd} \sim 0.006\;.
\label{calEOT}
\end{equation}

By expressing $b_\pm$ and $n_{J\pm}$ in terms of $b_j$ and $n_{Jj}$ (for
$j=1,2$) via the analog of Eq.\ (\ref{a12Def}), inserting these expressions
into Eq.\ (\ref{brevebpm}), then computing $\breve b_j$ via the analog of Eq.\
(\ref{a12Def}), we obtain
\begin{mathletters}
\begin{eqnarray}
&\null&
\breve b_1 = \left(1-{1\over2}{\cal E}_{\rm OTF}\right)b_1
-{i\over4}\sum_J({\cal E}_{J+}-{\cal E}_{J-})b_2
\label{breveb1'} \\
&&+{1\over2}\sum_J\left[(\sqrt{{\cal E}_{J+}}+\sqrt{{\cal E}_{J-}})n_{J1}
+i(\sqrt{{\cal E}_{J+}} -\sqrt{{\cal E}_{J-}})n_{J2}\right], \nonumber 
\end{eqnarray}
\begin{eqnarray}
& \null &
\breve b_2 = \left(1-{1\over2}{\cal E}_{\rm OTF}\right)b_2 
+{i\over4}\sum_J({\cal E}_{J+}-{\cal E}_{J-})b_1
\label{breveb2'} \\
&&+{1\over2}\sum_J\left[(\sqrt{{\cal E}_{J+}}+\sqrt{{\cal E}_{J-}})n_{J2}
-i(\sqrt{{\cal E}_{J+}}-\sqrt{{\cal E}_{J-}})n_{J1}\right]\;. \nonumber 
\end{eqnarray}
\label{brevebj'}
\end{mathletters}
Here
\begin{eqnarray}
{\cal E}_{\rm OTF} &\equiv& 
{1\over2}\sum_J({\cal E}_{J+} + {\cal E}_{J-}) \nonumber\\
&=& 
{\cal E}_{\rm OT} + {1\over2} ( {\cal E}_{{\rm I}+} + {\cal E}_{{\rm
I}-} + {\cal E}_{{\rm II}+} + {\cal E}_{{\rm II}-})
\nonumber\\
&\simeq& {\cal E}_{\rm OT} + \epsilon \sum_{J={\rm I,II}} \; \sum_{s=+,-}
{\gamma/\delta_J\over 1+(\xi_J + s \Omega/\delta_J)^2}
\label{calEfiltDef}
\end{eqnarray}
is the net, $\Omega$-dependent loss factor for the entire output optical train
including the filter cavities.  
From Eqs.\ (\ref{epsilonDef}), (\ref{calEOT}) and (\ref{calEfiltDef}) and
Fig.\ \ref{fig:Fig10}, we infer that
\begin{equation}
{\cal E}_{\rm OTF} \sim 0.009
\label{EOTF}
\end{equation}
with only a weak dependence on frequency, which we shall neglect.

In Eqs. (\ref{brevebj'}),
the terms $i\times($quantity linear in ${\cal E}_{J\pm})b_j$ [the $b_2$ term in
$\breve b_1$ and the $b_1$ term in $\breve b_2$]
will contribute amounts second order in the
losses ($\propto {\cal E}_J^2)$ to the signal and/or noise, and thus can be
neglected.  We shall flag our neglect of these terms below, when they arise. 

\subsection{Computation of noise spectra for variational-output and
squeezed-variational interferometers}
\label{sec:ComputationSpectra}

The output of a squeezed-variational interferometer or variational-output
interferometer is the frequency-dependent quadrature $b_\zeta$ depicted 
in Fig.\ \ref{fig:Fig11}.  This quantity,
when split into signal $\propto h$ plus noise $\propto \Delta \breve
b_\zeta$, takes the following form:
\begin{eqnarray}
&\null&
b_\zeta = \breve b_1 \cos\zeta + \breve b_2 \sin\zeta \nonumber\\
&&= \sin\zeta\left[\sqrt{2{\cal K}_*}\left(1-{1\over2}{\cal E}_{\rm OTF}\right)
{h\over h_{\rm SQL}}e^{i\beta_*} + {\Delta \breve
b_\zeta\over\sin\zeta}\right]\;;
\label{bzetaSignalNoise}
\end{eqnarray}
cf.\ Eqs.\ (\ref{bjFromajLossy}) and (\ref{brevebj'}).  Here we have omitted an
imaginary part of the factor $1-{1\over2}{\cal E}_{\rm OTF}$ 
[arising from the $b_2$ term in $\breve b_1$, Eq.\ (\ref{breveb1'})] 
because its modulus
is second order in the losses ($\propto {\cal E}_J^2$) and therefore it
contributes negligibly to the signal strength. 

Equation (\ref{bzetaSignalNoise}) implies that the gravitational-wave noise
operator is
\begin{equation}
h_n = \left(1+{1\over2}{\cal E}_{\rm OTF}+{1\over4}{\cal E}\right) 
{h_{\rm SQL}\over \sqrt{2{\cal K}}}e^{-i\beta_*} 
(\Delta \breve b_2 + \Delta \breve b_1 \cot\zeta)\;, 
\label{hnLossy}
\end{equation}
where we have used Eq.\ (\ref{KDefLossy}) for ${\cal K}_*$.

For a squeezed-variational interferometer, the dark-port input field $a_j$ 
is in a squeezed state, with squeeze factor $R$ and squeeze angle
$\lambda(\Omega)$ (which, after optimization, will turn out to be $\lambda =
\pi/2$ as for a lossless interferometer).  For a variational-output
interferometer, $a_j$ is in its vacuum state, which corresponds to squeezing
with $R=0$ so we lose no generality by assuming a squeezed input.  
Since all
the noise fields except $a_j$ are in their vacuum states, the light's full 
input state is
\begin{equation}
|{\rm in}\rangle = |0_n\rangle 
\otimes |0_{n_{\rm OT}}\rangle\otimes |0_{n_{\rm I}}\rangle \otimes |0_{n_{\rm
II}}\rangle \otimes S(R,\lambda)|0_a\rangle\;,
\label{inLossy}
\end{equation}
where the notation should be obvious.

The gravitational-wave
noise is proportional to
\begin{equation}
\langle{\rm in}|h_n h_{n'}|{\rm in}\rangle = \langle 0 | h_{ns}
h_{ns'}|0\rangle
\label{EVhn''}
\end{equation}
where $|0\rangle$ is the vacuum state of all the noise fields
$a$,
$n$, $n_{\rm OT}$, $n_{\rm I}$, and $n_{\rm II}$; and
$h_{ns}$ is the usual squeezed 
noise operator
\begin{equation}
h_{ns} = S^{\dag}(R,\lambda) h_n S(R,\lambda)\;.
\label{hns4}
\end{equation}
We bring this squeezed-noise operator into an explicit form by
(i) inserting Eq.\ (\ref{hnLossy}) into Eq.\ (\ref{hns4}), then
(ii) replacing the $\Delta \breve b_j$'s  
by expressions (\ref{brevebj'}) 
[with $\Delta$ put onto all the $b$'s, i.e.\ with the signal removed],
then (iii) replacing the $\Delta b_j$'s by expressions
(\ref{DeltabjLossy}), and then (iv) invoking Eqs.\ (\ref{Squeezea12}) for
the action of the squeeze operators on the $a_j$'s.  The result is
\begin{eqnarray}
&&
h_{ns} = \left(1-{1\over4}{\cal E}\right)\times\hbox{[Eq.\ (\ref{hns''})]}
\nonumber\\
&&+{h_{\rm SQL}\over\sqrt{2{\cal K}}} \left\{ \left(-{\cal
K}e^{i\beta}\sqrt{\epsilon/2} + \sqrt{{\cal E}}\cot\zeta\right)e^{i\beta}n_1 +
\sqrt{\cal E}e^{i\beta}n_2 \right. \nonumber\\
&&
+{1\over2}\sum_J \left[(\sqrt{{\cal E}_{J+}}+\sqrt{{\cal E}_{J-}})
\cot\zeta
-i(\sqrt{{\cal E}_{J+}}-\sqrt{{\cal E}_{J-}})
\right]n_{J1} 
\nonumber\\
&&
\left.
+{1\over2}\sum_J \left[\sqrt{{\cal E}_{J+}}+\sqrt{{\cal E}_{J-}}
+i(\sqrt{{\cal E}_{J+}}-\sqrt{{\cal E}_{J-}})
\cot\zeta\right]n_{J2} \right\}
\nonumber\\
\label{hnsLossySV} 
\end{eqnarray}
where we have omitted terms, arising from $b_2$ in Eq.\ (\ref{breveb1'})
and from $b_1$ in Eq.\ (\ref{breveb2'}), which contribute amounts 
${\cal O}({\cal E}_J^2)$ to $S_h$; 
and we have omitted a term\footnote{
\label{betastarnote} 
this term is an imaginary part, $2i(\beta_*-\beta){\cal K}
= -{1\over2}i\epsilon {\cal K} \sin2\beta$, of the quantity $\tilde{\cal K}$,
which enters Eq.\ (\ref{hns''}) via Eq.\ (\ref{tildeKtildePhiDef}).  Because
this imaginary part produces a correction to the loss-free part of $h_n$ that
is 90 degrees out of phase with the loss-free part and is of order $\epsilon$, 
it produces a correction to $S_h$ that is quadratic in $\epsilon$ and thus
negligible.
}
proportional to
$\beta_* - \beta$ which contributes an amount ${\cal O}(\epsilon^2)$. 

By virtue of Eq.\ (\ref{EVhn''}) and the argument preceding Eqs.\  
(\ref{Sa1a2Vacuum}), we can regard all of the quadrature noise operators
$a_j$, $n_j$, $n_{Jj}$ in this $h_{ns}$
as random processes with unit spectral densities and
vanishing cross-spectral densities.  Correspondingly, the 
gravitational-wave noise is the sum of the squared moduli of the
coefficients of the quadrature noise operators in Eq.\ (\ref{hnsLossySV}):
\begin{eqnarray}
S_h&=& {h_{\rm SQL}^2\over 2{\cal K}} \Big[ \left(1-{1\over2}{\cal E} \right)
\left(1+{\tilde{\cal K}}^2\right) \nonumber\\
&&\times \{e^{-2R} + \sinh 2R[1-\cos2(\tilde \Phi + \lambda)]\} 
\label{ShLossy} \\
&& + {\cal K}^2 {\epsilon\over2} +(1- \tilde{\cal K}\cot\zeta){\cal E}
+ (1+\cot^2\zeta) {\cal E}_{\rm OTF}\Big] \nonumber
\end{eqnarray}
where 
\begin{equation}
{\tilde{\cal K}} = {\cal K} - \cot\zeta\;,\quad
\tilde\Phi = {\rm arccot \tilde{\cal K}}
\label{tildeKtildePhiDef'}
\end{equation}
[Eq.\ (\ref{tildeKtildePhiDef})].
In Eq.\ (\ref{ShLossy}), the first two lines come from $a_1$ and $a_2$
[squeezed vacuum entering the dark port; cf.\ Eq.\  
(\ref{ShSqueezedVariational})] modified by losses in the arm cavities [the
factor $1-{\cal E}/2$)]; the first two terms on the third line come from $n_1$
and $n_2$ [shot noise due to vacuum entering at loss points in the arm 
cavities]; and the last
term comes from $n_{J1}$ and $n_{J2}$ [shot noise due to vacuum entering
at loss points in the output optical train, including the filters]. 

As for the lossless interferometer [Eqs.\ 
(\ref{lambdazetaOpt}) and (\ref{ShSqueezedVariationalOptimized})],
the noise (\ref{ShLossy}) is minimized by setting the input squeeze angle
$\lambda$ and output homodyne phase $\zeta$ to 
\begin{equation}
\lambda = \pi/2\;, \quad \zeta = \Phi \equiv {\rm arccot}{\cal K}\;
\label{lambdazetaOpt1}
\end{equation}
[aside from a neglible correction $\delta\zeta = ({\cal E}+2 {\cal E}_{\rm
OTF})e^{-2R}/$ $({\cal K} + {\cal K}^{-1})$].
This optimization produces 
$\tilde {\cal K} =0$ and $\lambda = \tilde\Phi = \pi/2$, so
\begin{eqnarray}
&\null& S_h = {h_{\rm SQL}^2\over 2} 
\label{ShLossyOptimized} \\
&& \times\left[{\left(1-{1\over2}{\cal E}\right) e^{-2R}
+ {\cal E} + {\cal E}_{\rm OTF} \over{\cal K}} 
+ {\cal K}\left({\epsilon\over2} + {\cal E}_{\rm OTF}\right) 
\right]\;. \nonumber
\end{eqnarray}

Note that the optimization has entailed a squeezed input with 
frequency-independent squeeze phase, as in the lossless interferometer; so no
filters are needed in the input.  The output filters must produce a FD homodyne
angle $\zeta = \Phi(\Omega)$ that is the same as in the lossless case and
therefore can be achieved by two long, Fabry-Perot cavities.

It is instructive to compare the noise (\ref{ShLossyOptimized}) for a 
lossy squeezed-variational interferometer with that 
(\ref{ShSqueezedVariationalOptimized}) for one without optical losses.
In the absence
of losses, the output's FD homodyne detection can completely remove the 
radiation-pressure back-action noise from the signal; only the shot noise,
$\propto 1/{\cal K}\propto 1/I_o$, remains.  Losses in the interferometer's 
arm mirrors prevent this back-action
removal from being perfect: they
enable a bit of vacuum field $n$ 
to leak into the arm cavities, and this field produces
radiation-pressure noise that remains in the output after the
FD homodyne detection
(the 
${\cal K} \epsilon/2$ 
term in Eq.\ (\ref{ShLossyOptimized})].  

The  ${\cal K} {\cal E}_{\rm OTF}$ noise in Eq.\ (\ref{ShLossyOptimized})
has the same dependence on laser power, $\propto {\cal K} \propto I_o$, as
the radiation-pressure noise.  Nevertheless, it is actually shot noise, not
radiation pressure noise.  
It is produced by the vacuum loss-noise fields 
that leak into the output signal light when it encounters each lossy 
optical element.  
Those fields' shot noise gets
weighted by  the factor $\cot\zeta = \cot\Phi = {\cal K}$ in the homodyne
process, which accounts for their proportionality to ${\cal K} \propto I_o$.

A reasonable estimate for the amount of input-light squeezing that might be 
achieved in LIGO-III is\cite{SqueezeAmount} 
\begin{equation}
e^{-2R} \simeq 0.1 \;.
\label{REstimate}
\end{equation}
By contrast, Eqs.\ (\ref{epsilonDef}), (\ref{epsilonJepsilon}), 
(\ref{calEDef}) and (\ref{calEfiltDef}) suggest
\begin{equation}
({\cal E} + {\cal E}_{\rm OTF}) \sim 0.01\;.
\label{calEEstimate}
\end{equation} 
This motivates our neglecting ${\cal E} +
{\cal E}_{\rm OTF}$ compared to $e^{-2R}$ in expression  
(\ref{ShLossyOptimized}),
and rewriting the noise (\ref{ShLossyOptimized}) as
\begin{equation}
S_h \simeq {h_{\rm SQL}^2\over 2} \left[
{e^{-2R}
\over{\cal K}} + {\cal K}\epsilon_* 
\right]\;,
\label{ShLossySV}
\end{equation}
where
\begin{equation}
\epsilon_* \equiv {\epsilon\over2} + {\cal E}_{\rm OTF} \sim 0.0010\;;
\label{epsilonstarDef}
\end{equation}
cf.\ Eqs.\ (\ref{epsilonDef}) and (\ref{EOTF}).

Equation (\ref{ShLossySV}) is our final form for the 
noise spectrum of a lossy squeezed-variational interferometer. When 
we set the input squeeze factor to unity, $e^{-2R} = 1$, it becomes
the noise spectrum for a lossy variational-output interferometer: 
\begin{equation}
S_h \simeq {h_{\rm SQL}^2\over 2} \left[
{1 \over{\cal K}} + {\cal K}\epsilon_*
\right]\;,
\label{ShLossyVO}
\end{equation}

Errors 
$\Delta\lambda = \lambda-\pi/2$
in the input squeeze angle
and $\Delta\zeta = \zeta-{\rm arccot}{\cal K}$ in the output homodyne phase
will increase the noise spectral density.  By performing a power series
expansion of expression (\ref{ShLossy}), we obtain for the noise increase
\begin{eqnarray}
\Delta S_h &=& {h_{\rm SQL}^2\over{\cal K}}
\left[ \sinh 2R \;\Delta\lambda^2 
- 2(1+{\cal K}^2)\sinh 2R \;\Delta\lambda \Delta\zeta \right. \nonumber\\
&& 
\quad\quad \quad 
\left. + {(1+{\cal K})^2 e^{2R}\over2} \;\Delta\zeta^2 \right] 
\label{deltaSh}
\\
&\simeq& {h^2_{\rm SQL}\over 2{\cal K}} e^{2R}[\Delta\lambda - (1+{\cal
K}^2)\Delta\zeta]^2\;, \nonumber 
\end{eqnarray}
where the second expression is accurate in the limit $e^{2R}\gg e^{-2R}$.
Numerical evaluations show that,
for $e^{-2R} =0.1$ and $\epsilon_* = 0.01$ (see above), and for 
${\cal K}\sim 1$ to $3$ (the range of greatest interest; cf.\ Sec.\ 
\ref{sec:Discussion}), $\Delta \sqrt{S_h}$ will be less than 
${1\over4}\sqrt{S_h}$ so
long as: (i) the input squeeze angle is accurate to $|\Delta\lambda| \alt 0.05$,
and (ii) the FD output homodyne phase is accurate to $|\Delta\zeta|
\alt 0.01$.  At ${\cal K} = 1$ the FD phase's required accuracy is reduced
to $|\Delta\zeta| \alt 0.04$. 
The FD phase $\zeta$ is 
determined by the filter cavities' half bandwidths $\delta_J$ 
and fractional frequency offsets $\xi_J$, 
and the local oscillator phase or equivalently the final, conventional 
homodyne detector's
homodyne phase $\theta$.  The filter cavities' half bandwidths $\delta_J$ (or
equivalently their finesses) are fixed by the mirror coatings.
Coating-produced errors in $\delta_J$ can be compensated to some degree by
tuning the fractional frequency offsets $\xi_J$ (via adjusting the mirror
positions) and by tuning the local oscillator phase or equivalently $\theta$.
Finesse errors as large as five per cent, $|\Delta\delta_J|/\delta_J \alt
0.05$,
can be compensated to yield the required $|\Delta\zeta|\alt0.01$ by tuning the
offsets and homodyne phase to one percent accuracy, $|\Delta\xi_J| \alt 0.01$,
$\Delta\theta\alt 0.01$
[Eqs.\ (\ref{PhiDesired}), (\ref{FilterRequirement}), (\ref{alphaJpm'})
and Fig.\ \ref{fig:Fig10}]. 
These requirements are challenging. 

\subsection{Computation of the noise spectrum for a squeezed-input
interferometer}
\label{sec:ComputationSISpectrum}

For a squeezed-input interferometer, as for squeezed-variational, the 
losses in the input optical train influence the noise only through
their impact on the squeeze factor $e^{-2R}\sim 0.1$ of the dark-port vacuum 
when it enters the arm cavities.  By contrast, losses in the arm cavities and
in the output optical train will produce noise in much the same manner as
they do for a squeezed-variational interferometer.  More specifically:

The effect of arm-cavity and output-train losses
on the squeezed noise operator $h_{ns}$ can be
read off of the squeezed-variational formula (\ref{hnsLossySV}) as follows:
(i) Set $\zeta=\pi/2$ so the quantity measured is $\breve b_2$ [no output
filtering; Eq.\ (\ref{bzetaSignalNoise})]; (ii) correspondingly set $\cot\zeta
= 0$, $\tilde {\cal K} = {\cal K}$, and $\tilde \Phi = \Phi \equiv {\rm
arccot}{\cal K}$ [Eqs.\ (\ref{tildeKtildePhiDef'})]; (iii) in the sum over $J$
include only $J={\rm OT}$ and not $J={\rm I, II}$ since there are no output
filters.  The result is
\begin{eqnarray}
&&
h_{ns} = \left(1-{1\over4}{\cal E}\right)\times\hbox{[Eq.\ (\ref{hnsExpand})]}
\nonumber\\
&&+{h_{\rm SQL}\over\sqrt{2{\cal K}}} \left( -{\cal
K}\sqrt{\epsilon/2}\;e^{2i\beta} n_1 
+\sqrt{\cal E}e^{i\beta}n_2 +\sqrt{{\cal E}_{\rm OT'}}\; n_{\rm OT'2}\right)\;.
\nonumber\\
\label{hnsSILossy}
\end{eqnarray}
Here the prime on the subscript OT indicates that we must omit losses due to
mode-matching into the output filters and mixing with the local oscillator, 
since there are no output filters or homodyne detection. 
Correspondingly,
\begin{equation}
{\cal E}_{\rm OT'} = \epsilon_{\rm bs} + \epsilon_{\rm circ} + \epsilon_{\rm
pd} \sim 0.003
\label{EOT'}
\end{equation}
is the net fractional photon loss in the output optical train.  

Treating the quadrature noise operators as random processes with unit spectral
density and vanishing cross spectral densities, we read off $S_h$ from Eq.\
(\ref{hnsSILossy}):
\begin{eqnarray}
S_h = {h^2_{\rm SQL}\over2} 
\Bigg[&&{{\cal E} + {\cal E}_{\rm OT'}\over{\cal K}} +
{\epsilon\over2}{\cal K} 
+ \left(1-{1\over2}{\cal E}\right)\left({1\over {\cal K}
} + {\cal K}\right) \nonumber\\
&&\times \{\cosh2R - \cos[2(\lambda+\Phi)]\sinh2R\} \Bigg]\;.
\label{ShSILossy}
\end{eqnarray} 
As in the lossles case, the noise is minimized by
squeezing the dark-port input at the FD angle 
$\lambda(\Omega) = - \Phi \equiv - {\rm
arccot}{\cal K}$ [Eq.\ (\ref{lambdaOpt})].  The result is
\begin{eqnarray}
S_h = {h^2_{\rm SQL}\over2}
&& \left[ \left(1-{1\over2}{\cal E}\right)\left({1\over {\cal K}
} + {\cal K}\right)e^{-2R} \right. \nonumber\\ 
&& \left. + {{\cal E} + {\cal E}_{\rm OT}\over{\cal K}} +
{\epsilon\over2}{\cal K} \right]\;.
\label{ShSILossyOpt}
\end{eqnarray} 
For our estimated squeezing $e^{-2R}\sim 0.1$ and losses ${\cal E}_{\rm OT'}
\sim {\cal E} \sim \epsilon \alt 0.003$ in the LIGO-III time frame, the
loss parameters are small compared to the squeeze, and thus contribute
negligibly to the noise, so $S_h$ is well approximated by the lossless formula
\begin{equation}
S_h \simeq 
{h_{\rm SQL}^2\over2}\left({1\over{\cal K}} + {\cal K}\right)e^{-2R}\;.
\label{ShLossySI}
\end{equation}
However, it is important to keep in mind that
the input squeeze factor $e^{-2R}$ is constrained not only by the 
physics of the squeezing appratus, but also by losses in the input optical
train and mode matching into the arm cavities.  

By expanding expression (\ref{ShSILossy}) in powers of $\Delta\lambda =
\lambda+{\rm arccot}{\cal K}$, we see that the fractional increase in noise due
to errors in the FD squeeze angle is
\begin{equation}
{\Delta \sqrt{S_h}\over \sqrt{S_h}} = 
e^{2R}\sinh 2R \; \delta \lambda^2 \simeq {e^{4R}\over2} \Delta\lambda^2\;.
\label{dShSI}
\end{equation}
For $e^{-2R} = 0.1$, this fractional noise increase will be less than 1/4
so long as $\Delta\lambda$ is less than $0.07$.  This translates into 
accuracies of $\sim 7$ per cent for the prefilter squeeze angle, 
$\sim 15$ per cent for the filter cavities' fractional frequency
offsets ($|\Delta\xi_J|\alt 0.15$), and $\sim 10$ per cent for the cavities' 
half
bandwidths or equivalently their finesses ($\Delta\delta_J/\delta_J \alt 0.1$).
These constraints are 
significantly less severe
than those for a squeezed-variational interferometer 
(end of
Sec.\ \ref{sec:ComputationSpectra}); but, as we shall see, the potential
performance of this squeezed-input interferometer is poorer by a 
factor $\sim 1.5$ -- 2
than that of the squeezed-variational one. 

\section{Discussion of the Interferometers' Noise Spectra}
\label{sec:Discussion}

The noise spectra for our three lossy QND interferometers, Eqs.\
(\ref{ShLossySV}), (\ref{ShLossyVO}) and (\ref{ShLossySI}), all have the same
universal form---a form identical to that for a conventional
broadband interferometer, Eq.\ (\ref{ShConventional}). Only the parameters 
$\mu$ and $\sigma$ characterizing the 
noise differ from one interferometer to another.
This universal form can be written as   
\begin{mathletters}
\begin{equation}
{\sqrt{S_h(\Omega)}\over h_{\rm SQL}(\Omega)} 
= \mu \sqrt{{1\over 2}\left( {\Omega_*^2\over\sigma^2} +
{\sigma^2\over\Omega_*^2}\right)}\;,
\label{ShUniversalA}
\end{equation}
where $\Omega_*$ is the following function of angular frequency
\begin{equation}
\Omega_* \equiv {\Omega\over\gamma}\sqrt{{1+\Omega^2/\gamma^2}\over 2}
\label{OmegaStarDef}
\end{equation}
\label{ShUniversal}
\end{mathletters}
and $h_{\rm SQL}(\Omega)$ is given by Eq.\ (\ref{hSQLDef}). 
Notice that $\Omega_* = 1$ when $\Omega=\gamma \simeq 100$ Hz; $\Omega_* =
(\Omega/\gamma)/\sqrt2$ when $\Omega\ll \gamma$, and $\Omega_* =
(\Omega/\gamma)^2/\sqrt2$ when $\Omega\gg \gamma$.

\begin{figure}
\epsfxsize=3.2in\epsfbox{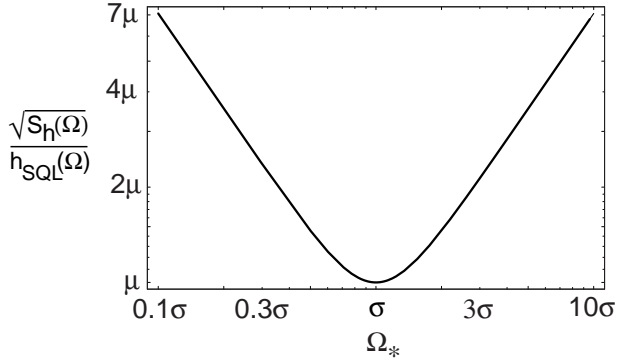}
\caption{Universal noise curve for conventional and QND interferometers [Eqs.\ 
(\protect\ref{ShUniversal})].
\label{fig:Fig12}
} 
\end{figure}

This universal noise curve is plotted as a function of $\Omega_*$ in Fig.\ 
\ref{fig:Fig12}.  Its two parameters are the minimum
value $\mu$ of the noise, i.e.\  
the minimum amplitude noise in units of the SQL,
and the dimensionless frequency 
$\sigma$ (in units of $\Omega_*$) at which the noise takes on this 
minimum value.  

\begin{figure}
\epsfxsize=3.2in\epsfbox{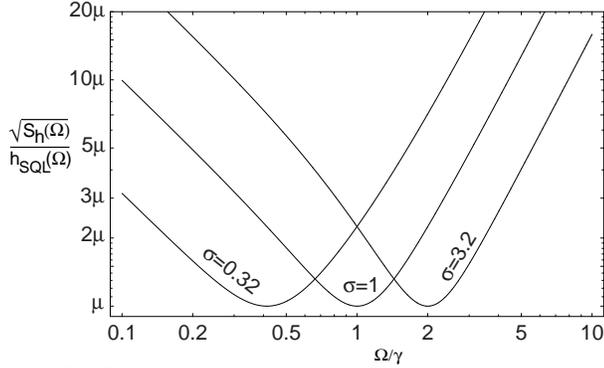}
\caption{Universal noise curve plotted as a function of angular frequency
$\Omega$ for various values of the dimensionless frequency parameter $\sigma$.
\label{fig:Fig13}
} 
\end{figure}

Figure \ref{fig:Fig13} shows this universal noise curve plotted as a
function of angular frequency $\Omega$.  Notice that, because of the relation
(\ref{OmegaStarDef}) between $\Omega_*$ and $\Omega$, the shape of the noise
curve depends modestly on the location $\sigma$ of its minimum. 

The values of the parameters $\mu$ and $\sigma$ for our various interferometer
configurations are shown in Table \ref{table:IFOParameters}.  Notice the
following details of this table:  (i) The minimum noise $\mu$
(the optimal amount by
which the SQL can be beat) is independent of the laser input power $I_o$ in all
cases; it depends only on the level of input squeezing $e^{-2R}$ and the level
of losses $\epsilon_*$.  (ii)
For our estimated loss level and squeeze level, the
squeezed-input interferometer and variational-output interferometer achieve the
same $\mu\simeq 0.32$, while the squeezed-variational interferometer achieves a
moderately lower $\mu\simeq 0.18$.  (iii) The frequency $\Omega_* = \sigma$ 
at which the minimum noise is achieved is proportional to 
$\sqrt{I_o/I_{\rm SQL}}$.
(Recall that $I_{\rm SQL}$ is the input power required for a conventional
interferometer to reach the SQL {\it at
the angular frequency} $\Omega = \gamma \simeq 2\pi \times 100$ Hz, i.e.\ at
$\Omega_* = 1$; to do so, the conventional interferometer must have $\sigma =
1$.)  (iv) For $I_o=I_{\rm SQL}$, the squeezed-input interferometer has
$\sigma=1$, but the variational-output and squeezed-variational interferometers
have $\sigma <1$, which means that the minimum of the noise curve is at
$\Omega<\gamma \simeq 100$ Hz.  To push $\sigma$ up to unity, i.e.\ to push 
the   noise-curve
minimum up to $\Omega=\gamma$, requires $I_o/I_{\rm SQL} = 1/\sqrt{\epsilon_*}
\simeq 10$ in a variational-output interferometer, and $I_o/I_{\rm SQL} =
\sqrt{e^{-2R}/\epsilon_*} \simeq 3.2$ in a squeezed-variational interferometer.

\begin{table}
\caption{The values of the parameters $\mu = ($minimum noise) and $\sigma
=($frequency of minimum) for various interferometer (``IFO'') configurations:
Conv = Conventional broadband [Eq.\ (\ref{ShConventional})], 
SI = Squeezed-Input [Eq.\ (\ref{ShLossySI})], VO = Variational-Output [Eq.\
(\ref{ShLossyVO})], and SV = Squeezed-Variational [Eq.\ (\ref{ShLossySV})].
The numerical values are for $e^{-2R} = 0.1$ and $\epsilon_* = 0.01$.
}
\label{table:IFOParameters}
\vskip15pt
\begin{tabular}{ccc}
IFO&$\mu$&$\sigma$\\
\tableline
\\
Conv.&1&$\sqrt{I_o/I_{\rm SQL}}$\\
SI&$\sqrt{e^{-2R}}\simeq 0.32$&$\sqrt{I_o/I_{\rm SQL}}$\\
VO&$\epsilon_*^{1/4}\simeq0.32$&$\sqrt{I_o/I_{\rm SQL}\over1/\sqrt{\epsilon_*}}
\simeq \sqrt{I_o/I_{\rm SQL}\over10}$\\
SV&$(e^{-2R}\epsilon_*)^{1/4} \simeq 0.18$&$\sqrt{I_o/I_{\rm
SQL}\over\sqrt{e^{-2R}/\epsilon_*}} \simeq \sqrt{I_o/I_{\rm SQL}\over3.2}$\\
\end{tabular}
\end{table}

\begin{figure}
\epsfxsize=3.2in\epsfbox{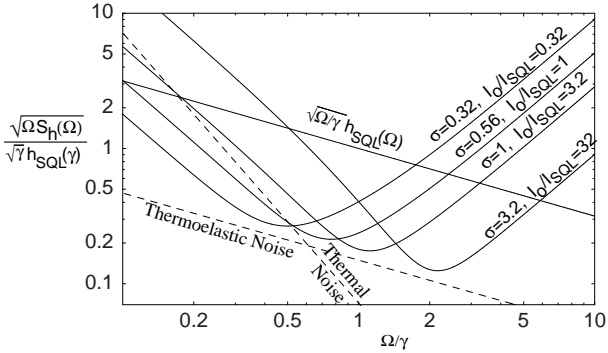}
\caption{Noise curves for SQL interferometers with noise minima $\mu = 0.18$
and various values of the frequency parameter $\sigma$.  The vertical axis is
weighted by $\sqrt{\Omega/\gamma}$ so the curves give an indication of the
relative noise in searches for waves from inspiraling binaries; see text.
The noise curves are labeled by the power $I_o/I_{\rm SQL}$ required by a 
squeezed-variational interferometer to achieve the given $\sigma$. 
\label{fig:Fig14}
} 
\end{figure}

The importance of pushing $\sigma$ up to unity or higher is explained in
Fig.\ \ref{fig:Fig14}.  This figure requires some discussion:  

The
most promising gravitational waves for LIGO are those from the last few minutes
of inspiral of black-hole/black-hole binaries, black-hole/neutron-star
binaries, and neutron-star/neutron-star binaries.  The amplitude
signal-to-noise ratio $S/N$ produced by these waves is given by
\begin{equation}
{S^2\over N^2} = 4 \int_0^{\infty} {|\tilde h|^2\over S_h} {d\Omega\over2\pi} =
4 \int_{-\infty}^{\infty} {|\Omega \tilde h|^2\over \Omega S_h} 
{d\ln\Omega\over 2\pi}\;,
\label{SoverN}
\end{equation}
where $\tilde h$ is the Fourier transform of the waveform $h(t)$.  For the
inspiraling binary $|\Omega \tilde h|$ is nearly independent of frequency
throughout the LIGO band \cite{300yrs}, 
so the signal-to-noise ratio is optimized by making
$\Omega S_h(\Omega)$ as small as possible over as wide a range of $\ln\Omega$
as possible.

Figure \ref{fig:Fig14} plots $\sqrt{\Omega S_h(\Omega)}$ as a
function of $\Omega/\gamma$ using logarithmic scales on both axes, and using
the minimum-noise parameter $\mu = 0.18$ corresponding to our fiducial
squeezed-variational interferometer (though the specific value of $\mu$ is
irrelevant to our present discussion).  From the shapes of the curves it should
be evident that {\it the larger is the frequency of the noise minimum, 
i.e.\ the larger is $\sigma$ at fixed $\mu$, the larger will be the $S/N$ for
inspiraling binaries.} 

A second factor dictates using large $\sigma$, in particular $\sigma \agt 1$. 
This is thermal noise in the interferometer's test-mass suspension fibers.
The thermal noise scales with frequency as $\sqrt{\Omega S_h^{\rm thermal}
(\Omega)} \propto \Omega^{-2}$ or $\propto \Omega^{-5/2}$ depending on the
nature of the dissipation \cite{Saulson}; see the steep dashed curve in Fig.\
\ref{fig:Fig14}.  It seems realistic to expect, in LIGO-III, that
this thermal noise will be at approximately the level shown in the figure, so
it compromises the performance of QND interferometers at $\Omega \alt
0.5 \gamma \simeq 50$ Hz \cite{WhitePaper,BraginskyThermalNoise}.  
Correspondingly, to
avoid the thermal noise significantly debilitating the $S/N$ for inspiraling
binaries, it will be necessary to have $\sigma\agt 1$.

Because $\sigma$ scales as $\sqrt{I_o/I_{\rm SQL}}$ for all interferometer
designs, large $\sigma$ entails large laser power.  In particular, $\sigma \agt
1$ requires $I_o \agt I_{\rm SQL}$; cf.\ Table \ref{table:IFOParameters}.
For our fiducial parameters (Table \ref{Parameters.tbl}), 
$I_{\rm SQL} = 10$ kW, which corresponds to an optical power circulating in
each of the interferometer's arm cavities
\begin{equation}
W_{\rm circ}^{\rm SQL} = {I_{\rm SQL}/2\over\gamma L/c} = {mcL\gamma^3\over
8\omega_o} = 0.62\; \hbox{MW}\;.
\label{WCircSQL}
\end{equation}
To construct mirrors capable of handling this huge power will be
an enormous technical challenge
(even though this is approximately the circulating power contemplated 
for LIGO-II).  To operate with a circulating power much larger than this might
not be possible.  Therefore, it may be important in
LIGO-III to achieve $\sigma \agt 1$ while keeping $I_o/I_{\rm SQL}$ not much
larger than unity.

The squeezed-input interferometer, with its $\sigma = \sqrt{I_o/I_{\rm SQL}}$
(Table  \ref{table:IFOParameters}) 
is the most attractive from this point of view [and also in terms of its
required filter and squeeze-phase accuracies; cf.\ end of Sec.\ 
\ref{sec:ComputationSISpectrum}]; and
the variational-output with its $\sigma = \sqrt{\sqrt{\epsilon_*}I_o/I_{\rm
SQL}} \simeq \sqrt{0.1 I_o/I_{\rm SQL}}$ is the least attractive.  The
squeezed-variational interferometer, with $\sigma =
\sqrt{\sqrt{\epsilon_*/e^{-2R}}I_o/I_{\rm SQL}} \simeq \sqrt{0.32 I_o/I_{\rm
SQL}}$ requires a modestly higher laser power to reach $\sigma =1$ than the
squeezed-input [and requires better filter and squeeze-phase accuracies], 
but it is capable of a lower noise minimum, $\mu \simeq
(e^{-2R}\epsilon_*)^{1/4} \simeq 0.18$ vs.\ $\mu = \sqrt{e^{-2R}}\simeq 0.32$
for squeezed-input.

This suggests an R\&D strategy:  Focus on input squeezing as a key foundation
for LIGO-III (it is needed both for squeezed-input and squeezed-variational
interferometers), and in parallel (i)
develop the technology and techniques for the FD homodyne detection required 
by squeezed-variational configurations, (ii) work to drive down optical losses
to the levels $\epsilon\sim 
\epsilon_{\rm circ} \sim \epsilon_{\rm bs} \sim \epsilon_{\rm mm} \sim
\epsilon_{\rm lo} \sim \epsilon_{\rm pd}\sim 0.001$ [Eq.\
(\ref{epsilonValues})], and (since ponderomotive squeezing, which
underlies all our QND interferometers, has never been
seen) (iii) carry out
experiments in a small test appratus to demonstrate ponderomotive squeezing
and to search for unexpected obstacles and imperfections in it.

If both input squeezing and FD homodyne detection can be implemented
successfully, then the squeezed-variational interferometer is likely to achieve
better performance than any other configuration discussed in this
paper, despite its apparent need for higher laser power (e.g.\ $I_o/I_{\rm SQL}
\simeq 3.2$ to achieve $\sigma=1$ compared to $I_o/I_{\rm SQL} = 1$ for
squeezed input, with our fiducial parameters).  If powers as high 
as $I_o/I_{\rm SQL} \simeq 3.2$ cannot be handled, then we can operate the
squeezed-variational interferometer with a lower power without much loss
of performance.  

Consider, for example, $\sqrt{S_h}/h_{\rm SQL}$ evaluated at
$\Omega = \sigma \simeq 2\pi \times 100$ Hz, as a function of $I_o/I_{\rm SQL}$
in a variational-output interferometer with our fiducial $e^{-2R} = 0.1$ and
$\epsilon_* = 0.01$.  The optimal $I_o/I_{\rm SQL} = 3.2$ produces
$\sqrt{S_h(\gamma)}/h_{\rm SQL} = 0.18$; pushing $I_o/I_{\rm SQL}$ down by a
factor 2, to 1.6, increases the noise at $\Omega = \gamma$ by only 10 percent,
to 0.20; pushing down all the way to $I_o / I_{\rm SQL} = 1$ increases the 
noise to
only $\sqrt{S_h(\gamma)}/h_{\rm SQL} = 0.23$, which is still significantly
lower noise than the optimized squeezed-input interferometer ($0.32$ at
$I_o/I_{\rm SQL} = 1$).

It is worth recalling that for noncosmological sources (sources at distance
$\ll 3$ Gpc), the volume of the
universe that can be searched for a given type of source scales as
the inverse cube of the amplitude noise, so a noise level $\sqrt{S_h}/h_{\rm
SQL} = 0.18$ corresponds to search-volume increase of $1/0.18^3 \simeq 180$
over a SQL-limited interferometer, i.e.\ over LIGO-II.
   
\section{Conclusions}
\label{sec:Conclusions}

In this paper we have explored three candidate ideas for QND LIGO-III
interferometers: squeezed-input, variational-output, and squeezed variational.
The squeezed-input and squeezed-variational interferometers both looks quite
promising.  For our estimated levels of optical loss and levels of squeezing,
and for an input laser power $I_o/I_{\rm SQL} = 1$ (the LIGO-II level), 
the squeezed-input interferometer 
could achieve a noise $\mu \simeq 0.32$ of the SQL, with a
corresponding increase ${\cal V} \simeq 1/0.32^3 \simeq 30$ over LIGO-II
in the volume of the universe that could be searched for a given source,
at non-cosmological distances.
The squeezed-variational interferometer could achieve $\mu \simeq 0.23$ 
of the SQL with a search-volume
increase over LIGO-II of ${\cal V}\simeq 80$.  If the optics can handle 
a laser power
$I_o/I_{\rm SQL} \simeq 3.2$, then the squeezed-variational interferometer
could
reach $\mu \simeq 0.18$ of the SQL and a search-volume increase of 
${\cal V} \simeq 180$.  These numbers
scale with the losses, squeezing, and laser power as shown in  Table
\ref{table:IFOParameters}.

The squeezed-input and squeezed-variational designs are therefore
sufficiently
promising to merit serious further study.  Some of the issues that need
theoretical analysis are:  
\begin{itemize}
\item How can one incorporate into these interferometer designs the various
light modulations that are required, in a real gravitational-wave
interferometer, to (i) make the interferometer be shot-noise limited (put
the gravitational-wave signal into $\sim 100$ Hz sidebands of a MHz 
modulation\footnote{\label{fn:modulation}
LIGO scientists are currently exploring the possibility of achieving 
shot-noise-limited performance in LIGO-II without this 
modulation/demodulation.
The modulation/demodulation may, in fact, be replaced in LIGO-II by homodyne 
detection at the interferometer output, making it more nearly like our
paper's LIGO-III designs. 
}),
(ii) control the mirror positions and orientations, ... 
\cite{Song}.

\item What accuracies and other characteristics are needed for the 
interferometers' new elements: the circulator, filter 
cavities,\footnote{The filter cavities will require
a mechanical stability far less than that of the arm
cavities, since the carrier power in the output light is small 
and filter mirror displacements of magnitude $\sim h L$
therefore do {\it not} imprint a significant signal on the light.} 
and input squeezing?  
How can these be achieved?  
For example, how stable must be 
the local oscillator for the conventional homodyne detector, and can it be
achieved simply by tapping some light off the interferometer's output 
or input beam?

\item If the filter cavities are placed in the same long vacuum tubes as the
interferometer's arm cavities (with their enormous circulating power), what 
will be the nature and level of noise 
due to scattering of light from the test-mass cavites to the filter
cavities?  (We thank Eanna Flanagan for raising this 
issue.)

\item 
Can the filter cavities be made to serve multiple purposes?  For example,
is it possible to use a single optical cavity for both filters, e.g., 
with the two filters based on two different polarization states (for which
the filter might be made to behave differently via birefringence), or with
the two filters based on different, adjacent longitudinal modes?  As another
example, could an output filter cavity be used 
as a source of ponderomotively squeezed vacuum for input into the
interferometer's dark port?\footnote{\label{foot:PondVacuum} For
ponderomotively squeezed vacuum, the squeeze
angle is frequency dependent, with $d\phi/d\Omega$ of the opposite sign to
that needed by a squeezed-input interferometer.  This must be compensated
by a filtering different from that discussed in 
Sec.\ \ref{sec:FDHomodyne}.}  

\item Signal recycling 
via resonant-sideband extraction (RSE) \cite{RSE}
is likely to be a standard tool in LIGO-II \cite{WhitePaper}.  How can one
best implement RSE simultaneously with the FD homodyne detection (and input
squeezing) of a variational-output (or squeezed-variational) interferometer?
\cite{Song}
How can one best achieve the FD homodyne's filtration 
[which will entail a different frequency dependence $\Phi(\Omega)$ from that
in this paper's non-RSE designs]?

\item In this paper's analysis we have made a number of simplifying
approximations [e.g., our approximating the phase of the coefficient of 
$f_j$ in Eq.\ (\ref{gj}) by $2\beta$
an approximation that fails by a frequency-dependent amount 
which can be nearly as large as one per cent].
At what level of sensitivity do these approximations become problematic (e.g.,
for our proposed two-cavity way of achieving the necessary FD homodyne
detection), and how can the resulting problems be overcome?  

\item Our analysis is based on the 
crucial assumption that the interferometer's output is strictly 
linear in its input \cite{TestMassQM}.
Matsko and Vyatchanin \cite{Nonlinearity}
have shown that this is not quite correct.  
In the interferometer's arms the back-action-induced mirror displacement $X$
produces a phase shift of reflected light given by 
$e^{-2 i\Omega X/c}$, which our linearized analysis approximates 
as $1-2i\Omega X/c$
[cf.\ Eq.\ (\ref{deltak})]; when the better approximation $1-2i\Omega X/c -
2(\Omega X/c)^2$ is used, the result is additional, nonlinear noise, which
limits the cancellation of the back-action noise by the shot noise and 
produces a limit \cite{Nonlinearity}
\begin{equation}
S_h^{\rm NL} \sim {h^2_{\rm SQL}\over{2\;{\cal N}_{\rm SQL}^{1/5}}} 
\sim 5\times10^{-5}
h_{\rm SQL}^2   
\label{ShNL}
\end{equation}
on the sensitivity that any of our QND designs can achieve.  Here
\begin{equation}
{\cal N}_{\rm SQL} = {I_{\rm SQL}\over\hbar\omega_o\gamma} =
{1\over2}\left({Tc/4\omega_o \over \sqrt{\hbar/m\gamma}}\right)^2 
\simeq 2\times 10^{20}
\label{NSQL}
\end{equation}
is the number of quanta entering a SQL interferometer in time 
$\gamma^{-1} \sim 2$ ms.  The nonlinear limitation (\ref{ShNL})
is sufficiently far below
the SQL that we need not be concerned about it.  Are there any other,
more serious sources of nonlinearity that might compromise the performance
of these interferometers?  

\end{itemize}

Experimental studies are also needed as foundations for any possible
implementation of variational-output or squeezed-variational interferometers
\cite{TowardSQL}.
Examples are
\begin{itemize}

\item
Studies of the debilitating effects of very high circulating 
powers, $W_{\circ} \sim $ a few MW, and how to control them.  

\item
A continuation of efforts to achieve large squeezing, robustly, via nonlinear 
optics \cite{SqueezeAmount}, and exploration of the possibility to do so 
ponderomotively 
\cite{Walther,Heidmann,Schiller,SchillerA}.$^{\ref{foot:PondVacuum}}$  

\item
A continuation of efforts to achieve low levels of losses in optical
cavities and interferometers,
so as to minimize the contamination of squeezed light by ordinary
vacuum\cite{XaioThesis}.

\item
Prototyping of FD homodyne detection by the technique proposed in this paper:
filtration followed by conventional homodyne detection.     

\end{itemize}

In the meantime, and in parallel with such studies, it is important to 
push hard on the effort to find practical QND designs that entail circulating
light powers well below 1 MW \cite{LowPowerQND}, and that might be much
less constrained by optical losses than the designs explored in this paper.

\section*{Acknowledgments}
For helpful discussions or email, one or more of the authors thank 
Vladimir Braginsky,  
Alessandra Buonanno,
Carlton Caves, 
Yanbei Chen,
Eanna Flanagan, Mikhail Gorodetsky, Farid Khalili,
Patricia Purdue, Stan Whitcomb,
Bill Unruh, and members of the Caltech QND Reading Group---most especially 
Constantin Brif, Bill Kells and John Preskill.
We also thank Buonanno and Chen for pointing out several errors in the
manuscript.
This paper was supported in part by 
NSF grants PHY--9503642 [SPV], PHY--9722674 [HJK], 
PHY--9732445 [ABM], 
PHY-9800097 [SPV] 
and PHY--9900776 [KST and YL], 
by the Office of Naval Research [HJK and ABM], 
by DARPA via the QUIC (Quantum Information and Computing) program administered
by ARO [HJK],
and by the Russian Foundation for 
Fundamental Research grants \#96-02-16319a and \#97-02-0421g [SPV].

\appendix

\section{Rotation and Squeeze Operators}
\label{app:2photon}

In this paper we make extensive use of squeeze operators and some
use of rotation operators.  In this appendix we list properties
of these operators that are be useful in verifying statements made
the text.  This appendix is based on the formalism for 2-photon quantum
optics developed by Caves and Schumaker 
\cite{CavesSchumaker,SchumakerCaves}. 

The rotation operator $R(\theta)$, which acts on the Hilbert space of the
modes with frequencies $\omega = \omega_o \pm\Omega$, is defined by
\begin{equation}
R(\theta) = \exp[-i\theta(a_+^{\dag} a_+ + a_-^{\dag} a_-)]
\label{RotationDef}
\end{equation}
(Eq.\ (4.33) of \cite{CavesSchumaker}); here $a_\pm$ are the annihilation
operators, and $a_\pm^{\dag}$ the creation operators for photons in these
modes.  This operator is unitary and has the inverse
\begin{equation}
R^{-1}(\theta) = R^{\dag}(\theta) = R(-\theta)\;.
\label{RInverse}
\end{equation}
The effect of a rotation on the modes' annihilation operators is
\begin{equation}
R(\theta) a_\pm R^{\dag}(\theta) = a_\pm e^{i\theta}\;
\label{Rotateapm}
\end{equation}
(Eq.\ (4.35) of \cite{CavesSchumaker}),
and its effect on the two-photon quadrature amplitudes [Eqs.\ (\ref{a12Def})] 
is
\begin{eqnarray}
R(\theta) a_1 R^{\dag}(\theta) &=& a_1\cos\theta - a_2 \sin\theta\;,
\nonumber\\
R(\theta) a_2 R^{\dag}(\theta) &=& a_1\sin\theta + a_2 \cos\theta\;
\label{Rotatea12}
\end{eqnarray}
(Eq.\ (4.36) of \cite{CavesSchumaker}).

The squeeze operator also acts on the Hilbert space of modes with 
frequencies $\omega=\omega_o \pm \Omega$, and is defined by
\begin{equation}
S(r,\phi) = \exp[r(a_+ a_- e^{-2i\phi} - a_+^{\dag} a_-^{\dag} 
e^{2i\phi})]
\label{SqueezeDef}
\end{equation}
(Eq.\ (4.9) of \cite{CavesSchumaker}; Eq.\ (1.8) of \cite{SchumakerCaves}).
This squeeze operator is unitary and its inverse is
\begin{equation}
S^{-1}(r,\phi) = S^{\dag}(r,\phi) = S(-r,\phi) = S(r,\phi+\pi/2)
\label{SqueezeInverse}
\end{equation}
(Eq.\ (1.9) of \cite{SchumakerCaves}).  
The effect of a squeeze on the modes' annihilation operators is
\begin{equation} 
S(r,\phi) a_\pm S^{\dag}(r,\phi) = 
a_\pm \cosh r + a_\mp^{\dag} e^{2i\phi}\sinh r
\label{Squeezeapm}
\end{equation}
(Eq.\ (4.10) of \cite{CavesSchumaker}).  From this equation and the definition
(\ref{a12Def}) of the quadrature amplitudes, we infer the effect of a squeeze
on those amplitudes
\begin{eqnarray}
S(r,\phi) a_1 S^{\dag}(r,\phi) &=& a_1(\cosh r + \sinh r\cos2\phi) \nonumber\\
&&+ a_2\sinh r \sin2\phi\;, \nonumber\\
S(r,\phi) a_2 S^{\dag}(r,\phi) &=& a_2(\cosh r - \sinh r\cos2\phi) \nonumber\\
&&+ a_1\sinh r \sin2\phi\;. 
\label{Squeezea12}
\end{eqnarray}

\section{Input-Output Relations for Interferometers}
\label{app:interferometer}
In this appendix we shall derive the input-output relations for
the fields $a_j$ and $b_j$ that enter and leave the interferometer's dark 
port.  From the outset we shall include optical losses in our 
derivation, thereby
obtaining the lossy input-output relations (\ref{bjFromajLossy}) and
(\ref{DeltabjLossy}); the lossless input-output relations (\ref{bjFromaj})
then follow by setting $\epsilon=0$.  

\subsection{Fields at beam splitter}

\begin{figure}
\epsfxsize=1.8in\center{\epsfbox{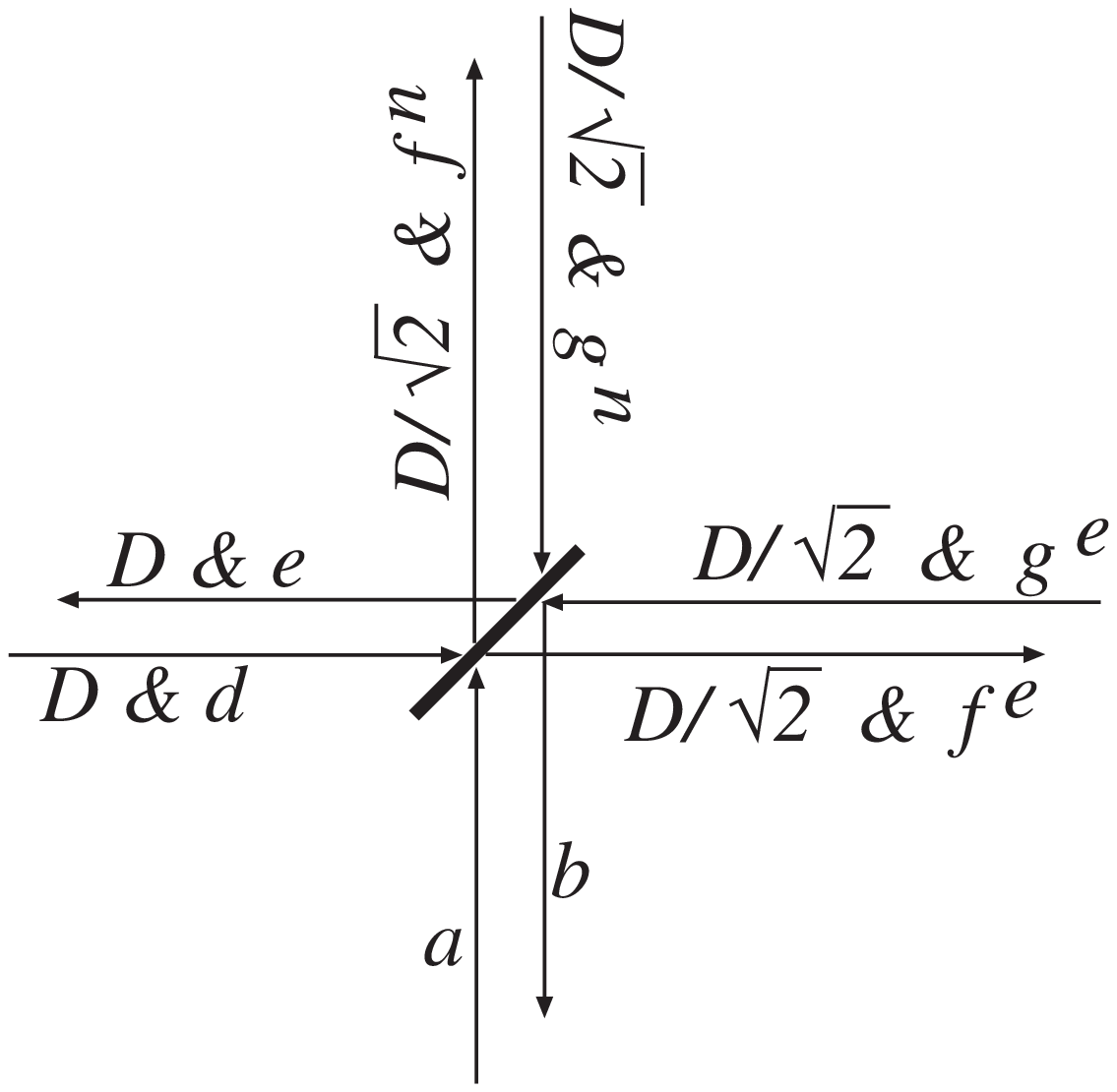}}
\caption{Field amplitudes entering and leaving the beam splitter
(which here is idealized as lossless).
The various amplitudes are defined in Eqs.\ 
(\protect\ref{EPlusInApp})--(\protect\ref{SplitterJunction}).
\label{fig:Fig15}
}
\end{figure}

We describe the field amplitudes entering and leaving the beam splitter by
the notation shown in Fig.\ \ref{fig:Fig15} (cf.\ Fig.\ \ref{fig:Fig3}.  
We idealize the beam splitter as lossless in this appendix, and deal with its
losses in the body of the paper in the manner sketched in Fig.\ 
\ref{fig:Fig11}.   
The amplitudes $D\;\&\;d$
of the field entering the beam splitter
from the laser are defined by the 
following formulae for
the positive-frequency part of the electric field
\begin{eqnarray}
&&E^{(+)}_{\rm in} = \sqrt{2\pi\hbar\omega_o\over{\cal A}c} 
\nonumber\\
&&\quad\times e^{-i\omega_o t}
\left[D+ \int_0^\infty \left(d_+ e^{-i\Omega t} + d_- e^{+i\Omega t} \right)
{d\Omega\over2\pi} \right]\;.
\label{EPlusInApp}
\end{eqnarray}
[cf.\ Eq.\ (\ref{EPlusIn1})]
and for the total electric field
\begin{eqnarray}
&&E_{\rm in} = \sqrt{4\pi\hbar\omega_o\over{\cal A}c} 
\nonumber\\
&&\quad\times\left\{ 
\cos(\omega_o t)
\left[ \sqrt2 D+ \int_0^\infty \left(d_1 e^{-\i\Omega t} + 
d_1^{\dag} e^{+i\Omega t}
\right) {d\Omega\over2\pi}\right]\right. \nonumber\\
&&\quad\quad + \left. \sin(\omega_o t)
\int_0^\infty \left(d_2 e^{-\i\Omega t} + 
d_2^{\dag} e^{+i\Omega t}
\right) {d\Omega\over2\pi} \right\}\;.
\label{EInApp}
\end{eqnarray}
Thus, $D$ is the classical amplitude of the laser light (carrier with frequency
$\omega_o$), $d_\pm$ are the annihilation operators for the $\omega_o\pm\Omega$
sidebands, and $d_1$ and $d_2$ are the quadrature amplitudes for the side
bands.  [Notice that the factor out front is a $\sqrt{2\pi}$ in Eq.\
(\ref{EPlusInApp}) but $\sqrt{4\pi}$ in (\ref{EInApp}), and notice the $\sqrt2
D$ in (\ref{EInApp}).]  The light power $I_o$ impinging on the beam splitter is
related to the classical amplitude 
$D$ by 
\begin{equation}
I_o = {\overline{E_{\rm in}^2}\over 4\pi} {\cal A} c = \hbar\omega_o D^2\;,
\label{IoD}
\end{equation}
where the overbar means time average.  
(Note that $D^2$ has dimensions Hz = 1/sec.)

For all other fields the classical amplitude and sideband amplitudes
are as indicated in the figure; for example, the field going toward the east
cavity has classical amplitude $D/\sqrt2$ and quadrature amplitudes $f^e_1$,
$f^e_2$.

With an appropriate choice of conventions
\cite{Siegman},
the fields' junction
conditions at the splitter are
\begin{eqnarray}
f_j^n = {d_j+a_j\over\sqrt2}\;, \quad
f_j^e = {d_j-a_j\over\sqrt2}\;, \nonumber\\
b_j = {g_j^n-g_j^e\over\sqrt2}\;, \quad
e_j = {g_j^n+g_j^e\over\sqrt2}\;.
\label{SplitterJunction}
\end{eqnarray}
Here $j=1$ or $2$.

\subsection{Arm cavities and fields}

The east and north arm cavities are presumed to be identical, with
power reflection and transmission coefficients $R$ and $T$ for the front
mirror, and
$\tilde R$ and $\tilde T$ for the back mirror.  The amplitude
reflection and transmission coefficients are chosen be real, with 
signs $\{+\sqrt{T},-\sqrt{R}\}$, $\{+\sqrt{\tilde T},-\sqrt{\tilde R}\}$ 
for light that impinges on a mirror from outside the cavity; and
$\{+\sqrt{T},+\sqrt{R}\}$, $\{+\sqrt{\tilde T},+\sqrt{\tilde R}\}$
for light that impinges from inside the cavity.

The dominant optical losses are for light impinging on mirrors from inside
the cavity (cf.\ Sec.\ \ref{sec:DissipationDescription}).  The influence of the
losses on the interferometer's signal and noise are independent of the physical
nature of the losses---whether it is light scattering off a mirror, absorption
in the mirror, or transmission through the end mirror.  (We ignore the 
effects of mirror heating.)  For computational simplicity, we model all the
losses as due to finite transmissivity 
${\cal L} = \tilde T \ne 0$
of the end mirror,
and correspondingly we set 
\begin{equation}
R+T= 1\;, \quad \tilde R + \tilde T = 1\;.
\label{RPlusT}
\end{equation}
The fractional loss of photons in each round trip in the cavity is then
$\tilde T$, and the net fractional loss of photons in the arm cavities is
\begin{equation}
\epsilon = {2{\cal L}\over T} = {2\tilde T \over T}\;
\label{epsilonApp}
\end{equation}
cf.\ Eqs.\ (\ref{calLDef}) and (\ref{epsilonDef}).  Recall that $T\simeq 0.033$
and 
$\epsilon \sim 0.0012$,
and also that $\Omega \sim \gamma = Tc/4L$ 
[Eqs.\ (\ref{gammaDef}), (\ref{epsilonDef})];
correspondingly, we shall make the approximations
\begin{equation}
\tilde T \ll T =4\gamma L/c \sim \Omega L/c \ll 1
\label{Tsmall}
\end{equation}
throughout our analysis.

\begin{figure}
\epsfxsize=3.2in\epsfbox{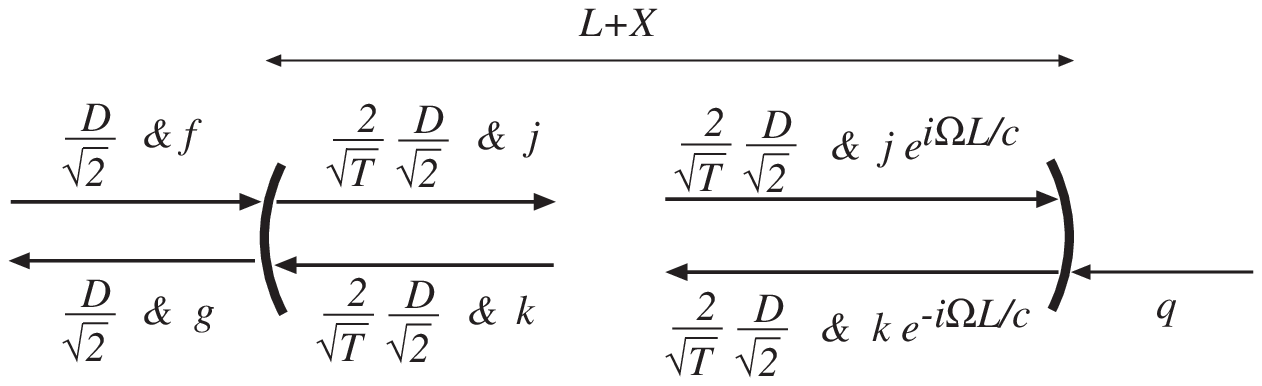}
\caption{Field amplitudes entering and leaving an arm cavity.
The cavity's front-port input and output amplitudes 
$D$, $f$ and $g$ are defined in Eqs.\
(\protect\ref{IoD}) and (\protect\ref{SplitterJunction}) 
and Fig.\ \protect\ref{fig:Fig15}, 
and it's back-port input $q$ is
defined in Eq.\ (\protect\ref{BackJunction}).
\label{fig:Fig16}
}
\end{figure}

Figure \ref{fig:Fig16} shows an arm cavity and the amplitudes of the
fields that impinge on or depart from its mirrors.  The amplitudes 
are those at the (front or back) mirror location, and the mirrors, like
the beam splitter, are idealized as infinitesimally thin.

For 
pedagogical simplicity, the distance from the beam splitter to the front mirror
of each arm cavity is set to an integral multiple of the carrier wavelength
and is assumed to be far smaller than $c/\Omega$ (the wavelength associated
with the sidebands).  This means that there are no net phase shifts of
the light in traveling between the beam splitter and the cavity's front
mirror; i.e., the field amplitudes $D/\sqrt2\;\&\;f$ (or $D/\sqrt2 \;\&\;g$)
arriving at (or departing from) the 
mirror are the same as those departing from (or arriving at) the beam splitter;
cf.\ Figs.\ \ref{fig:Fig16} and \ref{fig:Fig15}.  

The cavity's
length is adjusted to an integral number of carrier wavelengths so there is
no carrier phase shift from one end of the cavity to the other, and
inside the cavity the 
carrier
amplitude is amplified by the standard resonance factor $2/\sqrt{T}$.  
(Losses are small enough to be
of little importance for the carrier.)
Because the side bands
inside the cavity have a frequency dependence $g_j e^{-i\Omega t} \cos(\omega_o
t)$ at the 
front mirror location [cf.\ Eq.\ (\ref{EInApp})], they propagate down the 
cavity as $g_j e^{-i\Omega (t-z)} \cos[\omega_o (t-z)]$ and upon reaching the 
back mirror (where $cos[\omega_o(t-z)] = \cos[\omega t]$),  
they have acquired the phase shift 
indicated in the figure, $g_j e^{i\Omega L/c}$; and similarly for the $k_j$
field propagating in the other direction.  

The standard junction conditions at the front mirror imply that
\begin{equation}
j_j = \sqrt{T} f_j + \sqrt{R} k_j\;, \quad
g_j = - \sqrt{R} f_j + \sqrt{T} k_j\;.
\label{FrontJunction}
\end{equation}

We denote by $X(t)$ the change of arm length produced by radiation pressure
and the gravitational waves, and by $X$ its Fourier transform.  The oscillating
$X(t)$ pumps carrier light into the side bands.  More specifically, in
traveling from the front mirror $z=0$ to the perturbed position 
$z=L+X(t)$ of the back mirror, then reflecting and propagating to the 
unperturbed location $z=L$, the carrier field acquires the 
form\footnote{\label{fn:CarrierAttenuation}
Here we have neglected the attenuation of the carrier field 
due to the
arm-cavity losses.  This neglect is in the same spirit as our
ignoring attenuation in the input optics,
in the beam splitter, and in mode matching into the arm cavities.  
Including these attenuations would simply change $D$ in Eq.\
(\ref{CarrierBackMirror}) to $D \times (1-{1\over2}$power attenuation factor)
---i.e., $D\times(1-{1\over2}\epsilon)$ for the effect of arm-cavity losses. 
Equivalently, it would dictate replacing $I_o$ by $I_o(1-$power attenuation
factor$)$ in ${\cal K}$, ${\cal K}_*$, and all our formulas for the 
gravitational-wave noise.
}
\begin{eqnarray}
&&E_{\rm carrier} = 
\sqrt{4\pi \hbar\omega_o\over {\cal A}c} \sqrt{2\over T} \sqrt2 D
\cos\left(\omega_o[t-2X(t)/c]\right)
\nonumber\\
&&=\sqrt{4\pi \hbar\omega_o\over {\cal A}c}\sqrt{4\over T}D
\left( \cos\omega_o t + \sin\omega_o t {2\omega_o\over c}
\int_{-\infty}^{+\infty} X e^{-i\Omega t} {d\Omega\over 2\pi}\right)
\nonumber\\
\label{CarrierBackMirror}
\end{eqnarray}
Comparing with the standard expression for the field at the location of the 
unperturbed end mirror [Eq.\ (\ref{EInApp}) with the amplitude changes 
indicated in
the lower right of Fig.\ \ref{fig:Fig16}, 
$D\rightarrow
(2/\sqrt2)(D/\sqrt2)$ and $d_j \rightarrow k_j e^{-i\Omega L/c}$], we obtain
the following expression for the field fed from the carrier $D$ into the
sideband amplitudes $k_j$:
\begin{equation}
\delta k_1 = 0\;, \quad \delta k_2 = {2\over\sqrt{T}} D {2\omega_o\over c} X\;.
\label{deltak}
\end{equation}
This acts as a source term in the standard junction condition for the back
mirror: 
\begin{equation}
k_j e^{-i\Omega L/c} = \sqrt{\tilde R} j_j e^{i\Omega L/c} + \sqrt{\tilde T}
q_j + \delta k_j\;.
\label{BackJunction}
\end{equation}
Note that $q_j$ is the noise-producing vacuum fluctuation that leaks into the
cavity as a result of the optical losses.

\subsection{Cavity's internal field and radiation-pressure fluctuations}

By combining the front-mirror and back-mirror junction conditions
(\ref{FrontJunction}) and (\ref{BackJunction}) 
we obtain for the side-band amplitude in the cavity
\begin{equation}
j_j = {\sqrt{T} f_j +
\sqrt{R} e^{i\Omega L/c}\left(\sqrt{\tilde T}\;q_j + \delta k_j\right) 
\over 1-\sqrt{R\tilde R} e^{2i\Omega L/c}} 
\;.
\label{jj}
\end{equation}
Equations (\ref{Tsmall}) 
and $\sqrt{R} = \sqrt{1-T} = \sqrt{1-0.033} \simeq 1$
allow us to make the approximations $\sqrt{R}e^{i\Omega L/c} 
\simeq
1$
in the numerator and [using (\ref{gammaDef})] 
\begin{equation}
1-\sqrt{R\tilde R} e^{2i\Omega L/c} \simeq (2L/c)(\gamma_*-i\Omega)\;,\
\label{ResonantApproximation}
\end{equation}
\begin{equation}
\gamma_* \equiv \gamma(1+\epsilon/2)
\label{gammaStarDef}
\end{equation}
in the denominator 
(accurate to better than 1 per cent for all $\Omega$ of interest to us),
thereby bringing Eq.\ (\ref{jj}) into the form
\begin{equation}
j_j = {\sqrt{T}(f_j + \sqrt{\epsilon/2}\; q_j) + \delta k_j \over
(2L/c)(\gamma_* -i\Omega)}\;,
\label{jj1}
\end{equation}
where we have used $\tilde T = {1\over2}\epsilon T$.  
The cavity's 
internal electric field $E_{\rm int}$ 
is expression (\ref{EInApp}) with $D\rightarrow (2/\sqrt{T}) (D/\sqrt2)$
[Eq.\ (\ref{CarrierBackMirror})]
and $d_j \rightarrow j_j$ [expression (\ref{jj1})]; cf.\ Fig.
\ref{fig:Fig16}.  The power circulating in the cavity is this 
$(\overline{E_{\rm int}^2}/4\pi) {\cal A}c$, 
and consists of two parts, a steady classical piece 
\begin{equation}
W_{\rm circ} = {1\over 2} {4 D^2\over T}\hbar\omega_o = {2\over T} I_o =
{I_o/2\over \gamma L/c}\;,
\label{WcircApp}
\end{equation}
and a fluctuating piece 
\begin{equation}
\delta W_{\rm circ} = \int_0^\infty {\sqrt{I_o\hbar\omega_o} (f_1 +
\sqrt{\epsilon/2}\; q_1) \over (L/c)(\gamma_*-i\Omega)} e^{-i\Omega t} 
{d\Omega\over
2\pi} + {\rm HC}\;,
\label{deltaWcirc}
\end{equation}
where HC means Hermitian conjugate (adjoint) of the previous term.

\subsection{Mirror motion}

The circulating-power fluctuations (\ref{deltaWcirc}) produce a
fluctuating radiation-pressure (back-action) force 
\begin{equation}
F_{\rm BA} = 2 \delta W_{\rm circ}/c
\label{FBA}
\end{equation}
on each mirror.  This force is equal and opposite on the cavity's two mirrors
and, along with the gravitational waves, it produces the following acceleration
of the mirror separation:
\begin{equation}
{d^2 X(t)\over dt^2} = {1\over2} \eta_{ne} L {d^2 h(t)\over dt^2} + {4\delta
W_{\rm circ}(t) \over mc}\;.
\label{EOM}
\end{equation}
Here $h(t)$ is the gravitational-wave field (projected onto the
interferometer's arms), and $\eta_{ne}$ is $+1$ for the north arm and $-1$ for
the east arm (one arm is stretched while the other is squeezed).

Below we will need an expression for the (Fourier transform of the)
arm-length difference, $x=X_n - X_e$.  It can be obtained by Fourier
transforming the equation of motion (\ref{EOM}), solving for $X$ (i.e.,
$X_n$ or $X_e$), inserting expression (\ref{deltaWcirc}) for $\delta W_{\rm
circ}$, and then taking the difference of the north and east arms.  The result
is
\begin{equation}
x=Lh + x_{\rm BA}
\label{xhBAApp}
\end{equation}
[cf.\ Eq.\ (\ref{xhBA1})], where 
\begin{eqnarray}
x_{\rm BA} &=& {-4\sqrt{2I_o\hbar\omega_o}\;
(a_1 + \sqrt{\epsilon/2}\; n_1)\over
m\Omega^2 L (\gamma_*-i\Omega)} \nonumber\\
&=& - \sqrt{{\cal K}_*/2}\; Lh_{\rm SQL} (a_1 + \sqrt{\epsilon/2}\; n_1)
e^{i\beta_*}\;,
\label{xBA}
\end{eqnarray}
Here we have introduced the quadrature amplitude for the difference of the
arms' noise fields 
\begin{equation}
n_j \equiv {q_j^n - q_j^e \over \sqrt2}\;
\label{njDef}
\end{equation}
and have used Eq.\ (\ref{SplitterJunction}) for $f_1^n$ and $f_1^e$, and 
Eqs.\ (\ref{KDefLossy}), 
(\ref{betaDefLossy}), (\ref{ISQLDef}) 
and (\ref{hSQLDef})
for the coupling constant ${\cal K}_*$, the phase $\beta_*$, 
the SQL power $I_{\rm SQL}$ and the standard
quantum limit $h_{\rm SQL}$. 

Below we shall also need the following expression for the difference
of the two arms' sideband fields produced by the mirror motions' coupling to
the carrier:  
\begin{equation}
{\delta k_2^n - \delta k_2^e \over \sqrt2} = {2\sqrt2\over \sqrt{T}}
\sqrt{I_o\over\hbar\omega_o} {\omega_o x\over c}\;.
\label{Deltadeltakj}
\end{equation}
This follows from Eqs.\ (\ref{deltak}), (\ref{IoD}) and (\ref{xDef})

\subsection{Cavity output}
The field exiting from the (north or east) cavity is obtained by combining
Eqs.\ (\ref{FrontJunction}), (\ref{BackJunction}) and 
(\ref{jj}): 
\begin{equation}
g_j = {\sqrt{\tilde R} e^{2i\Omega L/c} - \sqrt{R} \over 1-\sqrt{R\tilde R}
e^{2i\Omega L/c}} f_j + 
{\left( \sqrt{T\tilde T} q_j + \sqrt{T}\delta k_j\right)e^{i\Omega L/c} \over 
1-\sqrt{R\tilde R} e^{2i\Omega L/c}}\;.
\label{gj}
\end{equation} 
Inserting Eq.\ (\ref{ResonantApproximation}) for the denominator and
analogous expressions for the numerator, and discarding terms that are higher
order than linear in the losses, we bring Eq.\ (\ref{gj}) into the form 
\begin{equation}
g_j = \left(1-{1\over2}{\cal E}\right)e^{2i\beta} f_j + \sqrt{\cal E}
e^{i\beta} q_j + \sqrt{(c/2L)^2 T\over \gamma_*^2+\Omega^2} e^{i\beta_*} \delta
k_j\;,
\label{gj1}
\end{equation}
where $\beta_*$ is given by Eq.\ (\ref{betaDefLossy}). 

\subsection{Beam splitter output}

By combining Eqs.\ (\ref{SplitterJunction}), (\ref{gj1}), and (\ref{njDef}), we
obtain for the dark-port output of the beam splitter
\begin{eqnarray}
b_j &=& \left( 1-{1\over2}{\cal E}
\right) a_j e^{2i\beta} + \sqrt{\cal E} n_1 e^{i\beta} \nonumber\\
&&+
\sqrt{(c/2L)^2 T \over \gamma_*^2+\Omega^2} \left( {\delta k_j^n - \delta
k_j^e\over\sqrt2}\right)e^{i\beta_*}\;.
\label{bj}
\end{eqnarray}
Inserting $\delta k_1^{n,e} = 0$ [Eq.\ (\ref{deltak})] and our expression 
(\ref{Deltadeltakj}) for the difference of the $\delta k_2$'s, and inserting
Eqs.\ (\ref{xhBAApp}) for $x$
and (\ref{KDefLossy}), (\ref{hSQLDef}), (\ref{gammaStarDef})
for ${\cal K}_*$, $h_{\rm SQL}$, $\gamma_*$,
we obtain 
for the output fields:  
\begin{mathletters}
\label{bjApp}
\begin{eqnarray}
b_1 &=& \left( 1-{1\over2}{\cal E} \right) a_1 e^{2i\beta} + \sqrt{\cal E} n_1
e^{i\beta}\;, \\ 
b_2 &=& \left( 1-{1\over2}{\cal E} \right) a_2 e^{2i\beta} + \sqrt{\cal E} n_2
e^{i\beta}\nonumber\\
&&+ \sqrt{2{\cal K}_*} \left({h+ x_{\rm BA}/L \over h_{\rm SQL}}\right)
e^{i\beta_*}\;.
\end{eqnarray} 
\end{mathletters}
By inserting expression (\ref{xBA}) for the back-action-induced mirror
displacement $x_{\rm BA}$,
we obtain the input-output relations
quoted in the text:
Eqs.\ (\ref{bjFromajLossy}) and (\ref{DeltabjLossy}) with losses, and
Eqs.\ (\ref{bjFromaj}) in the lossless limit

\section{Filter Parameters}
\label{app:filters}

In our discussion of FD homodyne detection 
[Sec.\ \ref{sec:FDHomodyne}], 
we derived the following requirement for the conventional homodyne phase 
$\theta$ and the filter parameters $\xi_J$ and $\delta_J$ (with $J={\rm I}$ 
and II):
\begin{eqnarray}
\tan\Phi(\Omega) &\equiv&
{\Omega^2(\gamma^2 + \Omega^2)\over \Lambda^4} \nonumber\\
&=& \tan \left( \theta - {\alpha_{{\rm I}+} + \alpha_{{\rm I}_-}
+ \alpha_{{\rm II}+} + \alpha_{{\rm II}_-}\over 2} \right)
\label{FilterRequirement}
\end{eqnarray}
[Eqs.\ (\ref{FilterParameterszeta}),   
(\ref{alphapmIandII}), and (\ref{PhiDesired})], where
\begin{equation}
\alpha_{J \pm} = \arctan(\xi_J \pm \Omega/\delta_J)\;
\label{alphaJpm'}
\end{equation}
[Eq.\ (\ref{alphaJpm})].
In this appendix,
we shall show that this requirement is satisfied by the parameter choices
asserted in the text: Eqs.\ 
(\ref{theta}) and (\ref{FilterParameters}).

We initially regard the parameters $\theta$, $\xi_J$ and $\delta_J$ as unknown.
By inserting Eq.\ (\ref{alphaJpm'}) into Eq.\ 
(\ref{FilterRequirement}) and invoking
some trigonometric identities, we obtain the requirement
\begin{eqnarray} && {(R_0 - I_0 \cot\theta) 
+ (R_2 - I_2\cot\theta)\Omega^2 + R_4\;\Omega^4 \over
(R_0\cot\theta + I_0) + (R_2\cot\theta +I_2)\Omega^2 + R_4\cot\theta\;\Omega^4}
\nonumber\\
&&= {\gamma^2\Omega^2 + \Omega^4 \over \Lambda^4}\;. 
\label{Requirement}
\end{eqnarray}
Here $R_0+ R_2\Omega^2 + R_4\Omega^4$ is the real part and 
$I_0+I_2\Omega^2$ is the imaginary part of 
$(1+i\tan\alpha_{{\rm I}+})(1+i\tan\alpha_{{\rm I}-})(1+i\tan\alpha_{{\rm II}+})
(1+i\tan\alpha_{{\rm II}-})$.
More specifically,
\begin{mathletters}
\label{RnIn}
\begin{eqnarray}
R_0 &=& 1-\xi_{\rm I}^2 - \xi_{\rm II}^2 
- 4 \xi_{\rm I} \xi_{\rm II} + \xi_{\rm I}^2 \xi_{\rm II}^2\;,
\label{R0}\\
R_2 &=& ({1-\xi_{\rm I}^2)/\delta_{\rm II}^2} 
+ ({1-\xi_{\rm II}^2)/ \delta_{\rm I}^2}\;,
\label{R2}\\
R_4 &=& 1/(\delta_{\rm I}^2 \delta_{\rm II}^2)\;,
\label{R4}\\
I_0 &=& 2(\xi_{\rm I} + \xi_{\rm II}) (1- \xi_{\rm I}\xi_{\rm II})\;,
\label{I0}\\
I_2 &=& 2\xi_{\rm II}/\delta_{\rm I}^2 
+ 2\xi_{\rm I}/\delta_{\rm II}^2\;. 
\label{I2}
\end{eqnarray}
\end{mathletters}

To get rid of the $\Omega^4$ term in the denominator of Eq.\
(\ref{Requirement}), we must set
\begin{equation}
\theta = \pi/2\;, \quad {\rm so } \cot\theta=0\;.
\label{theta'}
\end{equation}
(We cannot set $R_4=0$ since that would require an infinite bandwidth for one
or both of the filters.)  To get rid of the $\Omega^2$ term in the denominator
and the constant term in the numerator, and to make the $\Omega^2$ and
$\Omega^4$ terms in the numerator have the correct coefficients, we must set
\begin{mathletters}
\label{RnInset}
\begin{eqnarray}
I_2 &=& 0\;,
\label{I2set}\\
R_0 &=& 0\;,
\label{R0set}\\
R_2^2/(I_0 R_4) &=& \gamma^4/\Lambda^4 \equiv 4/P \;,
\label{R2I0R4set}\\
R_2/R_4 &=& \gamma^2\;.
\label{R2R4set}
\end{eqnarray}
\end{mathletters}
Here we have used definition (\ref{PQ}) of the constant $P$.

Equations (\ref{RnInset}) are four equations for the four unkown filter
parameters: the fractional frequency offsets $\xi_{\rm I}$, $\xi_{\rm II}$
and the half bandwidths $\delta_{\rm I}$, $\delta_{\rm II}$.  In
the next four paragraphs we shall explore the consequences of these four
equations, arriving finally at the solution (\ref{FilterParameters})
for $\xi_{\rm I}$, $\xi_{\rm II}$,
$\delta_{\rm I}$ and $\delta_{\rm II}$ given in the text.

Equation (\ref{I2set}) implies that
\begin{equation}
\delta_{\rm I}^2 / \delta_{\rm II}^2 = - \xi_{\rm II}/\xi_{\rm I}\;.
\label{d1overd2}
\end{equation}

Equation (\ref{R0set}) implies that
$(1-\xi_{\rm I} \xi_{\rm II})^2 = (\xi_{\rm I} + \xi_{\rm II})^2$. 
It turns out that one of the frequency offsets is positive and the other
is negative (cf.\ Fig.\ \ref{fig:Fig10}); 
we choose $\xi_{\rm I}$ to be the positive one.  It also
turns out that $\xi_{\rm I} + \xi_{\rm II}$ is positive
(cf.\ Fig.\ \ref{fig:Fig10}).  Consequently, we
can take the square root of the above equation to obtain
\begin{equation}
1-\xi_{\rm I} \xi_{\rm II} = \xi_{\rm I} + \xi_{\rm II}\;,
\label{xiIxiIIRelation}
\end{equation}
which enables us to express the frequency offsets in terms of each other:
\begin{equation}
\xi_{\rm I} = {1-\xi_{\rm II} \over 1+ \xi_{\rm II}}\;, \quad
\xi_{\rm II} = {1-\xi_{\rm I} \over 1+ \xi_{\rm I}}\;.
\label{xiIxiIIRelation'}
\end{equation}

Equation (\ref{R2I0R4set}), when combined with Eqs.\ (\ref{d1overd2})
and (\ref{xiIxiIIRelation}), implies that
\begin{equation} {8\over P} = 
{ \left[ \sqrt{-\xi_{\rm II}\over\xi_{\rm I}} (1-\xi_{\rm I}^2)
+ \sqrt{-\xi_{\rm I}\over\xi_{\rm II}} (1-\xi_{\rm II}^2) \right]^2 \over
(\xi_{\rm I} + \xi_{\rm II})^2 }\;.
\label{chain1}
\end{equation}
We shall now combine this equation with Eqs.\ (\ref{xiIxiIIRelation'}) to
obtain Eqs.\ (\ref{FilterParameters}) for the frequency offsets 
$\xi_{\rm I}$ and $\xi_{\rm II}$ in terms of $P= 4\gamma^4/\Lambda^4$.
Our first step is to define $A_\pm$ by Eqs.\ (\ref{xiI}) and (\ref{xiII}),
which are equivalent to 
\begin{equation}
A_+ \equiv {\xi_{\rm I}\over \xi_{\rm I}^2 -1}\;, \quad 
A_- \equiv {\xi_{\rm II} \over \xi_{\rm II}^2 -1}\;.
\label{ApAmDef}
\end{equation}
Note that the relation (\ref{xiIxiIIRelation'}) between $\xi_{\rm I}$ and
$\xi_{\rm II}$ is equivalent to
\begin{equation}
4A_+ A_- = 1\;.
\label{4ApAm}
\end{equation}
By using Eqs.\ (\ref{xiIxiIIRelation'}), (\ref{ApAmDef}) and (\ref{4ApAm}),
we can reexpress the right side of Eq.\ (\ref{chain1}) solely in terms of
$A_+$: 
\begin{equation}
{8\over P} = {(4 A_+^2 -1)^2 \over A_+(4A_+^2+1)}\;.
\label{chain2}
\end{equation}
It is convenient to define $Q$ by Eqs.\ (\ref{ApAm}), which are equivalent to
\begin{equation}
A_+ + A_- \equiv 2Q/P\;.
\label{QDef}
\end{equation}
Using Eqs.\ (\ref{4ApAm}) and (\ref{QDef}), we can rewrite (\ref{chain2})
in terms of $Q$ instead of $A_+$:
\begin{equation} {2\over P} = {2Q\over P} - {P\over2Q}\;,
\label{chain3}
\end{equation}
which can be solved for $Q$ as a function of $P$
\begin{equation}
Q= {1 + \sqrt{1+P^2}\over2}\;.
\label{QofP}
\end{equation}
This is the relation asserted in the text, Eq.\ (\ref{PQ}), and it completes
our derivation of Eqs.\ (\ref{PQ})--(\ref{xiII}) for the frequency offsets
$\xi_{\rm I}$ and $\xi_{\rm II}$ in terms of $P$.

Turn, finally, to the consequences of Eq.\ (\ref{R2R4set}), which says
\begin{equation}
\gamma^2 = \delta_{\rm I}^2 (1-\xi_{\rm I}^2)
+ \delta_{\rm II}^2 (1-\xi_{\rm II}^2)\;.
\label{chain5}
\end{equation}
By eliminating $\delta_{\rm II}$ with the aid of Eq.\ (\ref{d1overd2}), 
we obtain
\begin{equation}
\gamma^2 = \delta_{\rm I}^2 \xi_{\rm I} \left( {1-\xi_{\rm I}^2 \over \xi_{\rm
I}} - {1-\xi_{\rm II}^2 \over \xi_{\rm II}} \right)\;.
\label{chain6}
\end{equation}
Using Eqs.\ (\ref{ApAmDef}), 
(\ref{4ApAm}), and (\ref{ApAm}), we can rewrite this as
\begin{equation}
{\delta_{\rm I}\over\gamma} = \sqrt{P\over 8\xi_1\sqrt{Q}}\;,
\label{d2'}
\end{equation}
which is the formula for the half bandwidth $\delta_{\rm I}$ given
in the text, Eq.\ (\ref{deltaI}).  The corresponding formula for $\delta_{\rm
II}$, Eq.\ (\ref{deltaII}), follows directly from Eqs.\ (\ref{d2'}) and 
(\ref{d1overd2}).


\begin{references}
\bibitem{Nonlinearity} Equation 
(\ref{ShNL}) can be derived by the techniques 
used in Ref.\ \protect\cite{VM2}; 
see especially Eqs.\ (13)--(15) of \cite{VM2}.
\bibitem{TowardSQL}
The experimental challenge simply to reach the SQL for mirror position is 
daunting.  Two recent experiments
that addressed this challenge with small masses but
were impeded by thermal
noise are reported 
in Refs.\ \protect\cite{Heidmann} and
\protect\cite{Schiller}.
\end{references}
\end{document}